\documentclass[a4paper,11pt]{article}
\pdfoutput=1 \usepackage{amssymb}
\usepackage{amsmath}
\usepackage{braket}
\usepackage{mathtools}
\usepackage[usenames,dvipsnames,svgnames,table]{xcolor}
\usepackage[utf8]{inputenc}
\usepackage{color}
\usepackage{subcaption}
\usepackage{lmodern}
\usepackage{footnote}
\usepackage[normalem]{ulem}
\usepackage{glossaries-extra}
\setabbreviationstyle[acronym]{long-short}
\glssetcategoryattribute{acronym}{nohyperfirst}{true}

\usepackage{slashed}
\usepackage[pdftex]{graphicx}
\usepackage{multirow}
\usepackage{jheppub}
\usepackage{here}
\usepackage{hyperref}
\usepackage{verbatim}
\usepackage[justification=centering,singlelinecheck=false]{caption}
\usepackage{cleveref}
\usepackage{listings}
\usepackage{fancyvrb}
\usepackage{dsfont}
\usepackage{nicefrac,xfrac}

\Crefname{equation}{eq.}{eqs.}
\Crefname{section}{section}{sections}
\Crefname{figure}{figure}{figures}
\Crefname{appendix}{appendix}{appendices}

\newcommand{\cA}{\mathcal{A}}
\newcommand{\cB}{\mathcal{B}}

\newcommand{\cD}[0]{\mathcal D}
\newcommand{\cF}{\mathcal{F}}

\newcommand{\cK}[0]{\mathcal K}
\newcommand{\cL}[0]{\mathcal L}
\newcommand{\cM}[0]{\mathcal M}
\newcommand{\cO}[0]{\mathcal O}
\newcommand{\cR}[0]{\mathcal R}
\newcommand{\cS}[0]{\mathcal S}
\newcommand{\cT}[0]{\mathcal T}
\newcommand{\cY}[0]{\mathcal Y}

\newcommand{\df}[0]{\mathrm{df}}

\newcommand{\PV}[0]{{\mathrm{PV}}}

\newcommand{\bm}[0]{\boldsymbol}

\newcommand{\bcX}{\boldsymbol{\mathcal X}}
\newcommand{\bcY}[0]{\boldsymbol{\mathcal Y}}

\newcommand{\XL}[3]{\boldsymbol{\mathcal X}_{[kab]}^{(#1#2#3)\dagger}}
\newcommand{\XR}[3]{\boldsymbol{\mathcal X}_{[kab]}^{(#1#2#3)}}
\newcommand{\YL}[3]{\boldsymbol{\mathcal Y}^{[kab] \dagger}_{(#1#2#3)}}

\newcommand{\KSS}[0]{Kim:2005gf}

\newcommand{\HSQCa}[0]{Hansen:2014eka}
\newcommand{\HSQCb}[0]{Hansen:2015zga}

\newcommand{\BHSQC}[0]{Briceno:2017tce}

\newcommand{\largera}[0]{Romero-Lopez:2019qrt}

\newcommand{\HHanal}[0]{Blanton:2019vdk}
\newcommand{\isospin}[0]{Hansen:2020zhy}

\newcommand{\implement}[0]{Blanton:2021eyf}

\newcommand{\tetraquark}[0]{Hansen:2024ffk}
\newcommand{\Npp}[0]{Hansen:2025oag}
\newcommand{\multiplechannel}[0]{Draper:2024qeh}
\newcommand{\BSQC}[0]{Blanton:2020gha}

\newcommand{\BSnondegen}[0]{Blanton:2020gmf}
\newcommand{\BStwoplusone}[0]{Blanton:2021mih}

\newcommand{\Akakia}[0]{Hammer:2017uqm}
\newcommand{\Akakib}[0]{Hammer:2017kms}

\newcommand{\MD}[0]{Mai:2017bge}

\newcommand{\Maiaone}[0]{Mai:2021nul}
\newcommand{\HSrev}[0]{Hansen:2019nir}

\newcommand{\MDRrev}[0]{Mai:2021lwb}

\newacronym{CMF}{CMF}{center-of-momentum frame}

\DeclareFixedFont{\ttb}{T1}{txtt}{bx}{n}{9}
\DeclareFixedFont{\ttm}{T1}{txtt}{m}{n}{9}

\definecolor{deepblue}{rgb}{0,0,0.5}
\definecolor{deepred}{rgb}{0.6,0,0}
\definecolor{deepgreen}{rgb}{0,0.5,0}
\definecolor{jlab_red}{RGB}{192,39,45}
\definecolor{jlab_orange}{RGB}{249,102,0}
\definecolor{jlab_blue}{RGB}{47,122,121}
\definecolor{jlab_green}{RGB}{65,125,10}
\definecolor{jlab_blue}{RGB}{47,122,121}

\title{Accessing the Wess-Zumino-Witten term using Lattice QCD}

\author[a]{Fernando Romero-L\'opez}
\affiliation[a]{Albert Einstein Center, Institute for Theoretical Physics, University of Bern, 3012 Bern, Switzerland}

\author[b]{\!, Stephen R. Sharpe}
\affiliation[b]{Physics Department, University of Washington, Seattle, WA 98195-1560, USA}

\emailAdd{fernando.romero-lopez@unibe.ch}
\emailAdd{srsharpe@uw.edu}

\abstract{The Wess-Zumino-Witten interaction in chiral perturbation theory is a manifestation of chiral anomalies. 
One of its consequences is the presence of a nonvanishing $K \overline K \to \pi^+ \pi^0 \pi^-$ amplitude.
We derive the finite-volume formalism that allows one to access this, and related, amplitudes, using the spectra of two and three particles in finite volumes,
spectra that can be calculated using lattice QCD.
Specifically, we present the formalism for the systems $K \overline K + 3 \pi$ with $I=0$ and
$\pi K + \pi\pi K$ with $I=1/2$ and $3/2$.
In addition, we determine the threshold expansions for the $2\leftrightarrow 3$ and $3\leftrightarrow 3$ K matrices that enter the formalism, and calculate the leading order predictions from chiral perturbation theory for the coefficients that enter these expansions.
}
\allowdisplaybreaks

\begin{document}
\today
\maketitle
\flushbottom
\clearpage

\section{Introduction}
\label{sec:intro}

The presence of chiral anomalies~\cite{Adler:1969gk, Bell:1969ts,Bardeen:1969md} allows a class of transitions that would otherwise be forbidden. 
The best known of these is the $\pi^0\to\gamma\gamma$ process, which is the dominant decay mode of the neutral pion~\cite{ParticleDataGroup:2020ssz}. 
Other such processes occur purely within quantum chromodynamics (QCD). 
In chiral perturbation theory (ChPT), these arise from the Wess--Zumino--Witten (WZW) term~\cite{Wess:1971yu,Witten:1983tw}; see Ref.~\cite{Bijnens:1993xi} for a review. Examples include the transitions
$\pi^+ K^+ \to \pi^+\pi^0 K^+$ and $K^+K^- \to \pi^+\pi^0\pi^-$.

Lattice QCD (LQCD) enables nonperturbative calculations of hadronic transition and scattering amplitudes at low energies. While significant progress has been made toward determining certain anomalous processes, such as $\pi^0\to\gamma\gamma$~\cite{Feng:2012ck,Meyer:2013dxa,ExtendedTwistedMass:2023hin,Gerardin:2023naa}, no comparable progress has been made for anomalous processes involving only the strong interaction. Our aim here is to develop the formalism necessary to extract such two-to-three hadron scattering amplitudes (``$2\to3$'' amplitudes) in the isospin-symmetric limit using LQCD. Since LQCD simulations are currently restricted to Euclidean spacetime, an indirect method is required. This is provided by the finite-volume formalism, originally developed for elastic two-hadron scattering~\cite{Luscher:1986pf,Luscher:1991n1,Luscher:1991n2} and subsequently generalized in many ways, most recently to a variety of $3\to3$ amplitudes~\cite{\HSQCa,\HSQCb,\BHSQC,\Akakia,\Akakib,\MD,\isospin,\BSQC,\BSnondegen,\BStwoplusone,\Maiaone,Muller:2021uur,Jackura:2022gib,\tetraquark,\multiplechannel,Severt:2022jtg,Feng:2024wyg,Dawid:2023kxu,Jackura:2023qtp,Briceno:2024txg,Dawid:2023jrj,Sakthivasan:2026bph,\Npp,Alotaibi:2025pxz,Jackura:2025wbw}; see Refs.~\cite{\HSrev,\MDRrev,Romero-Lopez:2021zdo,Sharpe:2026mtt,Briceno:2024ehy} for recent reviews, and Refs.~\cite{Hansen:2020otl,Dawid:2025doq,Dawid:2025zxc,Briceno:2025yuq,Alharazin:2026lno,Yan:2025mdm,Yan:2024gwp,Feng:2026ixm} for recent applications.

In order to extend the finite-volume formalism to accommodate these anomalous transitions, one must account for both two- and three-particle states. A first step was taken in Ref.~\cite{\BHSQC}, using the generic relativistic field-theory (RFT) approach of Refs.~\cite{\HSQCa,\HSQCb}, but only for identical spinless particles. A major complication was the presence of $1\leftrightarrow2$ vertices in the generic effective field theory (EFT). These required the introduction of transition functions (also called cutoff functions) ensuring that two- and three-particle states sharing a common particle could not go on shell at the same kinematic point. In the present applications, however, this issue is avoided because $1\leftrightarrow2$ vertices are forbidden by the flavor and parity quantum numbers. This greatly simplifies the derivation. Indeed, we can use the formalism developed for three distinguishable particles in Refs.~\cite{\BSnondegen,Hansen:2020zhy} and combine it straightforwardly with the two-particle approach developed in Ref.~\cite{\KSS}.

We develop the new formalism for three processes in which the WZW vertex contributes at the lowest order at which it appears in ChPT, namely next-to-leading order (NLO): $\pi K\leftrightarrow \pi\pi K$ in the $I=3/2$ and $I=1/2$ channels, and $K\bar K\leftrightarrow3\pi$ in the $I=0$ channel. The $I=3/2$ channel is nonresonant and may therefore provide a particularly clean window into the WZW vertex. The other two channels are resonant: the $I=1/2$ channel contains $K^\ast$ states with $J^P=1^-$, while the $I=0$ channel contains the $\omega(782)$ and $\phi(1020)$ resonances. The present formalism allows resonances with both two- and three-meson decay modes to be studied rigorously, at least over a range of quark masses. Such resonant channels may, however, complicate the extraction of the WZW contribution. One might also consider $K\bar K\leftrightarrow3\pi$ in the $I=1$ channel. In this case, however, mixing with $\eta\pi$ complicates the analysis, and the WZW vertex first contributes only at next-to-next-to-leading order in ChPT. We therefore do not develop the formalism for this process.

This paper is organized as follows. We begin in \Cref{sec:prelim} with an overview of the three processes described above. In \Cref{sec:res}, we present the main results of this work: the finite-volume quantization conditions and corresponding integral equations for each of the three processes. Technical details and derivations are relegated to appendices, as described below. We conclude and provide an outlook in \Cref{sec:conc}.

The appendices contain definitions, derivations, and results needed for practical applications of the formalism. Relevant kinematic definitions are collected in \Cref{app:kinematics}. 
The subsequent two appendices derive the building blocks for the main derivation.
In \Cref{app:KKTOPT} we present a derivation of the two-particle formalism for the $K\overline K$ system with $I^G=0^-$.
While the result is well known, the derivation here is new as it uses time-ordered perturbation theory (TOPT)---an approach that we use in all derivations.
Similarly, in \Cref{app:TOPT}, we derive the formalism for three pions with $I=0$ using a TOPT-based approach, confirming earlier results using Feynman diagrams.
We then turn, in \Cref{app:deriv}, to the derivation of the formalism for $K\overline K\leftrightarrow3\pi$.
We do not provide details of the derivation of the results for the $\pi K\leftrightarrow \pi\pi K$ systems, since they follow by a straightforward generalization of those for $K\overline K\leftrightarrow3\pi$,
and, furthermore, many flavor factors can be taken from a recent analysis of the $\pi\pi N$ system~\cite{\Npp}.
In practical applications, the K matrices entering the formalism require parametrizations constrained by symmetries, which can be developed as threshold expansions. We present these expansions for all three processes in \Cref{app:threshold}. To guide future applications, we have also determined the corresponding coefficients at leading nontrivial order in chiral perturbation theory (ChPT); these results are collected in \Cref{app:chpt}.

\section{Preliminary comments for processes of interest}
\label{sec:prelim}

The WZW vertex involves fields with different flavors, and thus gives rise to the processes $K^+ K^- \leftrightarrow \pi^+ \pi^0 \pi^-$ and $K^0 \overline K^0 \leftrightarrow \pi^+ \pi^0 \pi^-$, as well as those obtained by crossing.
It leads to amplitudes that are completely antisymmetric under momentum exchanges.
Crossing leads to $1\leftrightarrow 4$ and $0 \leftrightarrow 5$ processes,
but these are not kinematically allowed for on-shell particles.
Thus we consider only $2\leftrightarrow 3$ processes, of which there are three types: $\pi K\leftrightarrow \pi\pi K$,  $K\overline K \leftrightarrow \pi\pi\pi$, and $\pi\pi \leftrightarrow \pi K\overline K $.
We focus on the first two, as they involve the smallest mass differences between the initial and final states. 
As noted above, both appear at NLO in ChPT, examples being $[ \pi K]_{I=3/2} \leftrightarrow \pi \pi K$, $[ \pi K]_{I=1/2} \leftrightarrow  \pi \pi K$,
and $[K \bar K]_{I=0} \leftrightarrow \pi^+ \pi^0 \pi^-$.
We consider these processes in turn, discussing the quantum numbers, kinematical constraints,
and the relative difficulty of performing LQCD calculations.

\subsection{$\pi K \leftrightarrow \pi \pi K$ with $I=3/2$}
\label{sec:KpKpp}

We choose $m=3/2$ for definiteness.
The $\pi K$ state is then $\pi^+(\bm k_1) K^+(\bm k_2)$,  while there are two independent flavor structures for $\pi\pi K$,
\begin{align}
\begin{split}
[[\pi\pi]_{2}K]_{3/2} &=
\sqrt{\frac45} \pi^+(\bm p_1) \pi^+(\bm p_2) K^0(\bm p_3)
\\
& \qquad - \sqrt{\frac1{10}} \left[ \pi^+(\bm p_1) \pi^0(\bm p_2) K^+(\bm p_3)
+ \pi^0(\bm p_1) \pi^+(\bm p_2) K^+(\bm p_3)
\right]
\end{split}
\label{eq:ppK3S}
\\
[[\pi\pi]_{1} K]_{3/2} &=
\sqrt{\frac1{2}} \Big[ \pi ^+(\bm p_1) \pi^0(\bm p_2) K^+(\bm p_3)
- \pi^0(\bm p_1) \pi^+(\bm p_2) K^+(\bm p_3)
\Big]\,,
\label{eq:ppK3A}
\end{align}
which are, respectively, symmetric and antisymmetric under pion exchange.
The subscripts on the left-hand side indicate total isospin, while the momentum labels are included for future use---note, in particular, that the kaon is chosen to have momentum $\bm p_3$.
The presence of these two structures implies that the three-particle part of the final quantization condition has an additional flavor label,
as will be made explicit in \Cref{sec:QCKpKpp3}.

\begin{figure}[h!]
\centering
\includegraphics[width=0.9\textwidth]{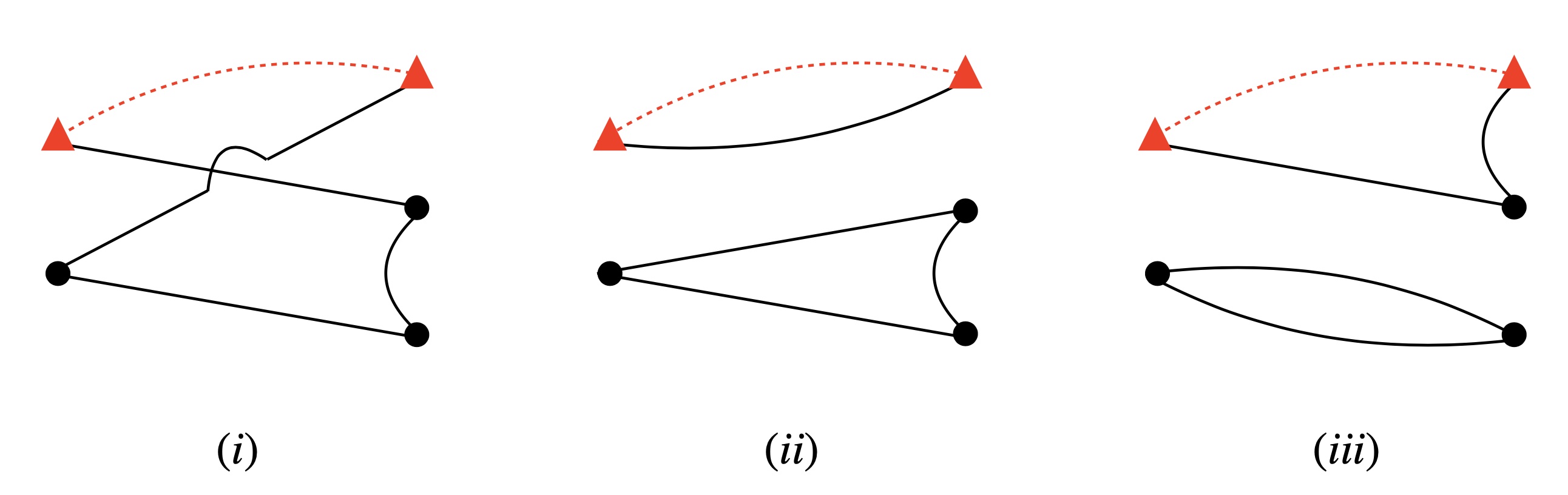}
\caption{Different topologies of quark contractions contributing to the
processes of interest. 
For $K\pi\leftrightarrow K \pi\pi(I=3/2)$ time varies horizontally, for
$K\overline{K} \leftrightarrow 3\pi$ time varies vertically,
while for $K\pi\leftrightarrow K \pi\pi(I=1/2)$ both directions are possible.
Dashed red lines represent strange quark propagators, while solid black lines are light (up or down) quark propagators.
Pions are represented by black circles; kaons by red triangles.
}
\label{fig:qcontract}
\end{figure}

In a LQCD calculation, the number and type of quark contractions are important factors in the complexity of the work.
The topologies that enter in the present case are shown in \Cref{fig:qcontract}.
All three appear in the process $K^+\pi^+ \leftrightarrow K^+\pi^+\pi^0$, while only the first and third are present for
$K^+\pi^+ \leftrightarrow K^0 \pi^+\pi^+$.
One needs ``all-to-all'' propagators for both topologies, which can be obtained using, e.g., distillation~\cite{HadronSpectrum:2009krc,Morningstar:2011ka}.
For the two-quark loop (``disconnected'') topologies, we stress that they are disconnected only in the $t$- or $u$-channels,
and thus should be relatively straightforward to calculate.

As can be seen from the second and third topologies in \Cref{fig:qcontract}, there are quark loops that involve three external particles. One might, therefore, be concerned that there would be three-particle vertices in an effective field theory of kaons and pions, in the present instance the processes $\pi \leftrightarrow 2\pi$ and $K\leftrightarrow K \pi$.
As noted in the introduction, such vertices would greatly complicate the derivation of the formalism.
Fortunately, however, both of these vertices are absent, the former due to G parity, which is exactly conserved since we work in the isospin limit, and the latter due to angular momentum and parity.
Nevertheless, the quark-disconnected contractions shown in the figure can be nonzero because of the presence of the second quark loop and the implicit gluon interactions.

We are interested in choices of overall spin-parity that can couple to the WZW term.
While $K^+\pi^+$ states can have any natural spin-parity, i.e $J^P=0^+$, $1^-$, $2^+, \dots$, the antisymmetry of the WZW vertex implies a coupling only to the odd $J$ channels, beginning with $J^P=1^-$.
The $\pi^+\pi^0 K^+$ states can have all values of $J^P$ except $0^+$, so the channel with the lowest $J$ for which mixing can occur has $J^P=1^-$.
One way to obtain a $J^P=1^-$ $\pi^+\pi^0 K^+$ state is to have
the $K\pi$ pair in the resonant $1^-$ $K^*$ subchannel, with this pair being in a $p$-wave relative to the remaining pion.

Since LQCD calculations are done in finite (usually cubic) boxes, and with total momentum $\bm P$ possibly nonzero, one loses full rotation invariance, and (in a moving frame)  symmetry under parity transformations. This means that one cannot project onto a single $J^P$ value, but rather onto an irreducible representation (irrep) of the appropriate subgroup of the cubic group.
Each subgroup irrep is subduced from an infinite tower of irreps of the full rotation group.
In order to access the WZW vertex, we need to choose irreps that are subduced from the $J^P=1^-$ channel, ideally as the lowest $J$ contributor.
The simplest examples are the $T_{1u}$ irrep in the rest frame,
which is subduced from $J^P=1^-, 3^-, 4^-,\dots$,
and the $E$ irrep in a moving frame with $\bm P \propto (0,0,1)$,
which is subduced from rotation-group states with helicities $\lambda = 1, 3,\dots$.

The formalism to be described below considers states in finite volume with total momentum $\bm P$ and energy $E$.
Throughout this work,
the corresponding center-of-momentum-frame (CMF) energy is denoted 
$E^\star = \sqrt{E^2 -\bm P^2}$.
The formalism for $\pi K\leftrightarrow \pi\pi K$ is valid only for 
\begin{equation}
\sqrt{M_K^2-M_\pi^2} <  E^\star < M_K + 3 M_\pi\,.
\label{eq:EstarrangeKp}
\end{equation}
This range is determined
such that the only kinematically-allowed on-shell intermediate states
are $K\pi$ and $K\pi\pi$.
The upper limit is set by the $\pi\pi\pi K$ threshold, while
the lower limit is set by the left-hand cut in $K\pi$ interactions due to two-pion exchange, below which the two-particle formalism breaks down.
The formalism is thus valid for a significant range both for physical  
and heavier-than-physical quark masses.

\subsection{$[\pi K]_{I=1/2} \leftrightarrow [ \pi \pi K]_{I=1/2}$}
\label{sec:KpKppI1}

Choosing $m=1/2$ for definiteness, the $\pi K$ state is
\begin{equation}
[\pi K]_{1/2}  = \sqrt{\frac23}  \pi^+ (\bm k_1)K^0(\bm k_2) - \sqrt{\frac13} \pi^0 (\bm k_1) K^+(\bm k_2)\,,
\label{sec:Kp1}
\end{equation}
while those for $\pi\pi K$ are (keeping momentum labels $\bm p_1$, $\bm p_2$, and $\bm p_3$ implicit)
\begin{align}
[[\pi\pi]_{0} K ]_{1/2} & = \sqrt{\frac13} \left( \pi^+\pi^- K^+
+ \pi^-\pi^+ K^+ + \pi^0\pi^0 K^+ \right)
\,,
\label{eq:ppK1S}
    \\
[[\pi\pi]_{1} K ]_{1/2} & =   
\sqrt{\frac13}\left( \pi^+\pi^0 K^0 -\pi^0\pi^+ K^0 \right)
- \sqrt{\frac16}\left(\pi^+\pi^- K^+ - \pi^-\pi^+ K^+ \right)
\,,
\label{eq:ppK1A}
\end{align}
which are again symmetric and antisymmetric, respectively.

The topologies of the contributing quark contractions are as for the $I=3/2$ channel,
with the only difference being that, in some cases of the third topology, the disconnection is in the $s$ channel, making the calculation more challenging.
This happens when the two pions in the $K\pi\pi$ state are contracted together.

As for the $I=3/2$ case, the lowest $J$ that can couple to the WZW vertex has $J^P=1^-$,
and to couple to this channel the same finite-volume irreps as discussed in the previous section are needed. Similarly, the absence of $1\to 2$ vertices carries over from the $I=3/2$ system.

The range of applicability of the formalism is the same as for $I=3/2$, \Cref{eq:EstarrangeKp}, aside from one caveat.
This concerns the lower limit in \Cref{eq:EstarrangeKp}. In general, the quantum numbers allow mixing with a single kaon state, which moves the lower limit up to $M_K < E^\star$.
However, if one chooses a finite-volume irrep that is not subduced from the quantum numbers of the kaon, $J^P=0^-$, then the lower limit is unchanged. 

The $I=1/2$ $\pi K+\pi\pi K$ channel contains a sequence of $K^*$ resonances in nature.
The lowest one, $K^*(892)$, decays primarily to $\pi K$.
The next resonance, the $K^*(1410)$, has been observed to decay to both channels.
This resonance does lie well above the $K\pi\pi\pi$ threshold, and thus beyond the range of strict applicability, although in practice the coupling to this four-particle channel may be weak.

\subsection{$[K \bar K]_{I=0} \leftrightarrow \pi^+ \pi^0 \pi^-$}
\label{sec:KKpppI0}

The WZW vertex couples to a fully antisymmetric three-pion state, 
which thus has $I^G=0^-$.
It consists of the flavor combination 
\begin{equation}
[[\pi \pi]_1 \pi]_0 = \cA \left\{ \pi^+(\bm p_1) \pi^0 (\bm p_2) \pi^-(\bm p_3)\right\}\,
\end{equation}
where the operator $\cA$ antisymmetrizes over the six momentum orderings, see \Cref{eq:ASstate}.
The antisymmetry implies no mixing with the $3\pi^0$ state.
It couples only to the $I=0$ component of the $K\bar K$ state, 
\begin{equation}
[K \overline K]_0 = \sqrt{\frac12}
\left[ K^+(\bm k_1) K^-(\bm k_2) + K^0 (\bm k_1) \bar K^0 (\bm k_2)\right]\,,
\end{equation}
Here we are using the sign convention in which the isodoublets of kaons are
$(K^+,K^0)$ and $(-\bar K^0, K^-)$, with the minus sign arising from the fact that
isodoublets at the quark level are
$(u,d)$ and $(-\bar d, \bar u)$.
The negative $G$ parity part of this state (which has $C=-$) is projected onto by
considering states that are antisymmetric under $\bm k_1 \leftrightarrow \bm k_2$ exchange.
Details are described in \Cref{app:deriv,app:TOPT}; we stress here that $G$ parity remains an exact symmetry in finite-volume. 
In infinite volume, the antisymmetry implies that the allowed values of $J^P$ are
$1^{-}, 3^{-}, \dots$.
These are also allowed for a three-pion state with $I=0$.
Thus we expect that the leading channel contributing to mixing is that with $J^P=1^-$,
which is the channel of the $\omega$ and $\phi$.
The appropriate finite-volume irreps to use to pick out this channel are the same 
as discussed for $\pi K\leftrightarrow \pi\pi K$ in \Cref{sec:KpKpp}.

All three topologies of the quark contractions shown in \Cref{fig:qcontract} also contribute here. The quark disconnections in diagrams (b) and (c) both occur in the $s$ channel, and will thus be more challenging to calculate.

For this process, the relevant $1\to 2$ transition would be $K\to 2 \pi$, or its charge conjugate, both of which are forbidden by strangeness conservation.

We turn finally to the kinematic range of applicability of the finite-volume formalism.
It must be the case that only $K\bar K$ and $3\pi$ on-shell states are allowed.
Although mixing with $4\pi, 6\pi, \dots$ is forbidden by $G$ parity,
mixing with $\pi K \overline K$, $\pi\pi\pi\eta$, and $5\pi$ states is allowed.
For physical quark masses, the five pion channel lies below the $K\overline K$ threshold, 
and the $\pi\pi\pi\eta$ channel lies close to it, so that, strictly speaking, the three-particle formalism is insufficient.
However, the coupling to both channels, as well as to $\pi K\overline K$ may be weak.
In particular, for the $5\pi$ channel, the lightest $I^G=0^-$ resonance observed to decay to this channel is the $\omega(1420)$.
We also note that the lightest $I^G(J^P)=0^-(1^-)$ resonance that decays to $\pi K\overline K$ is the $\phi(1680)$, suggesting a weak coupling to this channel.

In order to have strict applicability of the formalism, we need to work with unphysical quark masses such that the particle masses satisfy
\begin{equation}    
    {\rm max}(2M_K, 3M_\pi) < E^\star_{\rm max} = 
    {\rm min}(M_\pi + 2 M_K , 3 M_\pi+M_\eta, 5 M_\pi)\,.
\end{equation}
If this is satisfied, then the strict range of applicability is
\begin{equation}
 {\rm max}(2\sqrt{M_K^2 -  M_\pi^2},M_\pi) < E^\star < E^\star_{\rm max}\,.
    \label{eq:EstarrangeKKbar}
\end{equation}
The two quantities on the left-hand side of this inequality arise, respectively, from the left-hand cut in $K\overline K$ scattering due to $t$-channel two-pion exchange, and the corresponding left-hand cut in a pion pair in the $3\pi$ system.
As an example of a set-up in which there is a substantial window of applicability,
we consider the N200 CLS ensemble~\cite{Bruno:2014jqa}, for which $M_\pi=280\;$MeV, $M_K=460\;$MeV, and $M_\eta = 540\;$MeV~\cite{Bali:2021qem}.
The relevant kinematical quantities are then (in MeV)
\begin{equation}
\begin{gathered}
M_\pi=280\,,\ \ 2\sqrt{M_K^2 - M_\pi^2}= 730\,,\ \
3 M_\pi = 840 \,, \ \ 2 M_K = 920\,, 
\\ 
M_\pi + 2 M_K  = 1200\,, \ \
3 M_\pi + M_\eta = 1380\,, \ \ 5 M_\pi = 1400\,.
\end{gathered}\end{equation}
so that the allowed kinematical range is $730\;{\rm MeV} < E^\star < 1200\;$MeV.

\section{Results for quantization conditions and integral equations}
\label{sec:res}

\subsection{Overview}
\begin{figure}[h!]
\centering
\includegraphics[width=0.9\textwidth]{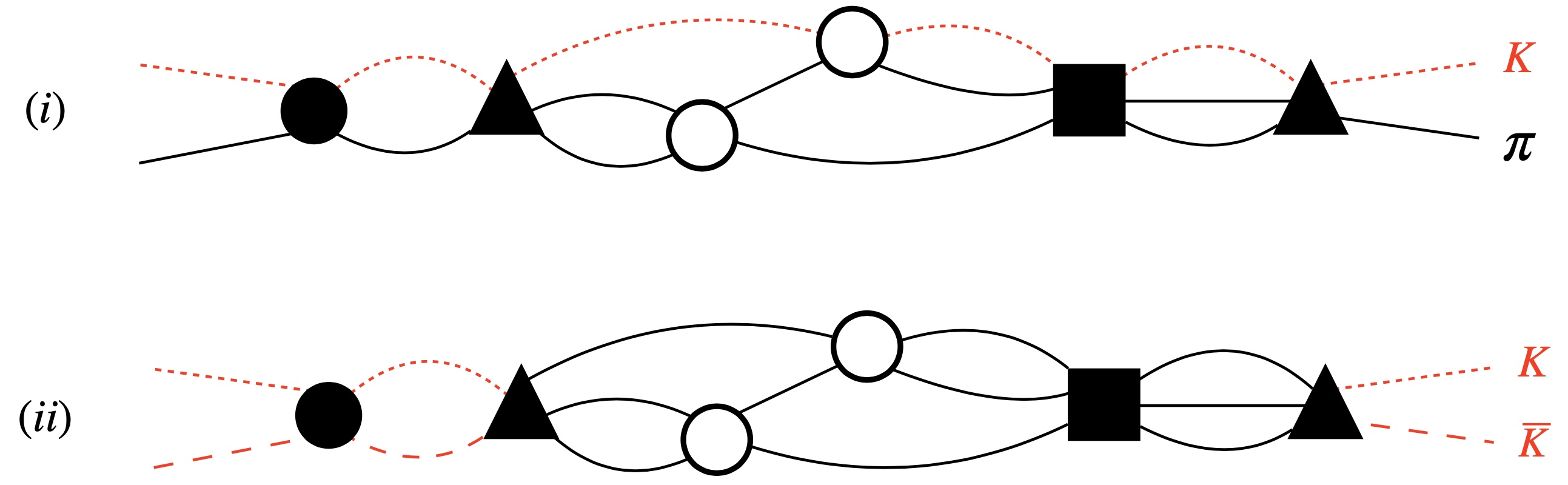}
\caption{Examples of contributions to correlation functions in a generic relativistic EFT: (i) for the $K\pi + K\pi\pi$ system; (ii) for the $K\overline K + 3\pi$ system.
Lines are fully-dressed propagators (normalized to have residues of free propagators): solid black lines are pions, long-dashed red lines are antikaons, and short-dashed red lines are kaons. Circles are $2\to 2$ Bethe-Salpeter kernels, with the filled ones being three-particle irreducible (3PI) in the $s$-channel, while the empty ones are 2PI.
Filled squares are 3PI $3\to 3$  kernels, while filled triangles are 3PI $2\to 3$ and $3\to 2$ kernels. The diagrams can be viewed as skeleton diagrams in either TOPT or relativistic PT.
}
\label{fig:corrfcn}
\end{figure}

The general strategy of the derivation is the same for all the processes we consider.
As noted in the introduction, the key simplification compared to the $2\leftrightarrow 3$ derivation of ref.~\cite{\BHSQC} is the absence of $1\leftrightarrow 2$ vertices in the EFT.
This implies that, when considering a  correlation function,
the two- and three-particle $s$-channel cuts lie in separate segments, connected by $2\to 3$
or $3\to 2$ kernels. This is illustrated in \Cref{fig:corrfcn}, which shows examples of skeleton diagrams contributing to correlators in which all two- and three-particle cuts are kept explicit. The upshot of this separation is that one can obtain the desired formalism by a straightforward combination of previous results for the two- and three-particle sectors.
This is explained in detail in \Cref{app:deriv} for the $K\bar K\leftrightarrow 3\pi$ process.
The generalization to the other processes is straightforward and we present only the final results.

The formalism consists of two parts: quantization conditions relating the finite-volume spectrum to (unphysical, but infinite volume) K matrices, and integral equations relating the K matrices to scattering amplitudes.

The quantization conditions for all three systems have the form\footnote{%
For the sake of brevity, we drop the subscript ``df'' on $\cK$, which has been used for the three-particle K matrix in the RFT literature since Ref.~\cite{\HSQCa}.
}
\begin{equation}
\det\left(\widehat {\bm 1} + \widehat \cF\, \widehat \cK \right) = 0 \,.
\label{eq:QC23}
\end{equation}
This general structure is familiar from previous RFT works, 
and, in particular, agrees with the form of the result for a system of $2+3$ scalars~\cite{\BHSQC}.
The ``hats'' over the objects in \Cref{eq:QC23} indicate a matrix structure in which there are blocks corresponding to two- and three-particle intermediate states.
The details differ according to the system, however, and we discuss them separately in the following. 
We also provide explicit expressions for the integral equations for all three systems.

\subsection{Results for $I=0$ $K\overline K \leftrightarrow 3\pi$}
\label{sec:QCKKppp}

For this system, the block form of the matrices is 
\begin{align}
    \widehat \cF = \begin{pmatrix} F^{(K\bar K)}_2 & 0 \\ 0 & F^{(3\pi)}_3 \end{pmatrix}\,,
    \quad 
\widehat \cK = \begin{pmatrix}
\cK_{22}^{K\bar K,I^G=0^-} & \cK_{23}^{K\bar K, 3\pi, I=0}
\\
\cK_{32}^{3\pi,K\bar K, I=0} & \cK_{33}^{3\pi,I=0}
\end{pmatrix}\,.
\label{eq:cFhatKdfhat}
\end{align}
The quantities in each of the blocks are themselves matrices.
The two-particle indices are $\{\ell,m\}$, corresponding to the angular momentum in the CMF. This is the standard choice for two-particle quantization conditions.
Similarly, indices in the three-particle sector are the standard $\{k\ell m\}$,
where $k$ denotes the spectator momentum $\bm k$, while $\{\ell, m\}$ describe the angular momentum of the non-spectator pair in its CMF.
$F_2$ and $F_3$ are matrices that appear separately in two- and three-particle quantization conditions. 
The definition of $F_2$ is given in \Cref{eq:F2def},
while that for $F_3$, given in \Cref{eq:F3res}, is repeated here in a more explicit notation, 
\begin{equation}
F^{(3\pi)}_3 = \frac{F^{(\pi,\pi\pi)}}6 - 
\frac{F^{(\pi,\pi\pi)}}2 \frac1{(\cK_{2,L}^{\pi,(\pi\pi,I=1)})^{-1} + \tfrac12 F^{(\pi,\pi\pi)} - G^{(\pi\pi,\pi)}} \frac{F^{(\pi,\pi\pi)}}2\,.
\label{eq:F3res3pi}
\end{equation}
The kinematic, finite-volume matrices $F$ and $G$ are defined, respectively, in \Cref{eq:Fdef,eq:Gdef}, while $\cK_{2,L}^{\pi,(\pi\pi,I=1)}$ is given in \Cref{eq:K2Lres}. 
The superscripts on these quantities indicate how the three particles are divided into a spectator and pair.
The result \Cref{eq:F3res3pi} has the standard RFT form for $F_3$, except for the minus sign multiplying the $G$ term.
The latter is a consequence of the antisymmetry of the $I=0$ three-pion state, as first noted in ref.~\cite{\isospin}.

The K matrices in $\widehat \cK$, \Cref{eq:cFhatKdfhat}, have subscripts indicating which blocks they connect.
The diagonal entries appear in the separate two- and three-particle formalisms---see \Cref{app:KKTOPT,app:TOPT}, respectively.
The new feature here, following ref.~\cite{\BHSQC}, is the presence of the offdiagonal elements $\cK_{23}$ and $\cK_{32}$, that connect the two- and three-particle sectors.
The K matrices describe interactions that do not involve on shell $K\overline K$ and $3\pi$ intermediate states. Given their construction, as outlined in \Cref{app:twotothreeAS}, $\widehat \cK$ is hermitian.

In \Cref{eq:cFhatKdfhat,eq:F3res3pi}, the presence of a cutoff function is implicit and defines the scheme of the K matrices. The cutoff function also renders all matrices entering the quantization condition finite-dimensional in the spectator-momentum index, $k$. Specifically, the cutoff function can be defined as
\begin{equation}
    H^{(i)}(\boldsymbol{p})=J\left(z_i(\boldsymbol{p})\right),
    \qquad
    z_i(\boldsymbol{p})=
    \left(1+\epsilon_H\right)
    \frac{\sigma_i(\boldsymbol{p})-\sigma_i^{\min}}
    {\sigma_i^{\mathrm{th}}-\sigma_i^{\min}}\,,
    \label{eq:cutoffdef}
\end{equation}
where $J(x)$ is a function as defined in eq. (2.23) of Ref.~\cite{Blanton:2021eyf}, $\epsilon_H$ is a small positive number, $\sigma_i(\boldsymbol{p})=(P-p_i)^2$ is the invariant mass squared of the pair, and $\sigma_i^{\mathrm{th}}$ is its value at the two-particle threshold. The cutoff function vanishes when $\sigma_i(\boldsymbol{p})<\sigma_i^{\min}$ and is unity when $\sigma_i(\boldsymbol{p})>\sigma_i^{\mathrm{th}}$. Different choices of $\sigma_i^{\min}$ are allowed, but it must lie above the first singularity of the two-to-two subprocess that is not accounted for explicitly; for example, $\sigma_\pi^{\min}>0$ for the $\pi\pi$ subsystem in the $I=0$ $3\pi$ case,
corresponding to the onset of the left-hand cut in $\pi\pi$ scattering.

Because the K matrices depend on the choice of cutoff functions, they are  unphysical. Nevertheless, they are Lorentz invariant, and satisfy (when expressed in momentum variables) complete antisymmetry with respect to interchanges of their momentum arguments. Aside from these properties, they are the dynamical quantities that one aims to determine by applying the quantization condition to the finite-volume spectrum.

We now describe the integral equations that remove the cutoff dependence by
relating the K matrices to the corresponding matrix of physical scattering amplitudes.  Many of their features carry over directly from the three-particle literature (see, in particular, Refs.~\cite{\BSnondegen,Draper:2023xvu,Dawid:2024dgy}), while those containing $2\leftrightarrow 3$ transitions are obtained by a minor generalization of the results of Ref.~\cite{Briceno:2017tce}.

The matrix of scattering amplitudes is 
\begin{equation}
\widehat \cM \equiv   
\begin{pmatrix} \cM_{22}(\bm k_1', \bm k_2'; \bm k_1,\bm k_2) 
& \cM_{23}(\bm k_1', \bm k_2'; \bm p_1, \bm p_2, \bm p_3)
\\
\cM_{32}(\bm p'_1, \bm p'_2, \bm p'_3; \bm k_1, \bm k_2) 
& \cM_{33}(\bm p'_1,\bm p'_2, \bm p'_3;\bm p_1, \bm p_2, \bm p_3)
\end{pmatrix}
\label{eq:Mhat}
\end{equation}
where we use notation for external momenta described in \Cref{app:kinematics}.
Here, and in the rest of the description of the integral equations, we are dropping superscripts, except in cases where they are needed to avoid ambiguity. They can be reinstated by matching the pattern in $\widehat\cK$, \Cref{eq:cFhatKdfhat}.
It follows from \Cref{eq:inteqfin,eq:Dhatfin,eq:Mhatdf3fin} that only $\cM_{33}$ contains divergences for physical momenta, and that these can be isolated into a ladder amplitude $\overline \cD$,
\begin{equation}
\cM_{33}(\bm p'_1,\bm p'_2, \bm p'_3;\bm p_1, \bm p_2, \bm p_3)
= 
\overline\cD(\bm p_1',\bm p'_2, \bm p'_3; \bm p_1, \bm p_2, \bm p_3) 
+ \cM_{\rm df,33}(\bm p'_1,\bm p'_2, \bm p'_3;\bm p_1, \bm p_2, \bm p_3)\,.
\label{eq:M33decomp}
\end{equation}
In the following, we express $\cM_{22}$, $\cM_{23}$, $\cM_{32}$, $\overline \cD$,
and $\cM_{\df,33}$ in terms of the K matrices.

The integral equations contain, rather than the amplitudes in \Cref{eq:Mhat}, asymmetric amplitudes in which one of the particles is singled out as a spectator, and the remaining pair are decomposed into spherical harmonics in their rest frame. 
This mirrors the form of the matrices entering the quantization conditions.
However, unlike these matrices, which have discrete momentum indices, the amplitudes in the integral equations depend on continuous (infinite-volume) momenta.
The angular momentum indices do, however, remain. These will be kept implicit in the integral equations below, with matrix multiplication implied for products.
The relationship between the asymmetric amplitudes and the desired physical amplitudes will be described below.

We are now ready to introduce the building blocks of the integral equations.
First we observe that the infinite-volume limit of $F_3^{(3\pi)}$ contains the physical $I=1$ two-pion scattering amplitude, which arises from a combination of $\cK_{2,L}^{\pi, (\pi\pi,I=1)}$ and $F^{(\pi,\pi\pi)}$ in \Cref{eq:F3res3pi}.
The amplitude depends on the spectator momentum $\bm p$, and is given by
\begin{equation}
\left[    \cM_2^{\pi\pi}(\bm p)\right]_{\ell' m'; \ell m}
= \delta_{\ell' \ell} \delta_{m' m} \cM_{2,\ell}^{\pi\pi, I=1}(q^\star(\bm p))\,.
\label{eq:M2pipiI1def}
\end{equation}
Here $q^\star(\bm p)$, the momentum of each member of the scattering pair in the pair rest frame, is given by \Cref{eq:qstardef} with flavor indices dropped and all masses set to $M_\pi$,
while $\cM_{2,\ell}$ is the partial-wave amplitude.

Another two-particle building block is $\cK_{22}^{K\bar K,I^G=0^-}$, the $22$ entry of $\widehat \cK$, which we shorten to $\cK_{22}^{K\bar K}$ in the following.
This is, by construction, an infinite-volume quantity.
We stress, however, that it is not directly related to the physical $K\overline K$ scattering amplitude since contributions from intermediate three pion states have been excluded as part of its definition.
Nevertheless, it is Lorentz invariant and admits a partial-wave expansion. 
Since it does not involve a spectator, it only carries angular-momentum indices,
and can be written
\begin{equation}
\left[    \cK_{22}^{K\bar K} \right]_{\ell' m'; \ell m}
= \delta_{\ell' \ell} \delta_{m' m} \cK_{22,\ell}^{K\bar K, I^G=0^-}(q^\star)\,.
\label{eq:K2KKbdef}
\end{equation}
Here $q^\star$ is the momentum of both kaons in the overall CMF,
and is given by \Cref{eq:qstar2} with both masses set to $M_K$.
We also note that, due to the negative G parity, $\cK_{22,\ell}^{K\bar K, I^G=0^-}$
vanishes for even values of $\ell$.

The other elements of $\widehat\cK$ in \Cref{eq:cFhatKdfhat} do involve one or two spectators, and their infinite volume forms are
$\cK_{23}(\bm k)_{\ell' m'; \ell m}\,,  \
\cK_{32}(\bm p)_{\ell' m';\ell m}\,,  \
\cK_{33}(\bm p,\bm k)_{\ell' m';\ell m}\,$.
We use the same symbols for the finite- and infinite-volume quantities,
with the difference determined by the presence or absence of momentum arguments.

The first integral equation is that for the asymmetric ladder amplitude, whose symmetric form is $\overline \cD$ in \Cref{eq:M33decomp}.
The infinite volume limit of \Cref{eq:DLuures} leads to 
\begin{equation}
    \cD(\bm p,\bm k) = \cM^{\pi\pi}_{2}(\bm p) G_\infty(\bm p, \bm k) \cM^{\pi\pi}_2(\bm k)
    + \cM^{\pi\pi}_2(\bm p) \int_{ r} G_\infty(\bm p, \bm r) \cD(\bm r, \bm k)\,,
    \label{eq:Dinfty}
\end{equation}
where the integral measure is relativistic, 
\begin{equation}
\int_{ r} \equiv \int \frac{d^3r}{(2\pi)^3 2\omega_r} \,.
\label{eq:measure}
\end{equation}
and $G_\infty$ is defined in \Cref{eq:Ginfty}. We note that integral equation \Cref{eq:Dinfty}, while of the standard RFT form,
differs from that which appears for bosons
(see, e.g. Ref.~\cite{\BSnondegen}) because here $G$ comes with a minus sign.

With $\cD$ in hand, we can define the left and right decoration operators, 
\begin{align}
    \cL(\bm p, \bm k) &= 
    \left[\frac13 - \cM_2^{\pi\pi}(\bm p) \rho_H(\bm p)\right] \overline\delta(\bm p-\bm k)
    - \cD(\bm p, \bm k) \rho_H(\bm k)\,,
    \label{eq:cLdef}
    \\
    \cR(\bm p, \bm k) &=
    \overline\delta(\bm p-\bm k)\left[\frac13 -\rho_H(\bm k) \cM_2^{\pi\pi}(\bm k)\right]
    - \rho_H(\bm p) \cD(\bm p, \bm k)\,,
    \label{eq:cRdef}
\end{align}
where $\overline \delta$ and $\rho_H$ are defined in \Cref{eq:deltabar,eq:rhoH}.

The core integral equations can now be stated. Following Ref.~\cite{Briceno:2017tce}, we introduce the intermediate quantities $\cT_{22}$, $\cT_{23}$, etc., which have the same index structure and momentum dependence as the elements of $\widehat \cK$.
These satisfy a set of nested equations, beginning with an integral equation for $\cT_{33}$,
\begin{align}
    \cT_{33}(\bm p, \bm k) &= \mathcal V(\bm p, \bm k) - 
    \int_r \int_s \mathcal V(\bm p, \bm r) \rho_H(\bm r) \cL(\bm r, \bm s) \cT_{33}(\bm s, \bm k)\,,
        \label{eq:T33res}
    \\
    \mathcal V(\bm p, \bm k) &= \cK_{33}(\bm p, \bm k) - 
    \cK_{32}(\bm p) \rho_2 \frac1{1 + \cK_{22}^{K\bar K} \rho_2} \cK_{23}(\bm k)\,,
    \label{eq:V33res}
\end{align}
where the two-body phase space factor $\rho_2$  is defined in \Cref{eq:rho2}.

The next equations in the chain are  for $\cT_{23}$ and $\cT_{32}$
\begin{align}
    \cT_{23}(\bm k) = \frac1{1 + \cK_{22}^{K\bar K}  \rho_2}
 \left[\cK_{23}(\bm k)  - \int_{r}  \int_s \cK_{23}(\bm r) \rho_H(\bm r) \cL(\bm r, \bm s)
 \cT_{33}(\bm s, \bm k)
    \right]\,, 
    \label{eq:T23res}
    \\
    \cT_{32}(\bm p) = 
 \left[\cK_{32}(\bm p)  - \int_{r}  \int_s 
 \cT_{33}(\bm p, \bm s) \cR(\bm s, \bm r) \rho_H(\bm r)\cK_{32}(\bm r)
    \right] \frac1{1 + \rho_2\cK_{22}^{K\bar K} } \,.
    \label{eq:T32res}
\end{align}
The last equation then yields $\cT_{22}$,
\begin{align}
    \cT_{22} = \frac1{1+\cK_{22}^{K\bar K} \rho_2} \left[
    \cK_{22}^{K\bar K} - \int_r \int_s \cK_{23}(\bm r) \rho_H(\bm r) \cL(\bm r, \bm s) \cT_{32}(\bm s)
    \right]\,.
    \label{eq:T22res}
\end{align}

The connection between $\cT$ and the scattering amplitudes involves three steps.
First, one must apply an external ``decoration'' of three-particle sides of the amplitudes, which adds final-state interactions due to two-particle scattering.
Second, one must convert from $\ell m$ indices to dependence on momenta, which is accomplished by the action of the operators $\bcX^\sigma$ and $\bcX_2$, defined in \Cref{eq:X2def,eq:XRdef}, and their conjugates.
Third, for the three-particle sides of amplitudes, one must sum over choices of spectator.
This is implemented by the operators $\bcX_A$ and $\bcX^\dagger_A$, defined in \Cref{eq:XAdef}.
The final results are
\begin{align}
    \cM_{22} &= \bcX_2 \circ \cT_{22} \circ \bcX_2^\dagger\,,
    \label{eq:M22final}
    \\
    \cM_{23} &= \bcX_2 \circ \int_r \cT_{23}(\bm r) \cL(\bm r, \bm k) \circ \bcX^\dagger_A\,,
    \label{eq:M23final}
    \\
    \cM_{32} &= \bcX_A \circ \int_r \cR(\bm p, \bm r) \cT_{32}(\bm r) \circ \bcX_2^\dagger\,,
    \label{eq:M32final}
    \\
    \cM_{\rm df,33} &= \bcX_A \circ \int_r \int_s \cR(\bm p, \bm r) \cT_{33}(\bm r, \bm s)
    \cL(\bm s, \bm k) \circ \bcX_A^\dagger \,,
    \label{eq:M33final}
\end{align}
where we leave the momentum dependence on the left-hand sides implicit.
In addition, we need the equation for $\overline \cD$,
\begin{align}
    \overline \cD &=  \bcX_A \circ \cD(\bm p, \bm k) \circ \bcX_A^\dagger\,.
    \label{eq:Dbarfinal}
\end{align}

The results above lead to scattering amplitudes that satisfy unitarity,
as shown in a simpler context in Ref.~\cite{Briceno:2019muc}.
They also allow one to see explicitly the relationship between the unphysical K matrix
$\cK_{22}^{K\bar K}$ and the $K \bar K $ scattering amplitude $\cM_{22}$, as we now explain.
First, we note that \Cref{eq:M22final} implies that $\cT_{22}$ is simply $\cM_{22}$  expressed in a different basis. Next, from \Cref{eq:T22res}, we observe that keeping only the first term in square brackets corresponds to the usual relation between K matrix and scattering amplitude, which we know is insufficient here due to the lack of $3\pi$ intermediate states in the K matrix. The second term in brackets corrects this, incorporating all $3\pi$ contributions. 
Similar discussions apply to the other K matrices.

In any practical application of the formalism, it will be necessary to parametrize the K matrices that appear in $\widehat \cK$ and $\cK_{2,L}^{\pi,(\pi\pi,I=1)}$
in a manner that is consistent with their symmetries. 
For $\cK_{2,L}^{\pi,(\pi\pi,I=1)}$, this is straightforward, because there are no constraints on the allowed intermediate states. The result is given in terms of phase shifts by \Cref{eq:K2Lres,eq:K2mod}, with only odd phase shifts contributing due to the antisymmetry of the $I=1$ $\pi\pi$ state.
The situation is more complicated for $\cK_{22}^{K\bar K,I^G=0^-}$.
A direct relation to the phase shift such as \Cref{eq:K22} is not valid, because the true relation involves integral equations as discussed above, see \Cref{eq:T22res}. However, one can still use a parametrization such as the effective range expansion (ERE), and it remains true that the projection onto negative $G$ parity implies that only odd partial waves contribute.

Turning to $\cK_{33}^{3\pi,I=0}$, a general parametrization in a threshold expansion is given in ref.~\cite{\isospin} (see also ref.~\cite{Baeza-Ballesteros:2024mii}), and this still holds in the presence of mixing with $K\bar K$, since it only depends on the symmetries and does not incorporate constraints from unitarity.
Parametrizations for the offdiagonal K matrices have not been considered previously, and we present the results in \Cref{app:threshold}.

An alternative approach is to use ChPT to determine the coefficients of the threshold expansion of the K matrices. This brings in the WZW term of ChPT, which provides a major motivation for this work. The lowest-order results are presented in \Cref{app:chpt}.

\subsection{Result for $I=3/2$ $\pi K\leftrightarrow \pi\pi K$}
\label{sec:QCKpKpp3}

The major change for this system compared to $K\bar K \leftrightarrow 3\pi$ is that the three-particle sector is enlarged by the addition of a flavor index.
This can take four values, a multiplicity that arises because there are two choices for the $\pi\pi K$ flavor wavefunction, as shown in \Cref{eq:ppK3S,eq:ppK3A}, for each of which there are two choices of spectator (kaon or pion).
The resulting flavor factors are nontrivial, but we can obtain most of them from those for the $\pi\pi N$ system (see Appendix A of ref.~\cite{\Npp}), since the presence of the nucleon spin has no impact on the flavor factors. 
The only new feature here is the presence of the offdiagonal K matrices.

The block forms of $\widehat \cF$ and $\widehat \cK$ change to 
\begin{equation}
    \widehat \cF = \begin{pmatrix} F^{(K\pi)}_2 & 0 \\ 0 & \widetilde F^{(K\pi\pi,I=3/2)}_3 \end{pmatrix}\,, \ \
\widehat \cK = \begin{pmatrix}
\cK_{22}^{K\pi,I=3/2} & \widetilde \cK_{23}^{K\pi, K\pi\pi, I=3/2}
\\
\widetilde \cK_{32}^{K\pi\pi,K\pi, I=3/2} & \widetilde \cK_{33}^{K\pi\pi,I=3/2}
\end{pmatrix}\,.
\label{eq:Kdfhatb}
\end{equation}
The tilde over several entries indicates that they are now matrices with the additional flavor index. 
Following eq.~(A.4) of ref.~\cite{\Npp}, the basis for this index space is 
\begin{equation}
 \left\{ [(\pi\pi)_2 K]_{3/2},\ [(\pi\pi)_1 K]_{3/2}, \
         [(K \pi)_{3/2} \pi]_{3/2},\ [(K\pi)_{1/2} \pi]_{3/2} \right\}\,,
    \label{eq:Kppbasis3}
\end{equation}
where the notation is as in \Cref{eq:ppK3S}, with the additional proviso that the third particle in each triplet is the spectator.
$F_2^{(K\pi)}$ is given by \Cref{eq:F2def}, while the $F_3$ term becomes
\begin{equation}
\widetilde F^{(K\pi\pi),I=3/2}_3 = \frac{\widetilde F^{(K\pi\pi)}}3 - 
{\widetilde F^{(K\pi\pi)}} \frac1{(\cK_{2,L}^{K\pi\pi,I=3/2})^{-1} + \widetilde F^{(K\pi\pi)} + \widetilde G^{(K\pi\pi),I=3/2}} {\widetilde F^{(K\pi\pi)}}\,,
\label{eq:F3resKpp3}
\end{equation}
where (see table~1 of ref.~\cite{\Npp})
\begin{align}
    \widetilde F^{(K\pi\pi)} &= {\rm diag}\left(F^{(K,\pi\pi)}, F^{(K,\pi\pi)}, F^{(\pi,K\pi)},F^{(\pi,K\pi)} \right)\,,
    \label{eq:FKpp}
    \\[0.5em]
    \widetilde G^{(K\pi\pi),I=3/2} &=
    \begin{pmatrix}
        0 & 0 & - \sqrt{\tfrac13} G^{(K\pi,\pi)} & - \sqrt{\tfrac53} G^{(K\pi,\pi)} 
        \\
        0 & 0 & - \sqrt{\tfrac53} G^{(K\pi,\pi)} &  \sqrt{\tfrac13} G^{(K\pi,\pi)} 
        \\
        - \sqrt{\tfrac13}G^{(\pi K,\pi)} & - \sqrt{\tfrac53}G^{(\pi K,\pi)} &
        - \tfrac23 P_\ell G^{(\pi\pi,K)} P_\ell &  \tfrac{\sqrt5}3 P_\ell G^{(\pi\pi,K)} P_\ell
        \\
        - \sqrt{\tfrac53}G^{(\pi K,\pi)} &  \sqrt{\tfrac13}G^{(\pi K,\pi)} &
        \tfrac{\sqrt5}3 P_\ell G^{(\pi\pi,K)} P_\ell & \tfrac23 P_\ell G^{(\pi\pi,K)} P_\ell
    \end{pmatrix}\,,
    \label{eq:GKpp3}
    \\[0.5em]
    \cK_{2,L}^{K\pi\pi,I=3/2} &= {\rm diag}\left(
    \tfrac12 \cK_{2,L}^{\pi\pi, I=2}, \tfrac12 \cK_{2,L}^{\pi\pi, I=1},
    \cK_{2,L}^{K\pi, I=3/2}, \cK_{2,L}^{K\pi,I=1/2}
    \right)\,,
    \label{eq:K2LKpp3}
\end{align}
with 
\begin{equation}
\left[P_\ell\right]_{k'\ell' m'; k\ell m} = \delta_{\bm k' \bm k} \delta_{\ell' \ell} \delta_{m' m} (-1)^\ell\,.
\label{eq:Pelldef}
\end{equation}
The  two-particle K matrices in \Cref{eq:K2LKpp3} can be expressed in terms of phase shifts as in \Cref{eq:K2Lres,eq:K2mod}. 
The kinematic factors $F^{(i,jk)}$ and $G^{(ij,k)}$ are, as above,
defined in \Cref{eq:Fdef,eq:Gdef}, respectively, although here we are making full use of the notational flexibility allowed by the superscripts.

We now turn to the integral equations, whose overall structure is the same as for $K\overline K\leftrightarrow 3\pi$.
The details differ, however, due to the addition of flavor factors.
In particular, all three-particle quantities now have a flavor index that takes four values.
We describe the resulting changes working through the equations in order.

The ladder amplitude, now a matrix in flavor space, satisfies a simple generalization of \Cref{eq:Dinfty}, 
\begin{equation}
    \widetilde \cD(\bm p,\bm k) = -\widetilde \cM_{2}(\bm p) 
    \widetilde G_\infty(\bm p, \bm k) 
    \widetilde \cM_2(\bm k)
    - \widetilde \cM_2(\bm p) 
    \int_{ r} \widetilde G_\infty(\bm p, \bm r) \widetilde \cD(\bm r, \bm k)\,,
    \label{eq:Dinfty3}
\end{equation}
where
\begin{equation}
    \widetilde \cM_2(\bm k) = {\rm diag}\left( 
    \tfrac12 \cM_2^{\pi\pi,I=2}(\bm k), \tfrac12 \cM_2^{\pi\pi,I=1}(\bm k), \cM_2^{K\pi,I=3/2}(\bm k),
    \cM_2^{K\pi, I=1/2}(\bm k)
    \right)
    \label{eq:M2tildeI3}
\end{equation}
with the individual entries defined analogously to \Cref{eq:M2pipiI1def}.
The matrix $\widetilde G_\infty(\bm p, \bm k)$ contains the same flavor factors as $\widetilde G^{(K\pi\pi),I=3/2}$,
\Cref{eq:GKpp3}, but the $G$ matrices are replaced by versions of $G_\infty(\bm p, \bm k)$, \Cref{eq:Ginfty}.
Specifically, $G_\infty^{(ij,k)}$ is obtained from $G^{(ij,k)}$ by dropping the $1/2\omega L^3$ factors on both ends.
$P_\ell$ can be implemented simply by multiplying by $(-1)^\ell$.
Finally, we note that $\widetilde G_\infty$ in \Cref{eq:Dinfty3} appears with opposite sign to $G_\infty$ in \Cref{eq:Dinfty}, corresponding to the fact that the three-particle $K\pi\pi$ amplitude is not completely antisymmetric. Some signs do occur, but these are explicitly contained in the components of the matrix $\widetilde G^{(K\pi\pi),I=3/2}$, \Cref{eq:GKpp3}.

The equations for the left and right decoration operators take the same form as \Cref{eq:cLdef,eq:cRdef},
except that all quantities are now flavor matrices and thus have a tilde.
Both $\widetilde \rho_H$ and $\widetilde \delta$ are diagonal matrices, with the former containing entries of the same form as in \Cref{eq:rhoH}, but with kinematical quantities appropriate to the choice of spectator, while the latter is proportional to the 4d identity matrix.

The equations for the $\cT_{ij}$, \Cref{eq:T33res,eq:V33res,eq:T23res,eq:T32res,eq:T22res}, also carry over, aside from the addition of tildes for each quantity involving a three-particle index.
The only other substantive change is that $\cK_{22}^{K \bar K}$ becomes $\cK_{22}^{K\pi, I=3/2}$.
The two-particle phase-space factor $\rho_2$ takes the same form, \Cref{eq:rho2}, except that $q^\star$ is now the CMF momentum appropriate to a $K\pi$ system. 

Finally, the equations for the scattering amplitudes and $\overline \cD$ become
\begin{align}
    \cM_{22} &= \bcX_2 \circ \cT_{22} \circ \bcX_2^\dagger\,,
    \label{eq:M22final3}
    \\
    \left[\cM_{23}\right]_y &= \bcX_2 \circ \int_r \widetilde \cT_{23}(\bm r) \widetilde \cL(\bm r, \bm k) \circ \bcX^{I=3/2\,\dagger}_y\,,
    \label{eq:M23final3}
    \\
    \left[ \cM_{32}\right]_x &= \bcX^{I=3/2}_x \circ \int_r \widetilde \cR(\bm p, \bm r) \widetilde \cT_{32}(\bm r) \circ \bcX_2^\dagger\,,
    \label{eq:M32final3}
    \\
    \left[\cM_{\rm df,33}\right]_{xy} &= \bcX^{I=3/2}_x \circ \int_r \int_s \widetilde \cR(\bm p, \bm r) \widetilde \cT_{33}(\bm r, \bm s)
    \widetilde \cL(\bm s, \bm k) \circ \bcX^{I=3/2\, \dagger}_y  \,,
    \label{eq:M33final3}
\\
    \left[{\overline \cD} \right]_{xy} &=  \bcX^{I=3/2}_x \circ \widetilde \cD(\bm p, \bm k) \circ \bcX^{I=3/2\, \dagger}_y\,.
    \label{eq:Dbarfinal3}
\end{align}
Here the labels $x,y$ can each take values $S$ or $A$, corresponding to choosing the external states to be symmetric [\Cref{eq:ppK3S}] or antisymmetric [\Cref{eq:ppK3A}] under pion exchange.
The operators that convert from the four-dimensional flavor space, \Cref{eq:Kppbasis3}, to the two external states are the same as for the $\pi\pi N$ system, and have been given in Ref.~\cite{\Npp}.
They are reproduced in \Cref{eq:X3S,eq:X3A}.

We next discuss the parametrization of the K matrices contained in $\widehat \cK$ in \Cref{eq:Kdfhatb}.
This is straightforward for $\cK_{22}^{K\pi,I=3/2}$, where one uses, in each partial wave, an analytic form such as the ERE. Since the scattering involves distinguishable particles, all partial waves contribute.
For the remaining K matrices, the parametrization is more complicated than in the $K\overline K \leftrightarrow 3\pi$ system, due to the additional flavor structure.
For $\widetilde \cK_{33}^{K\pi\pi,I=3/2}$ we can take over some of the results from refs.~\cite{\Npp,Draper:2024qeh}, while for the offdiagonal elements new results are needed.
Results for all three elements are collected in \Cref{app:threshold}. The corresponding forms at LO in ChPT are presented in \Cref{app:chpt:KpKpp3}.

We conclude this section by describing a subtle issue concerning the two different $\pi K$ scattering quantities that enter the formalism:
first, $\cK_{2,L}^{K\pi, I=3/2}$, which appears in \Cref{eq:K2LKpp3}, and, second,
$\cK_{22}^{K\pi,I=3/2}$, appearing in \Cref{eq:Kdfhatb}.
The former contains $\cK_2^{K\pi,I=3/2}$, which is algebraically related to the physical $\pi K$, $I=3/2$ scattering amplitude.
The latter, by contrast, is unphysical, as it does not account for intermediate $\pi\pi K$ states. This is analogous to the discussion concerning $\cK_2^{K\bar K, I^G=0^-}$ in the previous section.
However, in the present case, by solving the integral equations given above, one can, in principle, determine $\cK_{2}^{K\pi, I=3/2}$ in terms of $\cK_{22}^{K\pi,I=3/2}$ and the other K matrices.
This implies a complicated, nonlinear constraint on the parametrizations of the K matrices. 
In practice, however, following the discussion of \Cref{app:chpt}, the contribution of the $2\leftrightarrow 3$ K matrices may be small enough that one can treat them perturbatively, in which case a reasonable  approximation may be to set $\cK_{22}^{K\pi,I=3/2} \approx \cK_{2}^{K\pi, I=3/2}$.

\subsection{Result for $I=1/2$ $\pi K \leftrightarrow \pi\pi K$}
\label{sec:QCKpKpp1}

The results for this system are similar in form to those for $I=3/2$.
The flavor index takes four values, and the basis we use follows that of ref.~\cite{\Npp},
\begin{equation}
 \left\{ [(\pi\pi)_1 K]_{1/2},\ [(\pi\pi)_0 K]_{1/2}, \
         [(K \pi)_{3/2} \pi]_{1/2},\ [(K\pi)_{1/2} \pi]_{1/2} \right\}\,.
    \label{eq:Kppbasis1}
\end{equation}
The block forms of $\widehat \cF$ and $\widehat \cK$ become
\begin{equation}
    \widehat \cF = \begin{pmatrix} F^{(K\pi)}_2 & 0 \\ 0 & \widetilde F^{(K\pi\pi,I=1/2)}_3 \end{pmatrix}\,, \ \
\widehat \cK = \begin{pmatrix}
\cK_{22}^{K\pi,I=1/2} & \widetilde \cK_{23}^{K\pi, K\pi\pi, I=1/2}
\\
\widetilde \cK_{32}^{K\pi\pi,K\pi, I=1/2} & \widetilde \cK_{33}^{K\pi\pi,I=1/2}
\end{pmatrix}\,.
\label{eq:Kdfhatc}
\end{equation}
while $F_3$ becomes
\begin{equation}
\widetilde F^{(K\pi\pi),I=1/2}_3 = \frac{\widetilde F^{(K\pi\pi)}}3 - 
\widetilde F^{(K\pi\pi)} \frac1{(\cK_{2,L}^{K\pi\pi,I=1/2})^{-1} + \widetilde F^{(K\pi\pi)} + \widetilde G^{(K\pi\pi),I=1/2}} \widetilde F^{(K\pi\pi)}\,,
\label{eq:F3resKpp1}
\end{equation}
where $\widetilde F^{(K\pi\pi)}$ is unchanged from \Cref{eq:FKpp}, while
\begin{align}
    \widetilde G^{(K\pi\pi),I=1/2} &=
    \begin{pmatrix}
        0 & 0 &  \sqrt{\tfrac23} G^{(K\pi,\pi)} &  \sqrt{\tfrac43} G^{(K\pi,\pi)} 
        \\
        0 & 0 & \sqrt{\tfrac43} G^{(K\pi,\pi)} &  -\sqrt{\tfrac23} G^{(K\pi,\pi)} 
        \\
        \sqrt{\tfrac23}G^{(\pi K,\pi)} &  \sqrt{\tfrac43}G^{(\pi K,\pi)} &
        \tfrac13 P_\ell G^{(\pi\pi,K)} P_\ell &  -\tfrac{\sqrt8}3 P_\ell G^{(\pi\pi,K)} P_\ell
        \\
        \sqrt{\tfrac43}G^{(\pi K,\pi)} &  - \sqrt{\tfrac23}G^{(\pi K,\pi)} &
        -\tfrac{\sqrt8}3 P_\ell G^{(\pi\pi,K)} P_\ell & -\tfrac13 P_\ell G^{(\pi\pi,K)} P_\ell
    \end{pmatrix}\,,
    \label{eq:GKpp1}
    \\[0.5em]
    \cK_{2,L}^{K\pi\pi,I=1/2} &= {\rm diag}\left(
    \tfrac12 \cK_{2,L}^{\pi\pi, I=1}, \tfrac12 \cK_{2,L}^{\pi\pi, I=0},
    \cK_{2,L}^{K\pi, I=3/2}, \cK_{2,L}^{K\pi,I=1/2}
    \right)\,.
    \label{eq:K2LKpp1}
\end{align}
As for $I=3/2$, the K matrices in \Cref{eq:K2LKpp1} can be expressed in terms of phase shifts as in \Cref{eq:K2Lres,eq:K2mod}. 

The integral equations relating scattering amplitudes to K matrices take exactly the same form as for $I=3/2$. The differences lie in the definition of $\widetilde \cM_2$, which becomes
\begin{equation}
    \widetilde \cM_2(\bm k) = {\rm diag}\left(
    \tfrac12 \cM_2^{\pi\pi,I=1}(\bm k), \tfrac12 \cM_2^{\pi\pi,I=0}(\bm k), \cM_2^{K\pi,I=3/2}(\bm k),
    \cM_2^{K\pi,I=1/2}(\bm k)
    \right)
    \label{eq:M2tildeI1}
\end{equation}
in the flavor factors in $\widetilde G_\infty$, which are now the same as those in $\widetilde G^{(K\pi\pi), I=1/2}$, \Cref{eq:GKpp1}; in the replacement of $\cK_{22}^{K\pi,I=3/2}$ by $\cK_{22}^{K\pi,I=1/2}$ in the equations for the $\widetilde \cT_{ij}$; and in the replacement of $\bcX_i^{I=3/2}$ by $\bcX_i^{I=1/2}$ in \Cref{eq:M23final3,eq:M32final3,eq:M33final3,eq:Dbarfinal3}. The new operators are given in \Cref{eq:X1S,eq:X1A}.

The discussion in the previous section concerning the relation between the two different $K\pi$ scattering quantities appearing in the formalism carries over essentially verbatim to the $I=1/2$ system.
In particular, $\cK_{2,L}^{K\pi,I=1/2}$  contains $\cK_2^{K\pi, I=1/2}$,
which is physical and related to the phase shift, while $\cK_{22}^{K\pi,I=1/2}$ is unphysical and does not account for intermediate $\pi \pi K$ states. The former can be obtained by solving the integral equations if all other K matrices,
including $\cK_{22}^{K\pi,I=1/2}$, are known.
Furthermore, if $2 \to 3$ mixing is small, as predicted by ChPT, then a good approximation is to set $\cK_{22}^{K\pi,I=1/2} \approx \cK_2^{K\pi, I=1/2}$.

We close this section by noting that the parametrization of the K matrices is very similar to that for $I=3/2$, and is described in \Cref{app:threshold}. 
ChPT predictions are presented in \Cref{app:chpt:KpKpp1}.

\section{Conclusions and Outlook}
\label{sec:conc}

The chiral anomaly, represented in Chiral Perturbation Theory by the Wess-Zumino-Witten term, has important implications for the low-energy phenomenology of pseudo-Goldstone bosons, allowing processes that couple two- and three-particle states. In this work, we have developed the formalism needed to access these processes through lattice QCD calculations of coupled two-to-three scattering amplitudes, filling a gap in the toolkit for three-hadron spectroscopy.

Beyond its intrinsic interest, a formalism that accounts for $2\leftrightarrow3$ transitions also enables the study of certain QCD resonances, such as the $\phi(1020)$ and $K^\ast$ resonances. For certain choices of the quark masses, such as the CLS trajectory with $m_s+2m_l=\mathrm{const.}$~\cite{Bruno:2014jqa}, such mixing must be accounted for even for the $\omega(782)$ resonance.

Our results generalize those obtained previously
for a $2+3$ system composed of identical scalars~\cite{\BHSQC}.
For the systems we study, the two- and three-particle states involve nondegenerate particles, and the flavor quantum numbers lead to an additional index in the quantization conditions.
Furthermore, we have used TOPT to simplify the derivation.
As a side benefit, we have provided a TOPT derivation of the QC3 for three pions of any isospin, providing a check of the results of Ref.~\cite{\isospin}.

The next step is a numerical exploration of the quantization conditions and integral equations presented in this work. This includes efficiently constructing the matrices and projecting them onto irreps, expressing the K matrices in terms of the finite-volume $k\ell m$ variables in the quantization condition, performing the partial-wave projection of the integral equations, and finding parametrizations capable of describing the data. To facilitate the latter, we have provided generic threshold-expansion parametrizations of the K matrices and leading-order ChPT predictions for their coefficients.

A central practical challenge will be to solve the quantization conditions and integral equations for given parametrizations of the K matrices. 
In this regard, one must account for nonlinear constraints between the K matrices that arise from the integral equations, as discussed in \Cref{sec:QCKpKpp3,sec:QCKpKpp1}.
These constraints simplify in the limit that the  $2\leftrightarrow3$ coupling is small,
as suggested by ChPT. Whether this simplification is applicable will ultimately need to be determined by comparison with results from LQCD calculations.

The three systems considered in this work are amenable to lattice calculations in the near future. For instance, the nonresonant $I=3/2$ $K\pi\leftrightarrow K\pi\pi$ system could provide a particularly clean means of directly accessing the WZW contribution. Furthermore, resonances in the $I=1/2$ $K\pi\leftrightarrow K\pi\pi$ and $I=0$ $K\bar K\leftrightarrow\pi\pi\pi$ channels could be studied, extending the range of applicability of previous work in this direction~\cite{Yan:2024gwp}.

In summary, the present work opens a previously inaccessible corner of hadron spectroscopy to lattice QCD, enabling controlled studies of anomalous $2\leftrightarrow3$ transitions and the resonances to which they contribute.

\section*{Acknowledgements}
We gratefully acknowledge the contributions of Fabian M\"uller, who collaborated at an early stage in this work.

The work of FRL was supported in part by the Swiss National Science Foundation (SNSF) through grant No. 200021-236432. SRS acknowledges financial support through the U.S. Department of Energy Contract No. DE-SC0011637.  This work contributes to the goals of the USDOE ExoHad Topical Collaboration, contract DE-SC0023598.

This work was performed in part at Aspen Center for Physics, which is supported by National Science Foundation grant PHY-2210452. SRS thanks the Albert Einstein Center of the University of Bern for supporting a visit during which this work was completed.

\clearpage
\appendix

\section{Kinematic functions}
\label{app:kinematics}
In this appendix we collect standard results for the matrices that enter into the two- and three-particle formalism.
We need these both for the mass-degenerate $\pi^+\pi^0\pi^-$ and for $\pi\pi K$ systems.
Thus we quote a form that works in both cases.
We follow the conventions and notation of ref.~\cite{Hansen:2025oag}.
Throughout we assume a cubic box of side-length $L$.

\subsection{Quantization condition}

We begin with the two-particle $F$ function, also known as the L\"uscher zeta function~\cite{Luscher:1986n2}.
We assume two distinguishable particles labeled $i$ and $j$.
The result can be written
\begin{multline}
    \left[F_2^{(ij)}\right]_{\ell' m'; \ell m} 
    = \left[ \frac1{L^3} \sum_{\bm a} - \PV \int \frac{d^3 a}{(2\pi)^3} \right]
\\
\times \left[
\frac{\cY_{\ell' m'}^*(\bm a^{\star})}{q^{\star \ell'}}
\frac{h(a^\star)}{4\omega_i(\bm a) \omega_j(\bm b)
(E \!-\! \omega_i(\bm a) \!-\! \omega_j(\bm b))}
\frac{\cY_{\ell m}(\bm a^\star)}{q^{\star \ell}}
\right]
    \,.
    \label{eq:F2def}
\end{multline}
The momenta $\bm a$ and $\bm b= \bm P - \bm a$ are, respectively,
those of particles of types $i$ and $j$.
In the summation, $\bm a$ is summed over the finite-volume set.
The sum and the integral are regulated in the ultraviolet by the function $h(a^\star)$, a possible choice for which is presented in ref.~\cite{\Npp}.
The integral over the pole uses the PV prescription.
The superscript $\star$ indicates quantities associated with the CMF of the pair,
and should be distinguished from the superscript $*$ used for complex conjugation.
In particular, $q^\star$ is the relative momentum of each member of the pair in the CMF,
given by
\begin{equation}
    q^{\star 2} = \frac{\lambda(E^{\star 2},M_i^2,M_j^2)}{4 E^{\star 2}}\,,\quad
    E^{\star 2} = E^2 - \bm P^2\,,
    \label{eq:qstar2}
\end{equation}
with $\lambda(a,b,c)=a^2+b^2+c^2-2ab-2ac-2bc$ the standard triangle function.
The momentum $\bm a^\star$ is the spatial part of the four-vector $(\omega_i(\bm a),\bm a)$ boosted to the pair CMF.
Finally, harmonic polynomials are defined with nonstandard normalization,
\begin{equation}
    \cY_{\ell m} (\bm a) = \sqrt{4\pi} a^\ell Y_{\ell m}(\hat a)\,.
\end{equation}

We now turn to the three-particle $F$ function, 
which we define for a general system consisting of particles of types $i$, $j$, and $k$, 
\begin{align}
\left[F^{(i,jk)}\right]_{p'\ell'm';p\ell m}
&=
\delta_{\bm p'\bm p}\,
\frac{H^{(i)}(\bm p)}
     {2\omega_i(\bm p)L^3}
\left[
\frac{1}{L^3}\sum_{\bm a}
-\PV\int\frac{d^3a}{(2\pi)^3}
\right]
\nonumber\\
\times &
\left[
\frac{
\cY_{\ell'm'}^*\bigl(\bm a_i^\star(\bm p)\bigr)
}{
\bigl(q_i^\star(\bm p)\bigr)^{\ell'}
}
\frac{
h\bigl(\bm a_i^\star(\bm p)\bigr)
}{
4\omega_j(\bm a)\omega_k(\bm b)
\bigl(E-\omega_i(\bm p)-\omega_j(\bm a)-\omega_k(\bm b)\bigr)
}
\frac{
\cY_{\ell m}\bigl(\bm a_i^\star(\bm p)\bigr)
}{
\bigl(q_i^\star(\bm p)\bigr)^\ell
}
\right] .
\label{eq:Fdef}
\end{align}
The superscript on $F^{(i,jk)}$ indicates, first, that the spectator particle, which has momentum $\bm p$, is of type $i$, while $jk$ indicates the remaining pair. The first label $j$ designates the primary particle of the pair (momentum $\bm a$), 
and $k$ is the third particle (momentum $\bm b = \bm P - \bm p - \bm a$).
The integral over the pole uses the PV prescription, as for $F_2$,
but may be generalized with an
``$I_{\bm{PV}}$ term'' introduced in ref.~\cite{Romero-Lopez:2019qrt}.
The quantity $q_i^\star(\bm p)$ is the relative momentum of particles $j$ and $k$
in the pair CMF, 
\begin{equation}
    q_i^\star(\bm p)^2 = \frac{\lambda(\sigma_i(\bm p), M_j^2, M_k^2)}{4 \sigma_i(\bm p)}\,,
    \quad
    \sigma_i(\bm p) = (E - \omega_i(\bm p))^2 - (\bm P - \bm p)^2\,.
    \label{eq:qstardef}
\end{equation}
The momentum $\bm a_i^\star(\bm p)$ is the spatial part of the four-momentum $(\omega_j(\bm a),\bm a)$ after boosting to the pair CMF.
Finally, the function $H^{(i)}(\bm p)$, which depends on the pair invariant mass
$\sigma_i(\bm p)$, smoothly goes to zero as $|\bm p|$ becomes large.
Standard choices are described in the main text around \Cref{eq:cutoffdef}, and in ref.~\cite{\Npp}.

For odd values of $\ell'$ and/or $\ell$, the sign of $F^{(i)}$ depends on the choice of the primary flavor $j$, i.e. upon the chosen ordering of the particles in the list $\{i,j,k\}$. Our convention for the three pion case is $\{\pi^+,\pi^0,\pi^-\}$. Since all three particles are mass-degenerate all choices of $i,j,k$ lead to the same function. Therefore,  we often drop the superscript and denote the matrix $F$.

The general $G$-cut factor is given by
\begin{equation}
    \left[G^{(ij,k)}\right]_{p'\ell' m'; p \ell m} = 
    \frac{1}{2 \omega_i(\bm p') L^3}
\frac{\cY_{\ell' m'}^*(\bm p^{\star}_i(\bm p'))}{\big(q_i^\star(\bm p')\big)^{\ell'}}
\frac{H^{(i)}(\bm p') H^{(j)}(\bm p)}{b_{ij}^2 - M_k^2}
\frac{\cY_{\ell m}(\bm p'^{\star}_j(\bm p))}{\big(q_j^\star(\bm p)\big)^{\ell}}
    \frac{1}{2 \omega_j(\bm p) L^3}
    \,,
    \label{eq:Gdef}
\end{equation}
where $i$ ($j$) are the final (initial) spectator flavors, and $k$ denotes the type of the exchanged particle.
The four momentum of the exchanged particle is given by 
$b_{ij} = P - p' - p$, with $P^\mu=(E,\bm P)$, 
$p'^\mu=(\omega_i(\bm p'),\bm p')$, and $p^\mu=(\omega_j(\bm p), \bm p)$.
The arguments of the harmonic polynomials are defined analogously to
$\bm a^{\star}_i(\bm p)$ above. 
For example, $\bm p^{\star}_i(\bm p')$ is the spatial component of the four-momentum $p$ after boosting to the $jk$ CMF.
Note that for $G^{(ij,k)}$ no choice of ordering of particles is needed.
As for $F$, in the case of three pions we drop the superscript on $G$.

The matrix form of the two-particle K matrix that enters the three-particle quantization conditions is given by
\begin{equation}
  \left[\cK_{2,L}^{(i,jk)}\right]_{p'\ell'm';p\ell m}
  = \delta_{\bm p', \bm p} \delta_{\ell' \ell} \delta_{m' m} 2\omega_{i}(\bm p) L^3
\cK_{2,\ell}^{(jk)}(\bm p) \,,
\label{eq:K2Lres}
\end{equation}
where, as above, $i$ indicates the spectator, while $j$ and $k$ form the scattering pair.
The modified K matrix that enters this result is given by
\begin{equation}
    \left[\cK_{2,\ell}^{(jk)}(\bm p)\right]^{-1} =
    \frac{\eta_{jk}}{8\pi [q^\star_i(\bm p)]^{2\ell} \sqrt{\sigma_i(\bm p)}} \left\{
[q^\star_i(\bm p)]^{2\ell+1} \cot \delta^{(jk)}_\ell(q^\star_i(\bm p))
+ |q^\star_i(\bm p)| [1 - H^{(i)}(\bm p) ]
    \right\}\,,
    \label{eq:K2mod}
\end{equation}
where $\delta_\ell^{(jk)}$ is the usual scattering phase shift, while the symmetry factor is
$\eta_{jk}=1$ for distinguishable particles and $\eta_{jk}=1/2$ for identical particles or two-pion states of definite isospin.
An additional contribution is needed if the PV prescription in $F$ is generalized;
see refs.~\cite{\largera,Dawid:2024dgy}.

To represent $\widehat \cK$ in the finite-volume basis, we need the $\bm \cY$ operators,  which convert from the momentum to the $\{k\ell m\}$ basis. 
Using the notation and conventions of ref.~\cite{\Npp}, these operators act on a function of momenta $g(\bm p_1,\bm p_2, \bm p_3)$ as
\begin{align}
\left[{\boldsymbol {\mathcal Y}}_{{{\boldsymbol \sigma}}}^{[kab]} \circ g \right]_{k\ell m}
&=
\sqrt{\frac{1}{4\pi}} \int d\Omega_{a^\star} Y^*_{\ell m}(\hat {\bm a}^\star)
g(\bm p_1,\bm p_2,\bm p_3)\bigg|_{\bm p_{\sigma(1)}\to \bm k,\ \bm p_{\sigma(2)}\to \bm a, \ \bm p_{\sigma(3)}\to \bm b} \,.
\label{eq:YRdef}
\end{align}
In words, the first element of the permutation $\boldsymbol \sigma$ determines which of the three momenta is the spectator, whose momentum is then called $\bm k$,
while the second and third elements are then called $\bm a$ and $\bm b$, respectively.
The resulting function is then integrated over the direction of $\bm a$ after boosting to the CMF of the nonspectator pair, with the spherical harmonic picking out the $\ell m$ component.
The conjugate operator acts from the right as
\begin{align}
\left[g  \circ {\boldsymbol {\mathcal Y}}_{{{\boldsymbol \sigma}}}^{[kab] \dagger} \right]_{k\ell m}
&=
\sqrt{\frac{1}{4\pi}} \int d\Omega_{a^\star} Y_{\ell m}(\hat {\bm a}^\star)
g(\bm p_1,\bm p_2,\bm p_3)\bigg|_{\bm p_{\sigma(1)}\to \bm k,\ \bm p_{\sigma(2)}\to \bm a, \ \bm p_{\sigma(3)}\to \bm b} 
\,.
\label{eq:YLdef}
\end{align}

The corresponding operator for the two-particle sector acts on a function $g(\bm p_1, \bm p_2)$ and converts them to the $\{\ell m\}$ basis,
\begin{equation}
\left[{\boldsymbol {\mathcal Y}}_2 \circ g \right]_{\ell m}
=
\sqrt{\frac{1}{4\pi}} \int d\Omega_{a^\star} Y^*_{\ell m}(\hat {\bm a}^\star)
g(\bm p_1,\bm p_2)\bigg|_{\bm p_1 \to \bm a,\ \bm p_2 \to \bm b}\,.
\label{eq:cY2def}
\end{equation}
where $\bm a^\star$ is $\bm a$ after boosting the on-shell momentum to the overall CMF.
The conjugate operator is defined analogously.

\subsection{Integral equations}

For the integral equations described in \Cref{sec:QCKKppp}, we will need several quantities.
The first is $G_\infty$, which, up to a rescaling, is the infinite volume limit of the matrix $G$, \Cref{eq:Gdef}:
\begin{equation}
G_\infty(\bm p, \bm k)_{\ell' m' ; \ell m} 
= \frac{\cY^*_{\ell' m'}(\bm k^\star(\bm p)) }{q^\star(\bm p)^{\ell'}}
\frac{H(\bm p) H(\bm k)}{b^2 - M_\pi^2}
\frac{\cY_{\ell m}(\bm p^\star(\bm k))}{q^\star(\bm k)^\ell}\,.
\label{eq:Ginfty}
\end{equation}
The notation here is the same as for $G$, and is explained around \Cref{eq:Gdef}.
We define the Lorentz covariant delta function as
\begin{equation}
    \overline \delta(\bm p - \bm k)_{\ell' m';\ell m} 
    = \delta_{\ell' \ell} \delta_{m' m} 2 \omega_k (2\pi)^3 \delta^3(\bm p-\bm k)\,,
    \label{eq:deltabar}
\end{equation}
while the modified phase space factor is
\begin{equation}
\rho_H(\bm k)_{\ell' m'; \ell m}
=\delta_{\ell'\ell}\delta_{m' m} \eta
\left[ \frac{-i q^\star(\bm k) - |q^\star(\bm k)| (1 - H(\bm k))}{8 \pi \sqrt{\sigma(\bm k)}}
+ H(\bm k) \frac{1}{8 \pi \sqrt{\sigma(\bm k)}} \frac{C_{\rm PV}}{q^\star(\bm k)^{2 \ell}}\right]\,,
    \label{eq:rhoH}
\end{equation}
where $\sigma(\bm k)$ is defined in \Cref{eq:qstardef}.
The identical-particle factor is $\eta=1/2$ for the $K \overline K \leftrightarrow 3\pi$ formalism, but  otherwise equals unity. The term involving coefficient $C_{\rm PV}$ arises from a modification of the PV prescription introduced in Ref.~\cite{Romero-Lopez:2019qrt}.
The quantity $\rho_2$ is the two-particle phase space factor for distinguishable particles
\begin{equation}
\left[\rho_2\right]_{\ell' m'; \ell m} = \delta_{\ell' \ell} \delta_{m' m}
\frac{-i q^\star}{8 \pi E^\star}\,,
\label{eq:rho2}
\end{equation}
with $q^\star$ given by \Cref{eq:qstar2}.
This does not depend on a cutoff function, nor require a $C_{\rm PV}$ factor.

We now turn to the $\bcX$ operators,  introduced in ref.~\cite{\tetraquark}, that convert from the $\{k\ell m\}$ basis to the momentum basis for three on-shell particles. They are the inverse of the $\bcY$ operators discussed above. 
The operator $\bcX_{[kab]}^{\boldsymbol \sigma}$ acts on a vector $f_{k\ell m}$ as
\begin{equation}
\left[ \bcX_{[kab]}^{\boldsymbol \sigma} \circ f\right]( \{\bm p_1,\bm p_2, \bm p_3\} )
= \left[\sum_{\ell m} Y_{\ell m} (\hat a^\star) f_{k\ell m} \right]_{\bm k\to \bm p_{\sigma_1},
\bm a \to \bm p_{\sigma_2}, \bm b \to \bm p_{\sigma_3}}\,,
\label{eq:XRdef}
\end{equation}
where $\boldsymbol \sigma$ is a permutation of $\{1,2,3\}$.
In words, the sum over $\ell m$ yields a function of $\bm k$ and $\hat a^\star$.
The former is then equated to $\bm p_{\sigma_1}$ (the spectator momentum).
From $\hat a^\star$, we determine $\bm a^\star = q_{\sigma_1}^\star(\bm k) \hat a^\star $, 
which is the momentum of the primary member of the pair in the pair CMF, and then
boost to the ``lab frame'' (with total momentum $\bm P$) 
to obtain $\bm a$, which is equated to $\bm p_{\sigma_2}$.
Finally, $\bm b = \bm P - \bm k - \bm a$ is equated to $\bm p_{\sigma_3}$.
The result is a function of the three on-shell momenta $\{\bm p_1, \bm p_2, \bm p_3\}$.
The left-acting version $\boldsymbol{\mathcal X}_{[kab]}^{{\boldsymbol \sigma}\dagger}$ is defined analogously,
\begin{align}
\left[f \circ \boldsymbol{\mathcal X}_{[kab]}^{\boldsymbol \sigma \dagger} \right] (\{ \bm p_1, \bm p_2, \bm p_3\})
= \left[\sum_{\ell m} f_{k\ell m} Y_{\ell m}^*(\hat a^\star)
\right]_{\bm k\to \bm p_{\sigma_1} , \, \bm a\to \bm p_{\sigma_2}, \, \bm b \to \bm p_{\sigma_3}}
\,.
\label{eq:XLdef}
\end{align}

We also need the corresponding operators in the two-particle sector, which convert from the $\{\ell m\}$ to the $\{\bm k_1, \bm k_2\}$ basis for two-particle on-shell states. To our knowledge, these have not been explicitly written down before, although they are implicit in the two-particle quantization-condition literature. The operator $\bcX_2$ acts on a vector $g_{\ell m}$ as
\begin{equation}
    \left[\bcX_2 \circ g\right](\bm k_1,\bm k_2) = \left[\sum_{\ell m} Y_{\ell m}(\hat a^\star) g_{\ell m}\right]_{\bm a \to \bm k_1, \bm b \to \bm k_2}\,.
    \label{eq:X2def}
\end{equation}
Here, the sum over $\ell m$ leads to a function of $\hat a^\star$, which determines
$\bm a^\star=q^\star \hat a^\star$, where $q^\star$ is given in \Cref{eq:qstar2}.
Boosting to the lab frame then defines $\bm a$, while $\bm b = \bm P-\bm a$.
Note that the action of the operator $\bcX_2$ implicitly involves a choice as to which of the two particles is primary.
The conjugate operator acts analogously
\begin{equation}
    \left[g \circ \bcX_2^\dagger\right](\bm k_1,\bm k_2) = \left[\sum_{\ell m} Y^*_{\ell m}(\hat a^\star) g_{\ell m}\right]_{\bm a \to \bm k_1, \bm b \to \bm k_2}\,.
    \label{eq:X2dag}
\end{equation}
We also note that $\boldsymbol \cY_2 \circ \boldsymbol{\mathcal X}_2  $ and $\boldsymbol{\mathcal X}_2 \circ \boldsymbol \cY_2 $ act as the identity operator,
i.e. the two operators are inverses.

Finally, we collect a few definitions needed for the integral equations in the $\pi K +\pi\pi K$ systems:
\begin{align}   
\bcX^{I=3/2}_S &= \left( \sqrt2 \XR312, 0, -\sqrt{\frac16} [\XR132+\XR231],
- \sqrt{\frac56} [\XR132+\XR231] \right)\,,
\label{eq:X3S}
\\
\bcX^{I=3/2}_A &= \left( 0, \sqrt2 \XR312,  -\sqrt{\frac56} [\XR132-\XR231],
\sqrt{\frac16} [\XR132-\XR231] \right)\,,
\label{eq:X3A}
\\
\bcX^{I=1/2}_S &= \left(0, \sqrt2 \XR312,  \sqrt{\frac23} [\XR132+\XR231],
- \sqrt{\frac13} [\XR132+\XR231] \right)\,,
\label{eq:X1S}
\\
\bcX^{I=1/2}_A &= \left( \sqrt2 \XR312, 0, \sqrt{\frac13} [\XR132-\XR231],
\sqrt{\frac23} [\XR132-\XR231] \right)\,.
\label{eq:X1A}
\end{align}

\section{TOPT derivation for $I^G=0^-$ $K\bar K$ system}
\label{app:KKTOPT}

This is the first of three appendices in which we present the details of the derivation of the formalism for the $K\overline K + 3\pi$ system with $I^G=0^-$.
In this appendix we discuss the $I^G=0^-$ $K\overline K$ sector in isolation,
while \Cref{app:TOPT} discusses the $I=0$ $3\pi$ sector,
and \Cref{app:deriv} uses these building blocks to derive the final results given in the main text in \Cref{sec:QCKKppp}.
We use the variant of the RFT approach based on TOPT, first introduced in ref.~\cite{\BSQC}
for a system of identical scalars.
There have been many subsequent extensions of this TOPT approach---see refs.~\cite{\BSnondegen,\BStwoplusone,\tetraquark,Draper:2024qeh,\Npp}---and, along the way, the derivation has been somewhat streamlined and the notation clarified. To the extent consistent with readability, we refer to these earlier works for 
technical details, and adopt the improvements in notation.

Setting up a TOPT calculation in a generic EFT requires some technical preliminaries. These are summarized in \Cref{app:derivsetup}, and we do not repeat them here.

In order to ensure that the correct symmetry factors are obtained,
we sketch here a derivation of the relativistic two-particle formalism for multiple channels using the TOPT method. The final result is well known in the literature~\cite{\KSS,Hansen:2012tf}, 
but, to our knowledge, has not been derived in this manner previously. This formalism is only 
applicable if the quark masses are such that $2 M_K < 3 M_\pi$,
so we can choose a kinematic range in which there are no on-shell $3\pi$ states, namely
$E^\star < 3 M_\pi$.

As in all derivations in this paper, we begin with the finite-volume scattering amplitude, which we
denote $\widehat{\cM}_{22,L}$.\footnote{%
We use the subscript $22$ (indicating $2\to2$) to distinguish this from the related but different quantity $\cM_{2,L}$ that appears in the three-particle analysis.
} 
This is a matrix of amputated finite-volume correlators, calculated in TOPT,
in which the matrix indices are the momenta of the two particles, $\bm k_1$ and $\bm k_2$,
as well as an additional index $i$ denoting the flavor of the kaons, 
either $i=1$ for $K^+ K^-$
or $i=2$ for $K^0 \overline{K}^0$. Since, by assumption, we work in an energy range
such that we need only consider kaon-antikaon cuts, the form of this amplitude is easily shown to be
\begin{equation}
\widehat{\cM}_{22,L} = \widehat{\cB}_2 \frac1{1+ \widehat{D}_2 \widehat{\cB}_2}\,,
\end{equation}
where the flavor matrix of TOPT Bethe-Salpeter kernels is
\begin{equation}
\widehat{\cB}_2 = \begin{pmatrix}
\cB_2(K^+K^-\leftarrow K^+K^-) & \cB_2(K^0 \overline{K}^0\leftarrow K^+K^-) 
\\
 \cB_2(K^+K^- \leftarrow K^0\overline{K}^0)  &  \cB_2(K^0 \overline{K}^0\leftarrow K^0 \overline{K}^0)
 \end{pmatrix}\,,
 \label{eq:B2hat}
 \end{equation}
 while the matrix of cut factors is
 \begin{equation}
 \widehat{D}_2 = {\rm diag}\left( D_2, D_2\right)\,,
 \end{equation}
 with $D_2$ defined as
 \begin{equation}
     D_2 =\delta_{\bm k_1' \bm k_1} \delta_{\bm k_2' \bm k_2}
 \frac1{L^3} \frac1{2\omega_K(\bm k_1) 2 \omega_K(\bm k_2)}
\frac1{[E - \omega_K(\bm k_1) - \omega_K(\bm k_2)]}\,,
\label{eq:D2def}
 \end{equation}
 with on-shell energies defined in the standard way, $\omega_K(\bm k) = \sqrt{M_K^2 + \bm k^2}\,$.
 We note that the $\cB_2$ kernels here can contain intermediate $3\pi$ states, unlike those in \Cref{app:deriv}, since we assume kinematics such that these states cannot go on shell.
All Bethe-Salpeter kernels have an implicit dependence on $E$ and the initial and final momenta,
\begin{equation}
\cB_2 \equiv \cB_2(E; \bm k'_1, \bm k'_2; \bm k_1, \bm k_2)\,.
\end{equation}
There is no factor of $1/2$ appearing in $D_2$ because the particles are not identical.

We project onto the $I=0$ $K\overline K$ state, 
\begin{equation}
\ket{K (\bm k_1) \overline K (\bm k_2)}_{I=0} = 
\frac1{\sqrt2} \left[ \ket{K^+ (\bm k_1) K^- (\bm k_2}) 
+ \ket{K^0 (\bm k_1) \overline K^0 (\bm k_2)}\right]\,.
\label{eq:KKbarI0}
\end{equation}
using the vector
\begin{equation}
\bra{I=0} \equiv \tfrac1{\sqrt2} (1, 1)\,,
\end{equation}
which satisfies
\begin{equation}
\bra{I=0} \widehat{\cB}_2 = \cB_2^{I=0} \bra{I=0}\ \ {\rm and} \ \
\bra{I=0} \widehat{D}_2 = D_2 \bra{I=0}\,,
\label{eq:B2KKdef}
\end{equation}
as well as similar equations for the action of $\ket{I=0}$.
Then we find
\begin{equation}
\cM_{22,L}^{I=0} \equiv \bra{I=0} \widehat{\cM}_{22,L} \ket{I=0}
= \cB_2^{I=0} \frac1{1+ D_2 \cB_2^{I=0}}\,.
\label{eq:M22LI0}
\end{equation}

The next step is to project on shell, using a two-particle version of the argument given for three particles in sec. II.C of ref.~\cite{\BSQC}.
The required identity is
\begin{equation}
D_2 = F_2 + I_{F_2}\,,
\label{eq:D2decomp}
\end{equation}
where $I_{F_2}$ is an integral operator that connects the adjacent factors of $\cB_2^{I=0}$
into another infinite-volume quantity, while $F_2$ is a finite-volume function, given in \Cref{eq:F2def} (except here we drop the superscript for brevity).
$F_2$ implicitly projects adjacent kernels on shell.
The explicit form of $I_{F_2}$ can be reconstructed from the identity given in Appendix~A of ref.~\cite{\HSQCa}, but will not be needed.
The remainder of the steps are then identical to those in the Feynman diagram derivation of ref.~\cite{\KSS}, and lead to 
\begin{align}
\cM_{22,L}^{I=0} = \bcX_2 \circ
\cK_{22}^{I=0} \frac1{1+ F_2 \cK_{22}^{I=0}} \circ \bcX_2^\dagger\,, \text{ with } \,
\cK_{22}^{I=0} = \cB_2^{I=0} \frac1{1+ I_{F_2} \cB_2^{I=0}}\,.
\label{eq:M22K22I0def}
\end{align}
Both $\cM_{22,L}^{I=0}$ and $\cK_{22}^{I=0}$ are now (implicitly) on shell.
The former remains expressed in terms of the momenta $\bm k'_i$ and $\bm k_i$,
while the latter, as well as $F_2$, is a matrix in the $\{\ell m\}$ space describing the relative angular momentum of the pair in their CMF.
The operators $\bcX_2$ and $\bcX_2^\dagger$ convert from $\{\ell, m\}$ to momentum space,
and are defined in \Cref{eq:X2def,eq:X2dag}, respectively.
We stress that $\bcX_2 \cdot \cK_{22}^{I=0}\cdot \bcX_2^\dagger$ is a Lorentz-invariant quantity, despite being determined in TOPT, because it is obtained by a sum over all  diagrams. 
Thus $\cK_{22}^{I=0}$ is given in terms of standard phase shifts by
\begin{align}
\begin{split}
    \left[\cK_{22}^{I=0}\right]_{\ell' m';\ell m} &= \delta_{\ell' \ell} \delta_{m' m}
    \cK_{2,\ell}^{I=0}\,,
    \label{eq:K22}
    \\
    \cK_{2,\ell}^{I=0} &=\frac1{8\pi [q^\star]^{2\ell} E^\star_2} 
[q^\star]^{2\ell+1} \cot \delta_\ell(q^\star)\,,
\end{split}
\end{align}
where $q^\star$ and $E_2^\star$ are, respectively, the relative momentum and total energy in the CMF.

As a final step, we project onto the negative $G$-parity sector using
\begin{equation}
P_A \ket{K^+ (\bm k_1) K^- (\bm k_2)}
=
\frac12 \left[\ket{K^+ (\bm k_1) K^- (\bm k_2)} -  \ket{K^+ (\bm k_2) K^- (\bm k_1)} \right]\,,
\label{eq:PAdef}
\end{equation}
and similarly for the neutral kaons. We note that the projected states are not normalized. This choice is slightly more convenient technically, but does require that we are careful
with factors when we relate the two-particle components of $\widehat\cM_L$ to the predictions of ChPT in \Cref{app:chpt}.

$P_A$ acts in the $\{\ell m\}$ basis as the projector onto odd partial waves, e.g.
\begin{align}
\left[P_A \cK_{22}^{I=0} P_A \right]_{\ell' m'; \ell m}
&\equiv \left[\cK_{22}^{I^G=0^-}\right]_{\ell' m';\ell m}\,,
\label{eq:K22I0AS}
\\
&=\frac{1 - (-1)^{\ell'}}2 \left[\cK_{22}^{I=0}\right]_{\ell' m'; \ell m}
\frac{1 - (-1)^\ell}2 \,.
\label{eq:PAdefell}
\end{align}
Using the result that $F_2$ commutes with $P_A$,
because $(-1)^\ell F_2 (-1)^\ell = F_2$ up to exponentially-suppressed corrections~\cite{\HSQCa},
we finally obtain
\begin{equation}
\cM_{22,L}^{I^G=0^-} = \bcX_2 \circ
\cK_{22}^{I^G=0^-} \frac1{1+ F_2 \cK_{22}^{I^G=0^-}} \circ \bcX_2^\dagger\,.
\label{eq:M2LI0AS}
\end{equation}
From this, the two-particle quantization condition can be read off, 
\begin{equation}
    \det_{\ell m} \left( 1+ F_2 \cK_{22}^{I^G=0^-} \right)= 0,
\end{equation} 
which, as expected, is identical to the result derived using the Feynman diagram approach~\cite{\KSS}.

\section{TOPT derivation for $I=0$ three pion system}
\label{app:TOPT}

The QC for three distinguishable (and, in general, nondegenerate) spinless mesons was derived in Ref.~\cite{\BSnondegen}
using a TOPT-based method,
and led to a quantization condition that, in addition to the usual $\{k\ell m\}$ indices,
had a flavor index running over choices of the flavor of the spectator particle.
On the other hand, in Ref.~\cite{\isospin}, the QC for the general isospin three-pion state was derived, following the Feynman diagram approach of ref.~\cite{\HSQCa}, and,
for $I=0$, was found to involve a single flavor channel. This corresponds to the fact that there is a single
$I=0$ combination of three pions, namely
\begin{multline}
\ket{\pi^+(\bm p_1) \pi^0(\bm p_2) \pi^-(\bm p_3)}_{I=0}
= \frac1{\sqrt6} \cA 
\left[ \ket{\pi^+(\bm p_1) \pi^0(\bm p_2) \pi^-(\bm p_3)} \right] \equiv 
\\
\frac1{\sqrt6} \bigg[ \ket{\pi^+(\bm p_1) \pi^0(\bm p_2) \pi^-(\bm p_3)}
+ \ket{\pi^+(\bm p_2) \pi^0(\bm p_3) \pi^-(\bm p_1)}
+ \ket{\pi^+(\bm p_3) \pi^0(\bm p_1) \pi^-(\bm p_2)}
\\
- \ket{\pi^+(\bm p_2) \pi^0(\bm p_1) \pi^-(\bm p_3)}
- \ket{\pi^+(\bm p_3) \pi^0(\bm p_2) \pi^-(\bm p_1)}
- \ket{\pi^+(\bm p_1) \pi^0(\bm p_3) \pi^-(\bm p_2)}
\bigg]\,,
\label{eq:ASstate}
\end{multline}
which also defines the three-particle antisymmetrization operator $\cA$. 
The issue addressed in this appendix is how these two descriptions are related.
Specifically, we show how to obtain a one-dimensional quantization condition 
in the TOPT approach,\footnote{%
As noted in the previous appendix, we postpone a discussion of general technical considerations in setting up a TOPT calculation to \Cref{app:derivsetup}.
}
and this will be used in \Cref{app:deriv} as part of the generalized derivation including $2\leftrightarrow3$ transitions, the results of which are given in \Cref{sec:QCKKppp}.

We assume in this appendix that quark masses are chosen such that $3 M_\pi < 2 M_K$ and
$E^\star < 2 M_K$, so that intermediate $K \overline K$ states cannot go on shell.
This is the same kinematical range considered in ref.~\cite{\isospin}.
This implies that the three-particle Bethe-Salpeter kernels that appear can have
intermediate (off shell) $K \overline K$ states, unlike those in \Cref{app:deriv}.

\subsection{Asymmetric formalism}
\label{app:3piasym}

Since the $I=0$ three-pion state contains only distinguishable particles, i.e. $\pi^+\pi^0\pi^-$, we can directly take over results from ref.~\cite{\BSnondegen}.
We begin from the finite-volume three-particle amplitude
given in eq.~(43) of ref.~\cite{\BSnondegen}, a geometric series that sums to\footnote{%
We have dropped the bar on the $\cB_{2,L}$ and the superscript ``off'' on $\cM_{23,L}$ that are used in ref.~\cite{\BSnondegen}.
}
\begin{equation}
\cM_{23,L} = \left[\cB_3 + \sum_i \cB_{2,L}^{(i)} \right]
\frac1{1+ D_3 \left[
\cB_3 + \sum_i\cB_{2,L}^{(i)}\right]}\,.
\label{eq:M23L}
\end{equation}
The cut factor $D_3$ is composed of the TOPT energy denominator and propagator factors,  
\begin{equation}
    D_3 =  \delta_{\bm p'_1 \bm p_1} \delta_{\bm p'_2 \bm p_2} \delta_{\bm p'_3 \bm p_3}
\frac1{L^6}\frac1{2\omega_\pi(\bm p_1) 2 \omega_\pi(\bm p_2) 2 \omega_\pi(\bm p_3)} 
\frac1{[E - \omega_\pi(\bm p_1) - \omega_\pi(\bm p_2)- \omega_\pi(\bm p_3)]}\,.
\label{eq:D3def}
\end{equation}
The kernels $\cB_{2,L}^{(i)}$ are defined as
\begin{equation}
  \cB_{2,L}^{(i)}= \delta_{\bm p'_i, \bm p_i} 2\omega_{\pi}(\bm p_i) L^3
\cB_2^{(jk)} (E - \omega_\pi(\bm p_i); \bm p'_j, \bm p'_k; \bm p_j, \bm p_k)\,,
\end{equation}
with the superscript $i$ indicating which of the three pions is the spectator, while $jk$ indicates the scattering pair.\footnote{%
To completely define this kernel we need to pick a primary member of the pair, with respect to the direction of which a decomposition of the angular dependence 
into spherical harmonics will be made below. We will use a cyclic convention, in which
the three pions are ordered $\{\pi^+,\pi^0, \pi^-\}$ or cyclic permutations thereof,
such that the primary member of the pair follows directly after the spectator.}
$\cB_3$ is the three-particle TOPT Bethe-Salpeter kernel.
All quantities are matrices with indices given by the particle momenta,
$\{\bm p_1, \bm p_2, \bm p_3\}$.
Matrix multiplication thus leads to summing two independent momenta over the finite-volume set, the third constrained by $\sum_i \bm p_i = \bm P$.

We stress that $\cM_{23,L}$ contains not only the fully-connected $3\to 3$ amplitude,
but also $2\to 2$ amplitudes with each of the particles spectating.
This is shown diagrammatically in fig.~2 of ref.~\cite{\BSnondegen}. 
The reason for combining contributions in this way is that it simplifies intermediate expressions, 
although we will need to remove the additional contributions at the end.

To project onto the $I=0$ channel, we fully antisymmetrize the external momenta, 
using the operator $\cA$ defined in \Cref{eq:ASstate},
\begin{equation}
\cM_{23,L}^{I=0} = \frac{\cA}{\sqrt6} \cM_{23,L} \frac{\cA}{\sqrt6}\,.
\end{equation}
Note that $\cA$ can act both to the left and right.
We will need the following properties of $\cA$, as well as the corresponding two-particle
antisymmetrization operator $\cA_2 = 2 P_A$,
\begin{gather}
\cA^2 = 6 \cA\,,\ \ [\cA, D_3] = 0\,, 
\label{eq:Arel}
\\  
\cA \cB_3 = \cB_3 \cA = \tfrac16 \cA \cB_3 \cA = \cB_3^{I=0}\,,
\label{eq:B3rel}
\\
\cA_2^2 = 2 \cA_2\,,\ \ 
\cA_2 \cB_2^{(i)} = \cB_2^{(i)} \cA_2 = \tfrac12 \cA_2 \cB_2^{(i)} \cA_2 = \cB_2^{I=1}\,,
\label{eq:A2rel}
\\
\cA \sum_i \cB_{2,L}^{(i)} = \sum_i \cB_{2,L}^{(i)} \cA
= \frac14 \cA \cB_{2,L}^{I=1} \cA\,.
\label{eq:B2Lrel}
\end{gather}
To obtain the last two lines, we use the fact that the antisymmetric part of the scattering kernel for two different pions yields the $I=1$ kernel (with the normalization shown) irrespective of which pion pair is chosen.
In the final equality $\cB_{2,L}^{I=1}$ can have any choice for the spectator momentum;
for definiteness we take it to be $\bm p_3$.
Using these relations and straightforward algebra we find
\begin{align}
\cM_{23,L}^{I=0} &= \frac{\cA}2\,  ( \cB_{2,L}^{I=1}+ \tfrac19 \cB_3^{I=0} )
\frac1{1 + \frac14 D_3 \cA  ( \cB_{2,L}^{I=1} + \frac19 \cB_3^{I=0} )} \, \frac{\cA}2\,.
\label{eq:M23LI0b}
\end{align}
We note that this result has the same form as that for the fully symmetric $I=3$ state,
given in eq.~(16) of ref.~\cite{\BStwoplusone}. This can be seen using
\begin{equation}
\frac14 D_3 \cA = \frac{\cA}2 \frac{D_3}{3!} \frac{\cA}2\,,
\end{equation}
which follows from \Cref{eq:Arel}. In ref.~\cite{\BStwoplusone}, the corresponding expression is $\cS_{\rm ID} (D_3/3!) \cS_{\rm ID}$.
Like $\cA/2$, $\cS_{\rm ID}$ sums over the three choices of spectator, all with positive signs.
The $3!$ in the denominator can be understood as a symmetry factor for identical particles.

We can simplify the term involving $D_3 \cA$ in \Cref{eq:M23LI0b} by noting that,
when placed between two $\cB_{2,L}^{I=1}$ factors,
of the six terms it contains, two lead to $F$-type cuts (those in which the spectator momentum $\bm p_3$ is unchanged), while the remaining four lead to $G$-type cuts (those in which the spectator changes).\footnote{%
For more discussion of the difference between $F$- and $G$-type cuts, 
albeit in the context of identical particles,
see fig.~5 and the associated text in ref.~\cite{\BSQC}.
}
If there is a $\cB_3^{I=0}$ factor on either or both sides, $F$-type and $G$-type cuts are interchangeable, due to the antisymmetry of $ \cB_3^{I=0}$.
It follows that we can make the replacement
\begin{equation}
 \tfrac14 D_3 \cA  \to \tfrac12 D_3 + D_G\,,
 \label{eq:D3AS}
 \end{equation}
 where
 \begin{align}
 D_G &=  \delta_{\bm p'_1 \bm p_3} \delta_{\bm p'_2 \bm p_2} \delta_{\bm p'_3 \bm p_1}
\frac1{L^6}
\frac1{2\omega_\pi(\bm p_1) 2 \omega_\pi(\bm p_2) 2 \omega_\pi(\bm p_3)}
\frac1{[E - \omega_\pi(\bm p_1) - \omega_\pi(\bm p_2) - \omega_\pi(\bm p_3)]}\,.
\label{eq:DGdef}
\end{align}
This differs from $D_3$ [\Cref{eq:D3def}] only by the fact that the Kronecker deltas permute the first and third momenta.
The $D_G$ contribution comes with a positive sign, because signs arising from $\cA$
cancel those from the kernels.

It is instructive to compare the result \Cref{eq:M23LI0b} to eq.~(83) of ref.~\cite{\BSQC},
which we recall carries out the TOPT analysis for three identical scalars.
The latter equation is for an unsymmetrized version of $\cM_{23,L}$,
denoted $\cM_{2,L} + \cM_{3,L}^{(u,u)}$ in ref.~\cite{\BSQC}.
The symmetrization is carried out in eq.~(84) of that work, and involves summing over three choices of spectator momentum for both initial and final states.
This is replaced here by the factors of $\cA/2$ on the ends, which sum over the same three choices, although here with appropriate signs due to the antisymmetry.
Aside from these exterior factors, and a trivial algebraic reorganization,
the results have exactly the same form, except that
$\tfrac12 D_3 + D_G$ in \Cref{eq:M23LI0b} becomes $D_F+D_G$ in ref.~\cite{\BSQC}.
These factors are, however, the same, because $D_3= 2 D_F$,
since $D_F$ includes a symmetry factor of $1/2$ while $D_3$ does not.
Thus we find that the form of the $I=0$ three-pion result is essentially the same as that
for three identical particles, and, in particular, 
does not require additional flavor indices,
unlike the general case of three distinguishable particles~\cite{\BSnondegen}.

The next step is to project the cut factors and external legs on shell. This is done by adapting the identities given in eqs.~(52) and (57) of ref.~\cite{\BSQC} to the case of antisymmetric amplitudes, yielding
\begin{equation}
    D_3 = F + I_F\,, \ \ D_G = -G + I_G\,,\ \ \Rightarrow\ \ \tfrac12 D_3 + D_G = \tfrac12 F-G + I_{FG}\,,\ \ I_{FG} = \tfrac12 I_F + I_G\,.
    \label{eq:FGiden}
\end{equation}
Here $I_F$, $I_G$, and $I_{FG}$ act like integral operators, sewing adjacent kernels together to create new infinite-volume quantities. 
They are discussed in detail in ref.~\cite{\BSQC}, and we do not need the explicit forms here.
The matrices $F$ and $G$, to be discussed in more detail below, contain the singular parts of $D_3$ and $D_G$, and implicitly project adjacent kernels on shell.
Using \Cref{eq:FGiden}, and resumming in the standard RFT manner, one obtains\footnote{%
As noted above, for the sake of brevity we are not explicitly distinguishing between the
off-shell amplitude, which appears in the previous equations in this section, e.g.~\Cref{eq:M23LI0b}, and the on-shell amplitude appearing here. We recall that, in the context of TOPT, onshellness implies that the sum of the energies of the three particles equals $E$.
}
 \begin{align}
 \cM_{23,L}^{I=0} 
&=  \bcX_A \circ 
\cK_{23}^{(u,u), I=0} 
\frac1{1 + (\tfrac12 F-G) \cK_{23}^{(u,u), I=0} } 
\circ \bcX_A^\dagger\,,
\label{eq:M23LI0c}
\\
\cK_{23}^{(u,u), I=0} &= (\cB_{2,L}^{I=1}+ \tfrac19 \cB_3^{I=0} )
\frac1{1 + I_{FG} (\cB_{2,L}^{I=1}+ \tfrac19 \cB_3^{I=0} )} \equiv \cK_{2,L}^{I=1}+ \cK_{3}^{(u,u), I=0}\,,
\label{eq:Kdf23uuI0decomp}
\\
\cK_{2,L}^{I=1} &= \cB_{2,L}^{I=1} \frac1{1+ \frac12 I_F \cB_{2,L}^{I=1}}\,,
\label{eq:K2Ldef}
\end{align}
where
\begin{align}
\bcX_A &= \XR123+ \XR231 + \XR312\,,\quad
\bcX_A^\dagger = \XL123 + \XL231 + \XL312\,,
\label{eq:XAdef}
\end{align}
where $\bcX_{[kab]}^{\boldsymbol\sigma}$ and its conjugate are defined in \Cref{eq:XRdef,eq:XLdef}.
Aside from the exterior factors of $\bcX_A$, all quantities here are now matrices in $\{k\ell m\}$ space, where $k$ denotes the spectator momentum $\bm k$, while $\{\ell, m\}$ describe the angular momentum of the non-spectator pair in its CMF.
Explicit expressions for $F$ and $G$ are given in \Cref{eq:Fdef,eq:Gdef}, respectively,
with the superscripts in those equations dropped.
Comparing to eqs.~(71) and (85) of ref.~\cite{\BSQC}, we see that, aside from the $\bcX_A$ factors (to be discussed below) and the factor of two difference in the $F$ term (a change of convention, as noted above), the key difference is in the sign of the $G$ term.
The negative sign arises here because of a mismatch between the cyclic ordering that defines the primary member of the kernels and the transposition of momenta that occurs in $D_G$ and $G$.
In particular, the choice of primary member of the pair on one side of $G$ is opposite to that in the adjacent kernels, and to rectify this a sign must be introduced, which arises because the kernels are antisymmetric. Such rectification factors are present in the general form of $G$ for distinguishable particles of ref.~\cite{\BSnondegen}, and take the form of $(-1)^\ell$. Here, since $\ell$ is always odd, these factors are negative.

Returning to \Cref{eq:M23LI0c}, the kernel $\cK_{23}^{(u,u),I=0}$ is given the
$(u,u)$ superscript to indicate, following ref.~\cite{\HSQCa}, that it is not antisymmetric.
This is shown by the explicit expression for this quantity, \Cref{eq:Kdf23uuI0decomp}, 
in which the factors of $\cB_{2,L}^{I=1}$ pick out one of the three momenta.
The right-most equality of $\cK_{23}^{(u,u),I=0}$ given in \Cref{eq:Kdf23uuI0decomp}
picks out all the partially disconnected parts involving a single spectator and collects
them into $\cK_{2,L}^{I=1}$. 
This contains a modified two-particle K matrix and is given explicitly in \Cref{eq:K2Ldef}.
The remainder, $\cK_{3}^{(u,u)I=0}$, is a fully-connected three-particle $K$ matrix,
whose explicit form is given in eq.~(72) of ref.~\cite{\BSQC} but will not be needed here.
All we need to know is that, like $\cK_{2,L}^{I=1}$, it is nonzero only when both
$\ell'$ and $\ell$ are odd.

The final feature of \Cref{eq:M23LI0c} is the presence of the
operator $\bcX_A$ and its conjugate on the ends.
These are defined in \Cref{eq:XAdef}, and are composed of the operators 
given in \Cref{eq:XLdef,eq:XRdef}, which convert from the $\{k\ell m\}$ basis to functions of three on-shell momenta. The subscript $A$ indicates that these operators also antisymmetrize over choices of ordering of the external momenta, playing the role of the factors of $\cA/2$ in the original expression for $\cM_{23,L}^{I=0}$, \Cref{eq:M23LI0b}.
This is made explicit in \Cref{eq:XAdef}, where the three cyclic permutations of momentum assignments are summed.
The other three permutations of these assignments, which are odd permutations, would, if present, come with negative signs.
However, they are not needed, because of the antisymmetry of the kernels on which these operators act under the exchange of the second and third momenta.
We also stress that the presence of three, rather than six, terms in $\bcX_A$ matches the normalization of $\cA/2$, rather than $\cA$, as desired.

As an aside, we note that, at this stage of the derivation,
we can read off the asymmetric form of the $I=0$ three-pion quantization condition
from \Cref{eq:M23LI0c}, since $\cM_{23,L}$ is a form of finite-volume correlator.
The result is 
\begin{equation}
\det \left[1 + (\tfrac12 F-G) ( \cK_{2,L}^{I=1} + \cK_{3}^{(u,u), I=0} ) \right] = 0\,.
\label{eq:QCAS}
\end{equation}
This is an alternative quantization condition to the previously known one~\cite{Hansen:2020zhy} for the $I=0$ three-pion system.

\subsection{(Anti)symmetric form of the formalism}
\label{app:3piAS}

The final stage of the derivation is to use antisymmetrization identities to rewrite 
$\cM_{23,L}^{I=0}$ in terms of a three-particle K matrix, $\cK_{3}^{I=0}$, that is fully antisymmetric in the initial and final momenta.
It turns out that these identities are simple generalizations of those given for identical particles in Eqs.~(102)-(104) of Ref.~\cite{\BSQC}.
In particular, they hold with exactly the same form aside from the change $G\to -G$,
as long as one defines quantities appropriately.
To make this explicit, we recall some notation from ref.~\cite{\BSQC}.
In that work, $X^{(u)}$ and $Z^{(u)}$ indicate generic kernels (here either $\cK_{2,L}^{I=1}$ or $\cK_{3}^{(u,u),I=0}$) from which only the
right- and left-hand $\{k\ell m\}$ indices are considered.
These kernels are, as noted above, not antisymmetric,
in the sense that one momentum is picked out to be the spectator.
This is indicated by the superscript $(u)$.
Antisymmetrization involves combining with the two other choices of spectator,
for which the kernels are denoted with superscripts ${(s)}$ and $(\tilde s)$.
In the context of identical particles, their definitions are given in eqs.~(D1-D4) of ref.~\cite{\BSQC}, following the work of ref.~\cite{\HSQCa}.
The definitions we use here are such that 
\begin{equation}
\begin{gathered}
\XR123 \circ Z^{(u)} = \XR231 \circ Z^{(s)} = \XR312 \circ Z^{(\tilde s)}\,,
\\
X^{(u)}\circ \XL123 = X^{(s)}\circ \XL231 = X^{(\tilde s)}\circ \XL312\,,
\end{gathered}
\label{eq:kernelrel1}
\end{equation}
which are the analogs (using updated notation) of eqs.~(D3) and (D5) in ref.~\cite{\BSQC}.
Note that we use cyclic permutations of the momentum assignments,
as in \Cref{eq:XAdef}, so that all quantities have the same sign despite the overall antisymmetry.
The antisymmetric nature of the underlying kernels appears when the second and third momentum assignments are permuted, e.g.
\begin{equation}
\begin{gathered}
\XR123 \circ Z^{(u)} = - \XR132 \circ Z^{(u)}\,, \ \ 
\XR231 \circ Z^{(s)} = - \XR213 \circ Z^{(s)}\,, \\
\XR312 \circ Z^{(\tilde s)} = - \XR321 \circ Z^{(\tilde s)}\,,
\end{gathered}
\label{eq:kernelrel2}
\end{equation}
and similarly for the X kernels.
Combining \Cref{eq:kernelrel1,eq:kernelrel2}, and using the fact that the momentum labels are dummy variables that can be interchanged, one obtains 
\begin{equation}
    \XR123 \circ Z^{(s)} = - \XR132 \circ Z^{(\tilde s)}\,,
\label{eq:kernelrel3}
\end{equation}
which is the analog here of eq.~(D4) in ref.~\cite{\BSQC}.

These relations allow one to derive the following identities,
which hold up to exponentially-suppressed corrections
\begin{align}
X^{(u)} (\tfrac12 F-G) Z^{(u)} 
&= X^{(u)} \frac{F}2 Z_\Sigma + X^{(u)} \overrightarrow I_G Z^{(u)}\,,
\label{eq:ASidenta}
\\
&= X_\Sigma \frac{F}2 Z^{(u)} + X^{(u)} \overleftarrow I_G Z^{(u)}\,,
\label{eq:ASidentb}
\\
&= \frac13 X_\Sigma \frac{F}2 Z_\Sigma + X^{(u)} \overleftrightarrow I_{FG} Z^{(u)}\,,
\label{eq:ASidentc}
\end{align}
which are, respectively, the analogs of eqs.~(102-4) in ref.~\cite{\BSQC},
expressed in the notation of eqs.~(77-79) of ref.~\cite{\BStwoplusone}.
Here
\begin{equation}
X_\Sigma = X^{(u)} + X^{(s)}+ X^{(\tilde s)}\,,\quad
Z_\Sigma = Z^{(u)} + Z^{(s)}+ Z^{(\tilde s)}\,,
\end{equation}
are the antisymmetrized forms of the kernels,
while $\overrightarrow I_G$, $\overleftarrow I_G$ and $\overleftrightarrow I_{FG}$ 
(not to be confused with the operators $I_F$, $I_G$, and $I_{FG}$ introduced above)
are infinite-volume integral operators that are closely related to those in Ref.~\cite{\BSQC,\BStwoplusone}, but whose explicit form will not be needed.

Using these identities, and following the algebraic steps discussed in Ref.~\cite{\BSQC}, we find the standard RFT results for the finite-volume amplitude in their symmetric form, aside from the replacement $G \to -G$.
Specifically, to obtain the connected amplitude we must pull out the two-particle contribution, 
\begin{align}
{\cM}_{23,L}^{I=0} &= \bcX_A \circ \cM_{2,L}^{I=1} \circ  \bcX_A^\dagger + \cM_{3,L}^{I=0}\,,
\label{eq:M23Ldecomp}
\\
\cM_{2,L}^{I=1} &= \cK_{2,L}^{I=1} \frac 1{1 + \tfrac12 F \cK_{2,L}^{I=1}}\,,
\label{eq:M2LI1res}
\end{align}
leading to the following results
\begin{align}
{\cM}_{3,L}^{I=0} &= \bcX_A\circ \left[ {\cD}_L^{(u,u)} + {\cM}_{\df,3,L}^{(u,u)} \right] \circ \bcX_A^\dagger\,,
\label{eq:M3Lres}
\\
\cD_L^{(u,u)} &= \cM_{2,L}^{I=1} G \cM_{2,L}^{I=1} \frac1{1-G \cM_{2,L}^{I=1}}\,,
\label{eq:DLuures}
\\
\cM_{\df,3,L}^{(u,u)} &= (\tfrac13 - \cD_{23,L}^{(u,u)} \tfrac12 F) \cK_{3}^{I=0}
\frac1{1+ F_3 \cK_{3}^{I=0}} (\tfrac13 - \tfrac12 F \cD_{23,L}^{(u,u)})\,,
\label{eq:Mdf3Luures}
\\
\cD_{23,L}^{(u,u)} &= \cM_{2,L}^{I=1}+ \cD_L^{(u,u)}\,,
\label{eq:D23Luures}
\\
F_3 &= \frac{F}6 - \frac{F}2 \frac1{(\cK_{2,L}^{I=1})^{-1} + \tfrac12 F - G} \frac{F}2\,.
\label{eq:F3res}
\end{align}

An explicit expression for $\cK_{3}^{I=0}$ can be given, in terms of the asymmetric kernels and the integral operators, but is not illuminating. 
The only important property is that $\cK_{3}^{I=0}$ is completely antisymmetric when converted from $\{k\ell m\}$ to momentum variables, i.e.
\begin{equation}
    \XR123 \circ \cK_{3}^{I=0} \circ \XL123
    =  {\tt sgn}(\boldsymbol \sigma'){\tt sgn}(\boldsymbol \sigma) 
    \bcX_{[kab]}^{\boldsymbol \sigma'}\circ \cK_{3}^{I=0} \circ \bcX_{[kab]}^{\boldsymbol \sigma \, \dagger}\,.
\end{equation}
It follows from \Cref{eq:M3Lres} that the quantization condition can be written
\begin{equation}
 \det\left[1+ F_3 \cK_{3}^{I=0}\right] = 0\,,
 \end{equation}
which is the standard symmetric RFT form.

These results for $\cM_{3,L}^{I=0}$ and the quantization condition are exactly the same as those
found in ref.~\cite{\isospin} for the $I=0$ channel using the Feynman-diagram based method.
The integral equations for $\cM_3$ can now be obtained by taking the appropriate infinite-volume limit of \Cref{eq:M3Lres}, following ref.~\cite{\HSQCb}.
As observed in ref.~\cite{\BSQC}, assuming the invertibility of these equations, it then follows that the $\cK_3$ obtained using the TOPT approach is the same as that obtained using Feynman diagrams. This in turn implies that $\cK_{3}^{I=0}$ is Lorentz invariant.
Thus the discussion in secs.~3.1.2 and 3.2.1 of ref.~\cite{\isospin} concerning the parametrization of $\cK_{3}^{I=0}$ applies verbatim.

\section{Derivation for the $I=0$ $K\overline K \leftrightarrow 3\pi$ system}
\label{app:deriv}

In this appendix we derive the results quoted in \Cref{sec:QCKKppp} for the $I^G=0^-$ $K\bar K \leftrightarrow 3\pi$ system. 
We consider this system, rather than $K\pi\leftrightarrow K\pi\pi$, since its three-particle subsystem has only a single channel, so that issues arising from the inclusion of $2\to 3$ and $3\to 2$ transitions are not obscured by other technical details. We build upon the strategies followed in the derivation of the $2+3$ formalism for identical scalar particles given in ref.~\cite{\BHSQC}.

To reduce the level of technical detail in this appendix, we have provided descriptions of the derivation for uncoupled $K\overline K(I^G=0^-)$ and $3\pi(I=0)$ systems
in \Cref{app:KKTOPT,app:TOPT}, respectively. Many of the results from these two appendices can be taken over here with minor modifications, and this separation allows this appendix to focus on the new features associated with including $2\to 3$ transitions. We also have separated off definitions of key quantities entering the resulting formalism into \Cref{app:kinematics}.

\subsection{Setting up the TOPT calculation}
\label{app:derivsetup}

There are some subtleties with the use of TOPT,
as discussed in Appendix A of ref.~\cite{\BSQC}, partly based on the earlier discussion in
Appendix B of ref.~\cite{\BHSQC}.
It is shown in these works that, if one begins with a Feynman diagram representation of
finite-volume correlators, 
one can, in the context of determining finite-volume effects, absorb tadpole diagrams into
existing vertices, and similarly avoid a class of fake three-particle cuts. 
Furthermore, following
the discussion in Appendix B.1 of Ref.~\cite{\BHSQC}, one can use renormalization conditions such that propagators can be written in the free form but with physical masses. 
Therefore, all self-energy contributions can be absorbed into vertices. 
After these preliminaries using Feynman diagrams, one then converts to TOPT by integrating over the energy components in loops.
This leads to the usual rules for TOPT diagrams, 
except that now propagators involve physical masses, 
and there are no self-energy or tadpole diagrams. 
This is the starting point for the following analysis.

We stress that the above-described preliminaries are only sufficient in the absence of $1\leftrightarrow2$ transitions, as is the case here.
Otherwise, for example in the $N\pi+N\pi\pi$ system,
a more extended and elaborate discussion along the lines of that presented in Appendix B.2 of ref.~\cite{\BHSQC} is required.

Both the quantization condition and the integral equations relating K matrices to scattering amplitudes
can be derived from a quantity that we loosely describe as the finite-volume scattering amplitude~\cite{\HSQCb}.
Here this begins as a $2\times 2$ matrix of amputated, fully-connected correlation functions
joining the two initial states
\begin{equation}
\left( P_A \ket{K (\bm k_1) \overline K (\bm k_2)}_{I=0}, \ \  
\ket{\pi^+ (\bm p_1) \pi^0 (\bm p_2) \pi^- (\bm p_3)}_{I=0} \right)
\end{equation}
to the corresponding two final states, defined in \Cref{eq:KKbarI0,eq:PAdef,eq:ASstate}.
We denote this matrix of amplitudes $\widehat{\cM}_{L}$,
where the caret or hat is used for matrices that involve both two and three-particle states.
States are defined in finite volume, so that all momenta lie in the finite-volume set: $(2\pi/L) \bm n$, where $\bm n$ is a vector of integers.
The momenta satisfy
\begin{equation}
\bm k_1 + \bm k_2 = \bm P = \bm p_1 + \bm p_2 + \bm p_3\,,
\label{eq:momcons}
\end{equation}
where $\bm P$ is the total momentum, also drawn from the finite-volume set.
We stress that the correlator is considered in Minkowski, rather than Euclidean, time, and that this time has infinite range. After Fourier transformation, time dependence is converted into a dependence on the total energy $E$.

We analyze $\widehat{\cM}_L$ in an effective theory in which the degrees of freedom are pions and kaons, with no constraints on the form of vertices, and working diagrammatically to all orders in perturbation theory.
We note that, to obtain the finite-volume scattering amplitudes in TOPT
the initial states must always lie at the right end of the diagram
(corresponding to sending $t_i\to -\infty$),
while the final states must lie at the left end (corresponding to $t_f\to \infty$).
Amputation is achieved by removing propagators and energy denominators associated with the initial and final states.
No ``$Z$ factors'' are needed due to our renormalization conventions.
In this way, when the volume becomes infinite, we end up with scattering amplitudes having standard normalization, and which are equal to those obtained from the LSZ prescription applied to Feynman diagrams. In particular, they satisfy Lorentz invariance.
On the other hand, the amplitude in finite volume is a correlation function, albeit an unconventional one, and thus has poles at finite-volume energies.
These points are discussed in more detail in refs.~\cite{\HSQCb,\BSQC}.

The matrix $\widehat \cM_L$ divides into four blocks: the upper-left $2\to2$ block,
the lower-right $3\to3$ block, and two off-diagonal blocks corresponding to $2\to3$ and $3\to2$ processes. An example of the type of diagram contributing to the $2\to2$ block is shown in \Cref{fig:corrfcn}(ii).
The matrix indices in the two-particle block are $\{\bm k_1,\bm k_2\}$,
while those in the three-particle block are $\{\bm p_1,\bm p_2,\bm p_3\}$,
in both cases constrained by momentum conservation, \Cref{eq:momcons}.
Both diagonal blocks have been rederived using TOPT
in \Cref{app:KKTOPT,app:TOPT}, respectively.
This also serves to set our notation and define a number of quantities that enter the following discussion.

Following Ref.~\cite{\BSQC}, and as discussed in more detail in \Cref{app:TOPT},
we work with a modified object, $\widehat\cM_{23,L}$, instead of $\widehat\cM_{L}$.
The modification only affects the $3\to3$ block, 
to which we add all partially-connected diagrams in which a pair interacts
repeatedly with the third particle spectating. 
This change simplifies intermediate expressions.

\subsection{All orders expression for $\widehat\cM_{23,L}$}

By a straightforward extension of the arguments given in \Cref{app:KKTOPT,app:TOPT},
one can immediately write down an all-orders form for $\widehat \cM_{23,L}$ in the TOPT approach,
\begin{equation}
\widehat \cM_{23,L} =  \widehat \cA \widehat \cB \frac1{1 + \widehat D\widehat \cB} \widehat \cA\,.
\label{eq:MLhat}
\end{equation}
Here $\widehat \cB$ is a matrix of TOPT Bethe-Salpeter kernels
\begin{equation}
\widehat \cB = 
\begin{pmatrix}
\cB_2^{K\bar K,I=0} & \cB_{23} 
\\
\cB_{32} & (\cB_{2,L}^{(1)} +   \cB_{2,L}^{(2)} +   \cB_{2,L}^{(3)} +  \cB_3)
\end{pmatrix}\,,
\label{eq:Bhat}
\end{equation}
$\widehat D$ contains the two- and three-particle cut factors, $\widehat D = {\rm diag} ( D_2, D_3 )$, 
with $D_2$ and $D_3$ defined in \Cref{eq:D2def,eq:D3def}, respectively,
and $\widehat \cA$ is given by
\begin{equation}
\widehat \cA = {\rm diag}\left( P_A, \tfrac1{\sqrt6} \cA \right)\,.
\label{eq:Ahatdef}
\end{equation}
We stress that there are no symmetry factors in $\widehat D$, 
since the particles are distinguishable.
We also note that our convention is that the incoming states are on the right end of $\widehat\cM_{23,L}$, and outgoing states on the left end.

The Bethe-Salpeter kernels appearing in \Cref{eq:Bhat}
are sums over subsets of amputated TOPT diagrams with given initial and final states, and depend on the energy $E$ and the momenta of the incoming and outgoing particles.
Since the external states have $I^G=0^-$, and since these symmetries remain valid in finite volume, all the kernels appearing in $\widehat \cB$ are projected onto these quantum numbers. 
We stress that a kernel that is projected onto $I=0$ does not, in general, have the same normalization as the standard $I=0$ kernel, and only for the latter do we use the superscript ``$I=0$''.

The kernels contained in $\widehat \cB$ are all amputated TOPT Bethe-Salpeter kernels defined such that they have no $K\overline K$ or $3\pi$ $s$-channel intermediate states,
except for the $\cB_{2,L}^{(i)}$, which will be discussed separately below.
This implies that, up to terms exponentially suppressed in the volume, these kernels are infinite-volume quantities, since, for the kinematic range we consider, the only allowed on-shell states are $K\overline K$ and $3\pi$.
We note that, given our definition, the kernels are not strictly three-particle irreducible (3PI), since cuts involving the kinematically forbidden $K\overline K\pi$ states are allowed. An important remark is that the $\cB_3$ kernel in \Cref{eq:Bhat} is not the same as that with the same name in \Cref{eq:M23L}. The latter contains intermediate $K\bar K$ states, which cannot go on shell given the kinematical configuration considered in the previous appendix, while the former does not contain such intermediate states.

We now describe the kernels in more detail, beginning with $\cB_2^{K\bar K,I=0}$. The construction of this $I=0$ kernel from those involving particles of definite charge is 
described in \Cref{app:KKTOPT} (see \Cref{eq:B2hat,eq:B2KKdef}), and goes through unchanged.
The only difference here is that we use an explicit $K\bar K$ superscript to 
distinguish the kernel from that involving two pions.

The kernel $\cB_{32}$ connects an initial $K\bar K$ state to a three-pion final state. Useful results in the following are
\begin{equation}
\cB_{32} = \frac{\cA}6 \cB_{32} = \frac1{\sqrt 6} \cB_{32}^{I=0}\,,
\label{eq:AB32res}
\end{equation}
where the first equality follows from the fact that $\cB_{32}$ is (implicitly) projected onto $I=0$, and is thus completely antisymmetric on its three-pion side, while the second uses the standard normalization of the $I=0$ state given in \Cref{eq:ASstate} to define the $I=0$ kernel.
The definitions for $\cB_{23}$ are analogous.

The kernel $\cB_3$ connects three-pion states to themselves.
It satisfies the properties given in \Cref{eq:Arel,eq:B3rel}, from which one can show that
\begin{equation}
\cA \cB_3 = \frac{\cA}2 \frac{\cB_3^{I=0}}9 \frac{\cA}2 = \cB_3 \cA\,.
\label{eq:AB3res}
\end{equation}

Finally, $ \cB_{2,L}^{(i)}$ contains a  $2\pi \to 2\pi$ kernel that arises when pion $i$ is a spectator while the other two ($j$ and $k$) interact.
The form of this term is explained in Ref.~\cite{\BSQC} and is
\begin{equation}
  \cB_{2,L}^{(i)}= \delta_{\bm p'_i, \bm p_i} 2\omega_{\pi}(\bm p_i) L^3
\cB_2^{(jk)} (E - \omega_\pi(\bm p_i); \bm p'_j, \bm p'_k; \bm p_j, \bm p_k)\,,
\end{equation}
where $\{i,j,k\}$ is a cyclic permutation of $\{1,2,3\}$,
and $\cB_2^{(jk)}$ is the two-pion kernel in which $K\overline K$ intermediate states, as well as $4\pi$ and higher states, are allowed. 
Thus it is not strictly 2PI.
The properties that we need in the following are given in \Cref{eq:B2Lrel},
which we reproduce here using a notation in which the presence of $\pi\pi$ scattering is made explicit,
\begin{equation}
\cA \sum_i \cB_{2,L}^{(i)} = \frac{\cA}2 \cB_{2,L}^{\pi\pi,I=1} \frac{\cA}2 = \sum_i \cB_{2,L}^{(i)} \cA\,.
\label{eq:AB2Lres}
\end{equation}

Using \Cref{eq:AB32res,eq:AB3res,eq:AB2Lres}, we find that
\begin{equation}
\widehat \cM_{23,L} =  \widehat{ \cA}_{I=0} \, \widehat \cB_{I=0} 
\frac1{1 + \widehat{D}_{I=0}\, \widehat{ \cB}_{I=0}}\, \widehat{ \cA}_{I=0}\,,
\label{eq:MLhata}
\end{equation}
where the new matrices are
\begin{align}
\widehat{  \cB}_{I=0} &= 
\begin{pmatrix}
 \cB_2^{K\bar K,I=0} & \tfrac13 \cB_{23}^{I=0}
\\
\tfrac13 \cB_{32}^{I=0} & (\cB_{2,L}^{\pi\pi,I=1} +  \tfrac19 \cB_3^{I=0})
\end{pmatrix}\,,
\\
\widehat{ \cA}_{I=0} &= {\rm diag} \left( P_A, \ \tfrac12 \cA \right)\,,
\\
\widehat{ D}_{I=0} &= {\rm diag} \left( D_2, \tfrac1{4} D_3 \cA \right)
= {\rm diag} \left( D_2, \tfrac12 D_3 + D_G\right)\,.
\label{eq:Dtildehat}
\end{align}
The entries in the $2\to 2$ and $3\to 3$ blocks follow from the results in \Cref{app:KKTOPT,app:TOPT}, respectively, while those in the offdiagonal blocks are new.
The second equality in \Cref{eq:Dtildehat} uses the replacement \Cref{eq:D3AS},
with $D_G$ defined in \Cref{eq:DGdef}.
This replacement remains valid here due to the antisymmetry of $\cB_{23}$, $\cB_{32}$, and $\cB_{3}$ on their three-particle sides.

We now project on shell using the identities \Cref{eq:D2decomp,eq:FGiden}, and perform the standard rearrangement of the series, generalizing to the $2+3$ system the steps for the $2\to 2$ and $3\to 3$ blocks outlined in \Cref{app:KKTOPT,app:TOPT}, respectively.
The result is
\begin{align}
\widehat \cM_{23,L} &=  \widehat{X}_A \circ \widehat{\cM}_{23,L}^{(u,u)}\circ \widehat{ X}_A^\dagger\,,
\label{eq:MLhatb}
\\
\widehat{\cM}_{23,L}^{(u,u)} &= \widehat{\cK}_{23,L}^{(u,u)} \frac1{1 + \widehat{F}_G\, \widehat{\cK}^{(u,u)}_{23,L}}\,,
\label{eq:M23Luuhat}
\\
\widehat F_G &= {\rm diag} \left( F^{(K\bar K)}_2, \  \tfrac12 F - G \right)\,.
\label{eq:FGhat}
\end{align}
In these results, $\widehat \cM_{23,L}$ remains a function of momenta, though it now is (implicitly) on shell,
while $\widehat{\cK}_{23,L}^{(u,u)}$ and $\widehat F_G$ are matrices in which the
indices in the two- and three-particle sectors are $\{\ell m\}$ and $\{k \ell m\}$, respectively.
$\widehat F_G$ contains the finite-volume kinematic factors $F_2$, $F$, and $G$,
which are given, respectively, in \Cref{eq:F2def,eq:Fdef,eq:Gdef},
with the superscripts in the latter two definitions dropped here for the three-pion system.
The negative sign multiplying $G$ is explained following \Cref{eq:XAdef}, and follows from the antisymmetry of the three-particle kernels.
Finally, the matrix $\widehat{X}_A$ and its conjugate in \Cref{eq:MLhatb} contain operators that convert between the on shell $\{\ell m; k \ell m\}$ and momentum bases,
and are given by
\begin{equation}
    \widehat{X}_A = {\rm diag}(\bcX_2, \bcX_A)\,,\quad
    \widehat{X}_A^\dagger = {\rm diag}(\bcX_2^\dagger, \bcX_A^\dagger)\,,
    \label{eq:XAhat}
\end{equation}
where $\bcX_2$ and $\bcX_A$ are defined in \Cref{eq:X2def,eq:XAdef}, respectively.

We now turn to the matrix $\widehat{\cK}^{(u,u)}_{23,L}$ appearing in \Cref{eq:M23Luuhat}.
The off-shell version (with momentum coordinates) is given by
\begin{align}
\widehat{\cK}^{(u,u)}_{23,L} &= 
\widehat{P}_A \widehat{\cB}_{I=0} \frac1{1 + \widehat I_{FG} \widehat{\cB}_{I=0}} 
\widehat{P}_A \,,\quad
\widehat{P}_A = {\rm diag}(P_A,1)\,,
\label{eq:Kuu23Ldef}
\\
\widehat I_{FG} &= {\rm diag} \left( I_{F_2}, I_{FG} \right)\,,
\label{eq:IFGhat}
\end{align}
where the integral operators $I_{F_2}$ and $I_{FG}$ are defined in \Cref{eq:D2decomp,eq:FGiden}, respectively.
In \Cref{eq:M23Luuhat} we need the on-shell version, with $\{\ell m; k \ell m\}$ indices.
We also want at this stage to pull out the partially disconnected part of the $3\to 3$ amplitude. We can do this by decomposing the (implicitly on-shell) kernel as
\begin{align}
\widehat \cK_{23,L}^{(u,u)} &= \widehat \cK_{2,L} + \widehat \cK^{(u,u)}\,,
\\
\widehat \cK_{2,L} &= {\rm diag}\left(0, \,  \cK_{2,L}^{\pi\pi,I=1}\right)\,,
\\
\widehat \cK^{(u,u)} &=
\begin{pmatrix}
\cK_{22}^{K\bar K, I^G=0^-} & \cK_{23}^{(,u),I=0}
\\
\cK_{32}^{(u,),I=0} & \cK_{33}^{(u,u),I=0}
\end{pmatrix}
\,.
\label{eq:Kdfuu}
\end{align}
Here $\cK_{2,L}^{\pi\pi,I=1}$, defined in \Cref{eq:K2Ldef}, is given explicitly in \Cref{eq:K2Lres,eq:K2mod}.
This quantity is unaffected by the coupling of two- and three-particle sectors.
For the remaining, fully connected, part, $\widehat \cK^{(u,u)}$,
mixing between two- and three-particle sectors plays a central role.
Although in \Cref{eq:Kdfuu} we have given names to the diagonal components that mirror those appearing in \Cref{app:KKTOPT,app:TOPT} (with some minor changes), the kernels are different.

To explain this rather subtle point, we use the example of $\cK_{22}^{K\bar K, I^G=0^-}$.
First, we note that we have used the projectors $P_A$, which are present in \Cref{eq:Ahatdef} but not in \Cref{eq:XAhat}, to project this kernel onto negative $G$ parity, as noted explicitly in the superscript.\footnote{%
For the remaining kernels composing $\widehat \cK^{(u,u)}$ in \Cref{eq:Kdfuu}, this projection acts trivially since they involve three pions and thus automatically have negative $G$ parity. Thus we use simply $I=0$ in their superscripts.
}
Returning to the main point, we expand out \Cref{eq:Kuu23Ldef} to obtain
\begin{equation}
\begin{aligned}
\cK_{22}^{K\bar K,\,I^G=0^-} ={}&
\cB_2^{K\bar K,\,I^G=0^-}
- \cB_2^{K\bar K,\,I^G=0^-} I_{F_2} \cB_2^{K\bar K,\,I^G=0^-} \\
&+ \cB_2^{K\bar K,\,I^G=0^-} I_{F_2} \cB_2^{K\bar K,\,I^G=0^-} I_{F_2} \cB_2^{K\bar K,\,I^G=0^-}
+ \dots
\\
&- \frac{1}{9}\cB_{23}^{I=0} I_{FG} \cB_{32}^{I=0}
+ \frac{1}{9}\cB_{23}^{I=0} I_{FG} \cB_{32}^{I=0} I_{F_2} \cB_2^{K\bar K,\,I^G=0^-}
\\
&+ \frac{1}{9}\cB_2^{K\bar K,\,I^G=0^-} I_{F_2} \cB_{23}^{I=0} I_{FG} \cB_{32}^{I=0}
\\
&+ \frac{1}{9}\cB_{23}^{I=0} I_{FG}
\left(\cB_{2,L}^{\pi\pi,\,I=1}+\frac{1}{9}\cB_3^{I=0}\right)
I_{FG}\cB_{32}^{I=0}
+ \dots
\end{aligned}
\label{eq:K22KKexpand}
\end{equation}
The terms on the first two lines of the right-hand side form the standard two-particle geometric series corresponding to integrated $K\overline K$ intermediate states with the on-shell pole regulated with the PV prescription.
In \Cref{app:KKTOPT}, they lead to the expression in \Cref{eq:M22K22I0def} for $\cK_{22}^{I=0}$.
The terms on the last three lines of \Cref{eq:K22KKexpand}, by contrast, correspond to contributions involving virtual $3\pi$ intermediate states, as indicated by the appearances of the integral operator $I_{FG}$.
Similar contributions are also present in the kernel $\cK_{22}^{I^G=0^-}$ appearing in
\Cref{app:KKTOPT}, but there the $3\pi$ states are off shell, and thus can be fully included in the Bethe-Salpeter kernels, whereas here a separation must be made between the off-shell and on-shell contributions, with the former absorbed into the definition of $\cK_{22}^{I^G=0^-}$, while the latter is included explicitly through the matrix structure used here.  Therefore, the $\cB_2^{K\bar K,\,I^G=0^-}$ kernel of the pure $2\leftrightarrow 2$ formalism is not the same as the object with the same name in the coupled $2 \leftrightarrow 3$ formalism: they differ by the inclusion of intermediate $3\pi$ states.

Similar discussions hold for the expansions of the other kernels in \Cref{eq:Kdfuu}, so that, for example, $\cK_{33}^{(u,u),I=0}$ differs from the kernel $\cK_{3}^{(u,u),I=0}$ appearing in \Cref{app:TOPT} due to the differing treatments of intermediate $K\overline K$ states.
The offdiagonal kernels in \Cref{eq:Kdfuu} connect the two- and three-particle sectors and are new to this work.
The superscripts ``$(,u)$'' and ``$(u,)$'' indicate the lack of full antisymmetrization on their three-particle sides.  

We will not, in fact, need to use the explicit expressions for any of these kernels.
All we need to know is that these kernels are infinite-volume quantities that contain no singularities due to $K\overline K$ and $3\pi$ $s$-channel intermediate states.
Furthermore, due to the partial antisymmetry in the three-particle sector, and complete antisymmetry in the two-particle sector, all kernels are nonzero only if both $\ell'$ and $\ell$ are odd.

From \Cref{eq:MLhatb}, we can immediately write down the asymmetric form of the quantization condition:
\begin{equation}
\det \left(1 + \widehat F_G \widehat \cK_{23,L}^{(u,u)} \right) = 0 
\,.
\label{eq:QC23AS}
\end{equation}

\subsection{(Anti)symmetric form of formalism}
\label{app:twotothreeAS}

The final task of the derivation is to reexpress the results in terms of kernels whose three-particle sides are fully antisymmetric. This issue arises only in the three-particle sector and the methods used for the isolated $3\pi$ system, as described in \Cref{app:3piAS}, 
can, as explained below, be straightforwardly extended to include the $2\to 3$ and $3\to 2$ kernels.

The first step is to pull out the partially disconnected part of the $33$ element of $\widehat{\cM}_{23,L}$, as in \Cref{eq:M23Ldecomp,eq:M2LI1res},
\begin{align}
\widehat{\cM}_{23,L}^{(u,u)} &= \widehat{\cM}_{2,L}^{(u,u)} + \widehat{\cM}_{L}^{(u,u)} \,,
\label{eq:M23Luuhatdecomp}
\\
\widehat{\cM}_{2,L}^{(u,u)} &=
\begin{pmatrix} 0\quad & 0 \\ 0 & \quad\cK_{2,L}^{\pi\pi,I=1} \frac1{1+ \frac12 F \cK_{2,L}^{\pi\pi,I=1}} \end{pmatrix}\,.
\end{align}
The antisymmetrized, fully-connected finite-volume amplitude that we desire is then
\begin{align}
\widehat{\cM}_{L} &=  \widehat{X}_A \circ \widehat{\cM}_{L}^{(u,u)}\circ\widehat{X}_A^\dagger\,.
\end{align}
Using the result (C13) from ref.~\cite{\BSQC},\footnote{%
The derivation given in appendix C of ref.~\cite{\BSQC} relies on the invertibility of
$\widehat{\cM}_{2,L}^{(u,u)}$, which does not hold here. However, it is straightforward to adapt the derivation to the present case, e.g. by expanding out \Cref{eq:M23Luuhatdecomp,eq:M3Luuhatdecomp} and showing equivalence.
} 
we obtain an explicit expression for $\widehat{\cM}_{L}^{(u,u)}$,
\begin{align}
\widehat{\cM}_{L}^{(u,u)} &= \widehat \cD_{L} + \widehat{\cM}_{{\rm df},L}^{(u,u)} \,,
\label{eq:M3Luuhatdecomp}
\\
 \widehat{\cM}_{{\rm df},L}^{(u,u)} &= (1 - \widehat \cD_{23,L} \widehat F_G ) \widehat \cK^{(u,u)}
 \frac1{1 + (1 - \widehat F_G \widehat  \cD_{23,L}) \widehat F_G \widehat \cK^{(u,u)}}
 (1 - \widehat F_G \widehat \cD_{23,L})\,, \label{eq:Mhatdf3Luu}
 \\
  \widehat \cD_{L} &=  \widehat \cD_{23,L} - \widehat{\cM}_{2,L}^{(u,u)}\,,\quad
 \widehat \cD_{23,L} = \widehat{\cM}_{2,L}^{(u,u)} \frac1{1- \widehat{G} \widehat{\cM}_{2,L}^{(u,u)}}\,,\quad
 \widehat{G} = {\rm diag} \left( 0,\, G \right)\,.
  \label{eq:Dhat3L}
 \end{align}
 It is important to keep in mind that $\widehat{\cM}_{2,L}^{(u,u)}$, $\widehat \cD_{L}$, 
 $\widehat \cD_{23,L}$ and $\widehat G$ are nonzero only in the lower-right entry.
 The results in this entry for $\widehat \cD_{L}$ and $\widehat \cD_{23,L}$ are equivalent in form to those in the three-pion analysis, which are given in \Cref{eq:DLuures,eq:D23Luures}, respectively.

  Our aim now is to rewrite these expressions in terms of an antisymmetrized version of
 $\widehat \cK^{(u,u)}$. This antisymmetrization applies only for the parts of this matrix
 with $(u)$ superscripts, and has been described in detail in \Cref{app:3piAS} for the $I=0$ three-pion system.
 To apply it here, we simply need to generalize the antisymmetrization identities into the combined two- and three-particle matrix space,
 which is trivial as there is no need for further antisymmetrization in the two-particle sector.
 The identities \Cref{eq:ASidenta,eq:ASidentb,eq:ASidentc} apply to the three-particle sides of the $2\to3$ and $3\to 2$ kernels in $\widehat{\cK}^{(u,u)}$, as well as to both sides of the $3\to3$ kernel.
 We can therefore take over the algebraic steps presented in ref.~\cite{\BSQC}
 (and laid out in a notation closer to that used here in sec.~VIIIc of ref.~\cite{\BSnondegen}), and obtain the following results
 \begin{align}
\widehat{X}_A \circ \widehat \cM_{\df,L}^{(u,u)} \circ \widehat X_A^\dagger &=
\widehat{X}_A \circ \widehat \cL_L^{(u,)} \widehat \cK 
\frac1{1+ \widehat \cF\, \widehat \cK}
\widehat \cR_L^{(,u)} \circ\widehat X_A^\dagger\,,
\label{eq:Mhatdf3Luufin}
\\
\widehat \cL_L^{(u,)} &= {\rm diag}\left(1,\, \tfrac13 - \widehat \cD_{23,L} \tfrac12 F\right)\,,
\label{eq:LLuhat}
\\
\widehat \cR_L^{(,u)} &= {\rm diag}\left(1,\, \tfrac13 - \tfrac12 F \widehat \cD_{23,L} \right)\,,
\label{eq:RLuhat}
\end{align}
where $\widehat \cF$ and $\widehat \cK$ are given in \Cref{eq:cFhatKdfhat}.
The latter contains the quantities $\cK_{22}^{I^G=0^-}$,
$\cK_{23}^{I=0}$, $\cK_{32}^{I=0}$, and $\cK_{33}^{I=0}$, explicit expressions for which can be given in terms of the elements of $\widehat \cK^{(u,u)}$ and the integral operators appearing in the identities \Cref{eq:ASidenta,eq:ASidentb,eq:ASidentc}, but these are not illuminating.
The key point is that the entries in $\widehat \cK$ that couple to
three particles are (after combining with appropriate spherical harmonics) completely antisymmetric functions of the momenta.

From this result, we can read off the symmetric form of the quantization condition,
which is presented in the main text in \Cref{eq:QC23}.
This result, as well as that for $\widehat \cM_{L}$ given by the equations above, has exactly the same form as those in ref.~\cite{\BHSQC} [see Eqs.~(77) and (79) of that work],
aside from the presence of antisymmetrization operators here compared to symmetrization operators in ref.~\cite{\BHSQC}, and the change in sign of the $G$ term.
This is not a surprise---indeed, we could have guessed the form of the result here.
What we have achieved, however, is a simplified derivation, which, at the same time,
incorporates the additional features of having different particles in the two- and three-particle states, and the projection onto $I=0$ in the three-particle state.

The final form of the result for $\widehat \cM_{L}$ is
\begin{equation}
\widehat \cM_{L} =  \widehat{X}_A \circ \widehat \cD_{L} \circ\widehat X_A^\dagger
+ \widehat{X}_A \circ \widehat \cM_{\df,L}^{(u,u)} \circ\widehat X_A^\dagger\,,
\label{eq:M3Lhatfin}
\end{equation}
with the first term on the right-hand side given by \Cref{eq:Dhat3L}
and the second by \Cref{eq:Mhatdf3Luu}.
To obtain the integral equations relating $\widehat \cM$ to the K matrices, 
we follow ref.~\cite{\HSQCb} and insert $i\epsilon$ factors in the poles in $\widehat F_G$,
and then take the infinite-volume limit.
This leads to
\begin{align}
\widehat \cM &=  \lim_{\epsilon\to 0} \lim_{L\to\infty} \widehat \cM_{L}
\equiv \widehat \cD + \widehat \cM_{\rm df}\,,
\label{eq:inteqfin}
\\
\widehat\cD &= \lim_{\epsilon\to 0} 
\lim_{L\to \infty} \widehat{X}_A \circ \widehat \cD_{L} \circ\widehat X_A^\dagger\,,
\label{eq:Dhatfin}
\\
\widehat\cM_{\rm df} &= \lim_{\epsilon\to 0} \lim_{L\to\infty}
 \widehat{X}_A \circ \widehat \cM_{\df,L}^{(u,u)} \circ\widehat X_A^\dagger\,.
\label{eq:Mhatdf3fin}
\end{align}
Here $\widehat\cM_{\rm df}$ is the divergence-free amplitude, which is related to
$\widehat \cK$ by the $L\to\infty$ limit of \Cref{eq:Mhatdf3Luu}.
The detailed form of the integral equations that result is given in \Cref{sec:QCKKppp}. 
We stress that, since the elements of $\widehat \cM_{L}$ go over to Lorentz-invariant amplitudes when ${L\to\infty}$, and the integral equations themselves involve Lorentz-invariant measures, it follows (assuming the invertibility of the integral equations) that the K matrices contained in $\widehat \cK$ are themselves Lorentz invariant.

\section{General expressions for K matrices}
\label{app:threshold}

In this appendix we provide more detail on the  K matrices appearing in
\Cref{eq:cFhatKdfhat,eq:Kdfhatb,eq:Kdfhatc}, for many of which previous results are not available.
In particular, we provide the flavor factors in the $K\pi\leftrightarrow K\pi\pi$ systems and the threshold expansions for K matrices in all systems.

The threshold expansion that we use builds on that introduced in ref.~\cite{Blanton:2019igq}.
Although it will be expressed in terms of Lorentz invariants,
it is essentially an expansion in powers of the three-momentum of the particles in the overall CMF.
Looking at it in the latter fashion allows one to easily understand the power counting associated with new terms arising here that involve the Levi-Civita tensor.
One complication is the presence of two thresholds, namely those for the two- and three-particle systems, which are, in general, different.
Here we follow the approach of ref.~\cite{Draper:2024qeh}, which considered multiple three-particle channels, and simply treat all three momenta in the CMF as of the same order,
which we generically denote $|\bm p|$. For each quantity we work out the threshold expansion at least for the first two nontrivial orders.

With this introduction, we turn first to the $K\overline K\leftrightarrow3\pi$ system,
and consider the $2\leftrightarrow 3$ K matrices contained in \Cref{eq:cFhatKdfhat}.
These are related by PT symmetry,
\begin{equation}
\cK_{23}^{K\bar K, 3\pi, I=0}(\{k\},\{p\}) = \cK_{32}^{3\pi,K\bar K,I=0}(\{p\},\{k\}) \,,
\end{equation}
so we need only consider the $23$ matrix, i.e. that corresponding to $K\bar K \leftarrow 3\pi$.
We are writing these matrices as functions of the on-shell four-momenta of the particles, rather than the corresponding three-momenta, which is equivalent but more convenient in the following.
We are also using the collective notation $\{k\}=\{k_1,k_2\}$ and $\{p\}= \{p_1,p_2,p_3\}$.
Since the K matrices are Lorentz invariant, we expand them in terms of Lorentz-invariant combinations of the momenta with unknown coefficients. 
These combinations must be antisymmetric under the interchange $ k_1 \leftrightarrow k_2$, and also under permutations of the $p_i$. 
Furthermore, because of the intrinsic negative parity of all five particles, the overall parity of the Lorentz-invariant combinations must be negative.
This implies that there must be an odd number of Levi-Civita tensors.

The leading term involves a single such tensor and four momenta,
\begin{align}
\begin{split}
\cK_{23}^{(3)} &=  [k_1 k_2 p_1 p_2]_\epsilon + [k_1 k_2 p_2 p_3]_\epsilon + [k_1 k_2 p_3 p_1]_\epsilon\,,
\\
&= 3 [k_1 k_2 p_1 p_2]_\epsilon\,,
\end{split}
\label{eq:K23a}
\end{align}
where we are using the notation exemplified by
\begin{equation}
[k_1 k_2 p_1 p_2]_\epsilon \equiv
\epsilon_{\mu\nu\rho\sigma} k_1^\mu k_2^\nu p_1^\rho p_2^\sigma\,.
\label{eq:epdef}
\end{equation}
We use the convention $\epsilon_{0123}=-\epsilon^{0123} = 1$.
The second equality in \Cref{eq:K23a} follows from momentum conservation, $P=k_1+k_2=p_1+p_2+p_3$, and the antisymmetry of the Levi-Civita tensor.
This is, in fact, the form of the contribution predicted by the WZW term, as will be seen in \Cref{app:chpt} below.
In our threshold expansion, this term is of $\cO(|\bm p|^3)$, because, in the contraction with the Levi-Civita tensor, spatial components are used in three of the four four-momenta. 
Here and in the following the order in the threshold expansion is denoted by the superscript on the term.

Including two more factors of momenta, we find only a single term,
\begin{align}
\cK_{23}^{(5)} & = [k_1 k_2 p_1 p_2]_\epsilon (s - 9 M_\pi^2)\,,
\end{align}
where $s=P^2 = E^{\star 2}$, and, for definiteness, we are using the three-pion threshold.
This term is of $\cO(|\bm p|^5)$.
At the next order there are four terms of $\cO(|\bm p|^7)$, which we do not display.
\footnote{%
The counting of terms can be done using a group-theoretical analysis along the lines of that in Appendix B of ref.~\cite{Baeza-Ballesteros:2024mii}.}
Putting all this together, we find the threshold expansion
\begin{equation}
    \cK_{23}^{K\bar K, 3\pi, I=0}(\{k\},\{p\}) = c_{(3)}^{K\bar K,3\pi,I=0} \cK_{23}^{(3)}
    + c_{(5)}^{K\bar K,3\pi,I=0} \cK_{23}^{(5)} + \cO(|\bm p|^7)\,,
\end{equation}
where the constants $c_{(i)}$ are unknown, but are to be determined from fits to finite-volume spectra.

Now we turn to the $K\pi\leftrightarrow K\pi\pi$ processes.
For the $3\to 3$ K matrices contained in \Cref{eq:Kdfhatb,eq:Kdfhatc}, we can take over the results for the $N\pi\pi$ system presented in eqs.~(A.45-A.47) of ref.~\cite{\Npp}, with minor adaptations in notation.
Results for both $I=3/2$ and $1/2$ can be written in the same form
\begin{equation}
\left[ \widetilde {\cK}_{\rm 33}^{K\pi\pi,I}\right]_{i' k' \ell' m'; i k \ell m} =
\sum_{x,y\in S,A}
\boldsymbol {\mathcal Y}^{I; x}_{i';k'\ell' m'} 
\circ \cK_{33}^{K\pi\pi,I;\; x, y}(\{p'\},\{p\})
\circ \left[\boldsymbol {\mathcal Y}^{I; y}_{i;k \ell m}\right]^\dagger
\,.
\label{eq:KdfKppI}
\end{equation}
Here $\cK_{33}^{K\pi\pi,I; x,y}$ is an underlying K matrix, expressed as a function of initial and final four-momentum triplets. 
The two choices for $x$ and $y$ correspond to basis states that are symmetric  (``$S$'') and antisymmetric  (``$A$'') under interchange of the two pions. 
For $I=3/2$ these are, respectively, $[(\pi\pi)_2 K]_{3/2}$ and $[(\pi\pi)_1 K]_{3/2}$,
i.e. the states in \Cref{eq:ppK3S,eq:ppK3A}, respectively.
For $I=1/2$ they are $[(\pi\pi)_0 K]_{1/2}$ and $[(\pi\pi)_1 K]_{1/2}$,
i.e. the states in \Cref{eq:ppK1S,eq:ppK1A}, respectively.

Returning to \Cref{eq:KdfKppI}, the $\bcY$ operators convert from the momentum basis to the $\{k\ell m\}$ basis, which requires a choice of spectator, and thus expands the basis size from two to four. The explicit form of these matrices of operators is given in eqs.~(A.46) and (A.47) of ref.~\cite{\Npp}, which we reproduce here for completeness.
Collecting the $\left[\boldsymbol {\mathcal Y}^{I; y}_{i;k \ell m}\right]^\dagger$
into a row vector in which $i$ is the implicit label, and keeping the $\{k\ell m\}$ indices implicit, we have, for $I=3/2$,
\begin{align}
\begin{split}
\left[\bcY^{I=3/2;S}\right]^\dagger &= \left( \sqrt{\tfrac12}\YL312,\ 0,\ -\sqrt{\tfrac16}\YL132,\ -\sqrt{\tfrac56} \YL132 \right)\,,
\\
\left[\bcY^{I=3/2;A}\right]^\dagger &= \left( 0,\ \sqrt{\tfrac12} \YL312,\ -\sqrt{\tfrac56}\YL132,\ \sqrt{\tfrac16} \YL132 \right)\,,
\end{split}
\label{eq:Y3SA}
\end{align}
where the $\boldsymbol \cY^{[kab] \dagger }_{\boldsymbol \sigma}$ operators 
are defined in \Cref{eq:YLdef},
and, for $I=1/2$,
\begin{align}
\begin{split}
\left[\bcY^{I=1/2;S}\right]^\dagger &= \left( 0,\ \sqrt{\tfrac12} \YL312,\ \sqrt{\tfrac23}\YL132,\ -\sqrt{\tfrac13} \YL132 \right)\,,
\\
\left[\bcY^{I=1/2;A}\right]^\dagger &= \left( \sqrt{\tfrac12}\YL312,\ 0,\ \sqrt{\tfrac13}\YL132,\ \sqrt{\tfrac23} \YL132 \right)\,.
\end{split}
\label{eq:Y1SA}
\end{align}
The conjugate operators are defined analogously.

Next we consider the offdiagonal elements in \Cref{eq:Kdfhatb,eq:Kdfhatc}.
Here, for each isospin, there are two underlying K matrices, corresponding to whether the $K\pi\pi$ state is symmetric or antisymmetric under pion interchange.
Using the symmetrization operator for $2+1$ systems given in eq.~(80) of ref.~\cite{\BStwoplusone}, together with the flavor recoupling coefficients from the kaon spectator states to those with a pion spectator, we find
\begin{align}
\left[ \widetilde {\cK}_{23}^{K\pi,K\pi\pi,I}\right]_{\ell' m'; i k \ell m} &=
\sum_{y\in S,A}
\boldsymbol {\mathcal Y}_{2;\ell' m'} 
\circ \cK_{23}^{K\pi,K\pi\pi,I;\; y}(\{k\},\{p\})
\circ \left[\boldsymbol {\mathcal Y}^{I; y}_{i;k \ell m}\right]^\dagger
\,,
\label{eq:K23KppI}
\\
\left[ \widetilde {\cK}_{32}^{K\pi\pi,K\pi,I}\right]_{i' k' \ell' m'; \ell m} &=
\sum_{x\in S,A}
\boldsymbol {\mathcal Y}^{I; x}_{i';k'\ell' m'} 
\circ \cK_{32}^{K\pi\pi,K\pi,I;\; x}(\{p\},\{k\})
\circ \left[\boldsymbol {\mathcal Y}_{2; \ell m}\right]^\dagger
\,.
\label{eq:K32KppI}
\end{align}
In other words, the same $\bcY^{I;S}$ and $\bcY^{I;A}$ vectors that appear in $\cK_{33}$, \Cref{eq:KdfKppI}, also appear here.
What is new here is the need to introduce the operator $\boldsymbol {\mathcal Y}_2$, defined in \Cref{eq:cY2def}, which converts two-particle quantities from the momentum to the $\{\ell m\}$ basis.
In these results, there are only two independent underlying operators, because PT symmetry implies
\begin{equation}
    \cK_{23}^{K\pi,K\pi\pi,I;\; x}(\{k\},\{p\}) = 
    \cK_{32}^{K\pi\pi,K\pi,I;\; x}(\{p\},\{k\})\,.
    \label{eq:PTfor23}
\end{equation}
We provide the threshold expansion for these operators below.

To complete the specification of the $3\to 3$ K matrices, we need to determine the threshold expansions for the underlying K matrices in \Cref{eq:KdfKppI}.
These are constrained by Lorentz and PT invariance, as well as their symmetry properties under pion exchanges. The necessary analysis has been largely carried out in ref.~\cite{Draper:2024qeh}, which considered the kinematically similar $\pi\pi\eta$ channel.
To express the results we introduce the quantities
\begin{equation}
\begin{gathered}
M_{\rm tot}=2 M_\pi+M_K\,,\ \Delta = s  - M_{\rm tot}^2\,,\ 
t_{ij} =(p'_i - p_j)^2 - (M_i-M_j)^2\,,
\\
\Delta'_i = (P - p'_i)^2-(M_{\rm tot}- M_i)^2
= \sum_j t_{ij} + \Delta\,,\quad \sum_i \Delta'_i = \Delta\,,
\\
\Delta_j = (P - p_j)^2-(M_{\rm tot}- M_j)^2
= \sum_i t_{ij} + \Delta\,,\quad \sum_j \Delta_j = \Delta\,.
\end{gathered}
\end{equation}
All parity-even Lorentz invariants can be written in terms of the $t_{ij}$,
with $\Delta'_i$ and $\Delta_j$ being useful combinations.
These three quantities, together with $\Delta$, vanish at the $K\pi\pi$ threshold, and are of $\cO(|\bm p|^2)$, so can be used to develop the threshold expansion.

We choose the momentum labels such that $i=\{1,2,3\}$ corresponds to $\{\pi,\pi, K\}$, so that $S$ and $A$ refer to symmetry or antisymmetry under $1\leftrightarrow2$ exchange.
The counting of allowed terms is the same as for the $DD\pi$ system, which has the same symmetry structure, and was done in ref.~\cite{\tetraquark}. It was checked using a group-theoretical argument and extended to higher order in Appendix F of ref.~\cite{Draper:2024qeh}. We use a somewhat different choice of the basis of operators here.
We note that the results for $I=3/2$ and $1/2$ have the same form, differing only by the values of the unknown coefficients in the expressions below.
The $\cO(|\bm p|^2)$ results can be written
\begin{align}
\begin{split}
    \cK_{33}^{K\pi\pi,I;SS} &= c_{(0)}^{K\pi\pi,I;SS} + c_{(2)a}^{K\pi\pi,I;SS} \Delta
    + c_{(2)b}^{K\pi\pi,I;SS} t_{33}
    + c_{(2)c}^{K\pi\pi,I;SS} (\Delta'_3+\Delta_3)\,, % (t_{11}+t_{12} + t_{21} +  t_{22})\,,
\end{split}
\label{eq:Kdf3KppISS}
\\
\cK_{33}^{K\pi\pi,I;AA} &= c_{(2)}^{K\pi\pi,I;AA} ( t_{11}- t_{12} -t_{21} + t_{22})\,,
\label{eq:Kdf3KppIAA}
\\
\cK_{33}^{K\pi\pi,I;SA} &= c_{(2)a}^{K\pi\pi,I;SA} (\Delta_1-\Delta_2) +
c_{(2)b}^{K\pi\pi,I;SA} (t_{31}-t_{32}) \,, 
\label{eq:Kdf3KppISA}
\\
\cK_{33}^{K\pi\pi,I;AS} &= c_{(2)a}^{K\pi\pi,I;SA} (\Delta'_1-\Delta'_2) +
c_{(2)b}^{K\pi\pi,I;SA} (t_{13}-t_{23}) \,. 
\label{eq:Kdf3KppIAS}
\end{align}
Note that PT symmetry implies that the constants entering the last two equations are the same, as shown.

Finally, we describe the threshold expansion for $\cK_{23}^{K\pi,K\pi\pi,I;\; x}(\{k\},\{ p\})$,
with $x=S,A$. Here, unlike the $K\overline K\leftrightarrow 3\pi$ case discussed above,
there is no symmetry constraint on the two-particle side. The only constraint is the symmetry or antisymmetry under pion exchange on the three-particle side.
As for the $K\overline K\leftrightarrow 3\pi$ system, parity implies that there must be an odd number of Levi-Civita tensors. We have determined the allowed terms by first working in terms of three-momenta in the overall CMF, where one can enumerate rotation invariants straightforwardly using angular momentum addition, and then converting to Lorentz invariant forms. 
For the antisymmetric case, we find a single term at $\cO(|\bm p|^3)$, which has the same form as that for $K\overline K \leftrightarrow 3\pi$ in \Cref{eq:K23a}, and four terms at $\cO(|\bm p|^5)$,
\begin{align}
\begin{split}
\cK_{23}^{K\pi,K\pi\pi,I\;;A}
&= [k_1 k_2 p_1 p_2]_\epsilon \bigg[
c_{(3)}^{K\pi,K\pi\pi,I\;;A}
+ c_{(5)a}^{K\pi,K\pi\pi,I\;;A}\Delta
\\
&\quad
+ c_{(5)b}^{K\pi,K\pi\pi,I\;;A}\Delta_3
+ c_{(5)c}^{K\pi,K\pi\pi,I\;;A}
(k_1\cdot k_2-M_\pi M_K)
\\
&\quad
+ c_{(5)d}^{K\pi,K\pi\pi,I\;;A}
\left[
(k_1-k_2)\cdot p_3-(M_\pi-M_K)M_K
\right]
+ \cO(|\bm p|^7)
\bigg]\,.
\end{split}
\label{eq:K23KpKppA}
\end{align}
All except the ``$(5)d$'' term are antisymmetric under $k_1\leftrightarrow k_2$, and thus couple only to the odd $K\pi$ partial waves, while the $(5)d$ term couples only to even waves.
For the symmetric case, there are no terms at $\cO(|\bm p|^3)$, and two at $\cO(|\bm p|^5)$, 
\begin{multline}
\cK_{23}^{K\pi,K\pi\pi,I\;;S} =  [k_1 k_2 p_1 p_2]_\epsilon \bigg[
c_{(5)a}^{K\pi,K\pi\pi,I\;;S} (k_1+k_2)\cdot(p_1-p_2)
\\
+ c_{(5)b}^{K\pi,K\pi\pi,I\;;S} (k_1 - k_2) \cdot(p_1-p_2)
+ \cO(|\bm p|^7)
\bigg]\,.
\label{eq:K23KpKppS}
\end{multline}
The first term couples only to odd $K\pi$ partial waves, while the second only to even waves.

\section{ChPT predictions for K matrices}
\label{app:chpt}

In this appendix we present results from ChPT for all the K matrices that enter the three quantization conditions presented in the main text. 
We work only at the lowest order in which contributions occur in ChPT, which can be either LO or NLO.
A few of the required results can be obtained from the literature, but most are new.
Since the amplitudes involve kaons, we necessarily use $SU(3)$ ChPT, for which convergence is poorer than in the $SU(2)$ case.
Nevertheless, our results should give a guide to the values of the various K matrices.

The quantization conditions contain K matrices of two types: those in $F_3$ (which involve only two particles) and those in $\widehat\cK$
(which involve two and three particles).
The former do not depend on cutoff functions and can be expressed in terms of phase shifts, and in this sense are physical.
At LO in ChPT they are equal to the corresponding two-particle amplitudes, for which the results are well known (and are reproduced below).
The K matrices in $\widehat\cK$ are cutoff-dependent and thus unphysical.
In principle, they must be calculated using the integral equations that relate $\widehat \cK$ to scattering amplitudes. As will be discussed below, at the lowest order in the chiral expansion, a simpler analytical connection is possible. This method was first used in Ref.~\cite{\HHanal,\implement} to determine the three-particle K matrices involving pions and kaons at LO, and extended to NLO in Refs.~\cite{Baeza-Ballesteros:2023ljl,Baeza-Ballesteros:2024mii} for the three-pion system.
Here we generalize the method to include the $2\leftrightarrow3$ K matrices.

We explain the method for the $K \overline K \leftrightarrow 3\pi$ system,
for which $\widehat \cK$ is given in \Cref{eq:cFhatKdfhat}.
In general, the integral equations relating K matrices to scattering amplitudes,
given by the $L\to\infty$ limit of \Cref{eq:Mhatdf3Luu,eq:M3Lhatfin}, are nontrivial.
For example, $\cM_{22}^{I^G=0^-}$, the $[K \bar K]_{I^G=0^-} \to [K \bar K]_{I^G=0^-}$ amplitude, is related not only to $\cK_{22}^{K\bar K,I^G=0^-}$, but also to the other elements of  $\widehat \cK$,
and the relations are nonlinear.
This represents the physical feature that intermediate $3\pi$ states, which can, in general, go on shell, contribute to the $K\bar K$ amplitude.
However, as we will see, such nonlinear effects enter only at high orders in ChPT, 
and are absent at leading nontrivial order.

To understand this claim, we observe that elements of the scattering amplitude matrix scale in terms of the $SU(3)$ ChPT decay constant $F$ (not to be confused with the kinematic function of the same name entering the quantization condition),
\begin{equation}
\widehat \cM \equiv \begin{pmatrix} \cM_{22}^{I^G=0^-} & \cM_{23}^{I=0} \\
\cM_{32}^{I=0} & \cM_{33}^{I=0} \end{pmatrix}
 \sim \begin{pmatrix} 1/F^2 & 1/F^5 \\ 1/F^5 &  1/F^4 \end{pmatrix}\,.
\end{equation}
Note, in particular, that the offdiagonal elements, which are of NLO in ChPT, are suppressed by $1/F^2$ compared to their natural size of $1/F^3$.
The same scaling holds for the divergence-free amplitude $\widehat \cM_{\rm df}$, defined by \Cref{eq:M3Lhatfin,eq:Mhatdf3fin}.
Next we consider the integral equations relating $\widehat \cM_{\rm df}$ and $\widehat \cK$, given by the $L\to\infty$ limit of \Cref{eq:Mhatdf3Luu}.
It is straightforward to see that the leading contributions to the elements of $\widehat \cK$ must scale in the same way as those of $\widehat\cM_{\rm df}$, and, furthermore, must be equal
\begin{align}
\begin{split}
     \cK_{22}^{K\bar K, I^G=0^-} &= \cM_{2}^{K\bar K,I^G=0^-, LO} [1 + \cO(1/F^2)]\,, 
     \quad
\cK_{23}^{I=0}  = \cM_{23}^{I=0,NLO} [1 + \cO(1/F^2)]\,, \\
\cK_{32}^{I=0} & = \cM_{32}^{I=0,NLO} [1 + \cO(1/F^2)] \,, \quad
\cK_{33}^{I=0}  = \cM_{\df,3}^{I=0,LO} [1 + \cO(1/F^2)]\,.
\end{split}
\label{eq:KfromM}
\end{align}
Analogous relations hold for the $K\pi\leftrightarrow K\pi\pi$ processes.
We note that, since the scattering amplitudes on the right-hand side of these relations are physical quantities with no cutoff dependence (as will be seen explicitly below), it follows that the lowest nontrivial results for the corresponding K matrices are also cutoff-independent.

A corollary of the scaling just described is that the contributions of three-pion intermediate states to the $K\bar K$ amplitude enter at order $1/F^{10}$, i.e. at N$^4$LO in ChPT, since two transitions between two- and three-particle states are needed.
Similarly, the contribution of two-particle intermediate states to the $3\to 3$ amplitude scales as $1/F^{10}$,
and is thus subleading compared to the leading behavior of $1/F^4$.
Thus the two- and three-particle sectors decouple until high orders in ChPT.

In the following we collect and/or determine the ChPT results for the amplitudes appearing on the right-hand sides of the relations in \Cref{eq:KfromM}.

\subsection{Results needed for $K\overline K \leftrightarrow 3\pi$}

We begin with the two-particle K matrices.
That which enters in $F_3$ is given at LO by 
\begin{align}
\cK_2^{\pi\pi,I=1,LO} &= \cM_2^{\pi\pi,I=1,LO} = \frac{1}{F^2} (t-u)\,.
\label{eq:K2pp1}
\end{align}
For the two-particle K matrix entering $\widehat \cK$, we need
\begin{align}
\cK_{22}^{K \bar K, I^G=0^-} \approx \cM_{2}^{K \bar K, I^G=0^-, LO} = \frac{3}{4 F^2} (t-u)\,.
\label{eq:K22KKb0}
\end{align}

As seen from \Cref{eq:KfromM}, the $2\leftrightarrow 3$ K matrices that we need are given, at leading nontrivial order, by the corresponding scattering amplitudes.
These appear at NLO in ChPT, and arise from the WZW term.
The required amplitude is
\begin{equation}
\cM_{23}\left(K(k_1) \bar K(k_2) \leftarrow \pi^+(p_1) \pi^0(p_2) \pi^-(p_3) \right)
= C_{WZW} \frac1{\pi^2 F^5} [k_1 k_2 p_2 p_3]_\epsilon\,,
\label{eq:WZWamp}
\end{equation}
where we are using the notation of \Cref{eq:epdef}, and the coefficient is 
\begin{equation}
C_{WZW} = \frac34\,.
\end{equation}
Note that the result is the same for $K^+ K^-$ and $K^0 \overline K^0$ final states.
Using the normalized final and initial states given, respectively, in \Cref{eq:ASstate,eq:KKbarI0}, we find
 \begin{align}
\cK_{23}^{I=0}(\{k_i\};\{p_i\}) &= \frac{\sqrt{12} C_{WZW}}{\pi^2 F^5}
[k_1 k_2 p_2 p_3]_\epsilon \left[1 + \cO(1/F^2)\right]
\,.
 \label{eq:M23LO}
 \end{align}
 The factor of $\sqrt2\sqrt6=\sqrt12$ arises from wavefunctions of the $I=0$ $3\pi$ and $K\bar K$ states. 
 
Finally, it is simple to see that~\cite{Baeza-Ballesteros:2024mii}
\begin{equation}
\cM_{\df,3}^{I=0,LO} = 0\,,
\end{equation}
because the amplitude begins at quadratic order in the threshold expansion~\cite{\isospin},
whereas a putative LO chiral contribution would occur at linear order. The lowest nonzero order, NLO, is given in Ref.~\cite{Baeza-Ballesteros:2024mii}.

\subsection{K matrices for $K\pi \leftrightarrow K\pi\pi$ with $I=3/2$}
\label{app:chpt:KpKpp3}

For the $\cK_2$ factors entering $F_3$ we need [see \Cref{eq:K2LKpp3}]
\begin{align}
\cK_{2}^{\pi\pi,I=2,LO} &= \cM_2^{\pi\pi,I=2,LO} = \frac{1}{F^2} (2M_\pi^2-s)\,,
\label{eq:K2pp2}
\\
\cK_{2}^{K\pi,I=3/2,LO} &= \cM_2^{K\pi,I=3/2,LO} = \frac{1}{2F^2} (M_\pi^2+M_K^2-s)\,,
\label{eq:K2Kp3}
\\
\cK_{2}^{K\pi,I=1/2,LO} &= \cM_2^{K\pi,I=1/2,LO} = \frac{1}{F^2} (s + 3 t/4 - M_\pi^2-M_K^2)\,,
\label{eq:K2Kp1}
\end{align}
as well as \Cref{eq:K2pp1}.
For the $22$ component of $\widehat \cK$, defined in \Cref{eq:Kdfhatb}, we need \Cref{eq:K2Kp3}.

For the offdiagonal K matrices, the ChPT amplitude comes from the WZW term,
which implies that the two-pion subchannel in $K\pi\pi$ must be antisymmetric,
and thus have $I=1$.
This implies that
\begin{equation}
\cK_{23}^{K\pi,K\pi\pi,I=3/2;S} = 0\,.
\end{equation}
To obtain the nonzero contribution, we need the decomposition for the $m=3/2$ state 
with an $I=1$ pion pair, 
\begin{equation}
    [(\pi \pi)_1 K]_{3/2} = \frac1{\sqrt2}
    \left\{ \pi^+(p_1) \pi^0(p_2) K^+(p_3) - \pi^0(p_1) \pi^+(p_2) K^+(p_3) \right\}\,,
\end{equation}
as well as the form of the corresponding $K\pi$ state,
$[\pi K]_{3/2} = \pi^+(k_1) K^+(k_2)$.
The WZW vertex contributes
\begin{equation}
    \mathcal M_{23}(\pi^+(k_1) K^+(k_2) \leftarrow  \pi^+(p_1) \pi^0 (p_2)K^+(p_3) )= \frac{C_{WZW}}{\pi^2 F^5} [k_1 k_2 p_1 p_2]_\epsilon\,.
\end{equation}
From this we find, using the notation of \Cref{eq:K23KppI}
\begin{equation}
\cK_{23}^{K\pi,K\pi\pi,I=3/2;A} = \frac{\sqrt2 C_{WZW}}{\pi^2 F^5} [k_1 k_2 p_1 p_2]_\epsilon \left[1 + \cO(1/F^2) \right]\,. 
\end{equation}

\begin{figure}[h!]
\centering
\includegraphics[width=0.9\textwidth]{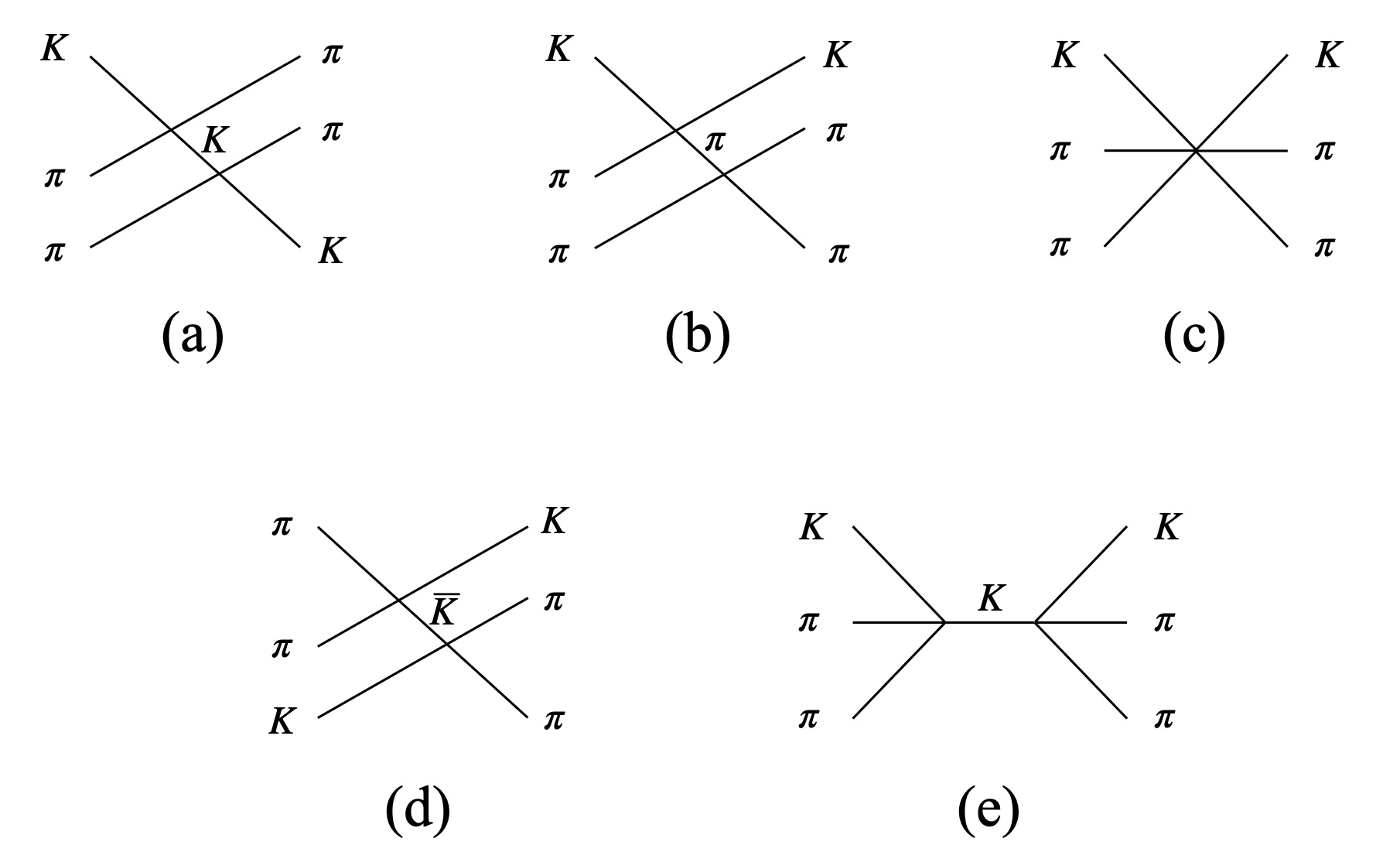}
\caption{Classes of diagrams contributing to $\cM_{\rm df,3}$ at LO in ChPT.
}
\label{fig:Mdfdiag}
\end{figure}

We now turn to the $3\to 3$ K matrices, $\cK_{33}^{K\pi\pi,I=3/2;x,y}$,
where $x,y=S,A$, corresponding to symmetric/antisymmetric pion pairs
[see \Cref{eq:KdfKppI}].
From \Cref{eq:KfromM}, we obtain these by calculating the corresponding LO $\cM_{\df,3}$. 
The method for doing so was worked out for $3\pi^+$ in Ref.~\cite{Blanton:2019vdk}, generalized to maximal isospin $K\pi\pi$ in Ref.~\cite{Blanton:2021eyf}, and generalized to three pions of all isospins in Ref.~\cite{Baeza-Ballesteros:2024mii}.
The five types of diagram that contribute are shown in \Cref{fig:Mdfdiag}.
Diagrams (a)-(b) require a subtraction to remove the divergence from the exchanged particle, and this is explained for the $K\pi\pi$ system in Ref.~\cite{Blanton:2021eyf}. The contact term, diagram (c), is present for all systems. Diagram (d) is new for the present system; it is not, however, divergent since the required on-shell cut involves three kaons. Diagram (e) (``$s$-channel one-particle exchange'') contributes only to $I=1/2$, and is similar to a diagram appearing for three pions with $I=1$.

The calculations are straightforward but tedious, as there are many choices of charge assignments that contribute. 
We have done two independent calculations using {\sf Mathematica}.

The final result is not a polynomial in the Mandelstam variables due to the subtraction in diagrams of classes (a) and (b), as well as the contribution from (d). 
Rather than quote the full, rather cumbersome result, we expand to linear order in the Mandelstam variables and quote results for the coefficients appearing in the threshold expansions described in \Cref{app:threshold}; see, in particular, \Cref{eq:Kdf3KppISS,eq:Kdf3KppIAA,eq:Kdf3KppISA}.   
In the $I=3/2$ channel, and to the quoted order, only classes (a-c) contribute, and the results are collected in \Cref{tab:Kpp3Chpt}. The total contributions are given in \Cref{tab:Kpp3Chpttot}.

\begin{table}
    \centering
    \tiny
    \begin{tabular}{l||c|c|c}
         &  $(a)$ & $(b)$  & $(c)$ \\[5pt]\hline\hline
     $c_{(0)}^{K\pi\pi,I=3/2;SS}$\vphantom{\rule{0pt}{12pt}}&
$\frac{M_K M_\pi (3 M_K-2 M_\pi)}{M_K+M_\pi}$ &
$-\frac{M_\pi \left(6 M_K^2+2 M_K M_\pi+11 M_\pi^2\right)}{3 (M_K+M_\pi)}$ &
$\frac{2}{3} M_\pi (M_\pi-3 M_K)$
     \\[5pt]\hline
     $c_{(2)a}^{K\pi\pi,I=3/2;SS}$ \vphantom{\rule{0pt}{12pt}}  &  
$\frac{3 M_K^3-M_K^2 M_\pi+4 M_K M_\pi^2-2 M_\pi^3}{4 (M_K+M_\pi)^3}$ &
$-\frac{2 M_K^3+6 M_K^2 M_\pi+11 M_K M_\pi^2-3 M_\pi^3}{4 (M_K+M_\pi)^3}$ &
$-\frac{1}{2}$
     \\[5pt] \hline
     $c_{(2)b}^{K\pi\pi,I=3/2;SS}$ \vphantom{\rule{0pt}{12pt}}   &  
$\frac{79 M_K^2+68 M_K M_\pi-56 M_\pi^2}{144 (M_K+M_\pi)^2}$ &
$-\frac{19}{36}$ &
$-\frac{1}{3}$
     \\[5pt]\hline
     $c_{(2)c}^{K\pi\pi,I=3/2;SS}$ \vphantom{\rule{0pt}{12pt}}   & 
$-\frac{5 \left(19 M_K^3+3 M_K^2 M_\pi+12 M_K M_\pi^2-8 M_\pi^3\right)}{144 (M_K+M_\pi)^3}$ &
$\frac{14 M_K^3+12 M_K^2 M_\pi-3 M_K M_\pi^2-31 M_\pi^3}{24 (M_K+M_\pi)^3}$ &
$\frac{7}{18}$
     \\[5pt]\hline
     $c_{(2)}^{K\pi\pi,I=3/2;AA}$  \vphantom{\rule{0pt}{12pt}}   &  
$-\frac{13 M_K^2+8 M_K M_\pi+4 M_\pi^2}{144 (M_K+M_\pi)^2}$ &
$-\frac{4 M_K+M_\pi}{12 (M_K+M_\pi)}$ &
$-\frac{5}{36}$
     \\[5pt]\hline
     $c_{(2)a}^{K\pi\pi,I=3/2;SA}$  \vphantom{\rule{0pt}{12pt}}   &  
$-\frac{\sqrt{5} \left(13 M_K^3+15 M_K^2 M_\pi+12 M_K M_\pi^2-2 M_\pi^3\right)}{48 (M_K+M_\pi)^3}$ &
$\frac{\sqrt{5} \left(M_K^3-5 M_\pi^3\right)}{24 (M_K+M_\pi)^3}$ &
$\frac{\sqrt{5}}{24}$
     \\[5pt]\hline
     $c_{(2)b}^{K\pi\pi,I=3/2;SA}$  \vphantom{\rule{0pt}{12pt}}   &  
$\frac{\sqrt{5} \left(5 M_K^2+10 M_K M_\pi+2 M_\pi^2\right)}{48 (M_K+M_\pi)^2}$ &
$\frac{\sqrt{5} (M_K+4 M_\pi)}{24 (M_K+M_\pi)}$ &
$\frac{\sqrt{5}}{24}$
    \end{tabular}
    \caption{Contribution from the classes of diagrams shown in \Cref{fig:Mdfdiag} to the zeroth- and second-order threshold expansion coefficients of $\cK_{33}^{K\pi\pi, I=3/2}$. Diagram (d) contributes only at quadratic order, while (e) does not contribute at any order.
    }
    \label{tab:Kpp3Chpt}
\end{table}

\begin{table}
    \centering
    \begin{tabular}{l||c}
Coefficient         &  Total \\[5pt]\hline\hline
     $c_{(0)}^{K\pi\pi,I=3/2;SS}$\vphantom{\rule{0pt}{18pt}}&
$-M_\pi (M_K+3 M_\pi)$
     \\[5pt]\hline
     $c_{(2)a}^{K\pi\pi,I=3/2;SS}$ \vphantom{\rule{0pt}{18pt}}  &  
$-\frac{M_K^2+12 M_K M_\pi+M_\pi^2}{4 (M_K+M_\pi)^2}$
     \\[5pt] \hline
     $c_{(2)b}^{K\pi\pi,I=3/2;SS}$ \vphantom{\rule{0pt}{18pt}}   &  
$-\frac{5 (M_K+2 M_\pi)^2}{16 (M_K+M_\pi)^2}$
     \\[5pt]\hline
     $c_{(2)c}^{K\pi\pi,I=3/2;SS}$ \vphantom{\rule{0pt}{18pt}}   & 
$\frac{5 \left(M_K^2+4 M_K M_\pi-2 M_\pi^2\right)}{16 (M_K+M_\pi)^2}$
     \\[5pt]\hline
     $c_{(2)}^{K\pi\pi,I=3/2;AA}$  \vphantom{\rule{0pt}{18pt}}   &  
$-\frac{(3 M_K+2 M_\pi)^2}{16 (M_K+M_\pi)^2}$
     \\[5pt]\hline
     $c_{(2)a}^{K\pi\pi,I=3/2;SA}$  \vphantom{\rule{0pt}{18pt}}   &  
$-\frac{\sqrt{5} \left(3 M_K^2+2 M_\pi^2\right)}{16 (M_K+M_\pi)^2}$
     \\[5pt]\hline
     $c_{(2)b}^{K\pi\pi,I=3/2;SA}$  \vphantom{\rule{0pt}{18pt}}   &  
$\frac{\sqrt{5} (M_K+2 M_\pi) (3 M_K+2 M_\pi)}{16 (M_K+M_\pi)^2}$
    \end{tabular}
    \caption{Total LO contribution to the threshold expansion coefficients of $\cK_{33}^{K\pi\pi, I=3/2}$.}
    \label{tab:Kpp3Chpttot}
\end{table}

\subsection{K matrices for $K\pi \leftrightarrow K\pi\pi$ with $I=1/2$}
\label{app:chpt:KpKpp1}

For the $\cK_2$ factors entering $F_3$ we need [see \Cref{eq:K2LKpp1}]
\begin{align}
\cK_{2}^{\pi\pi,I=0,LO} &= \cM_2^{\pi\pi,I=0,LO} = \frac{1}{F^2} (2s - M_\pi^2)\,,
\label{eq:K2pp0}
\end{align}
as well as \Cref{eq:K2pp1,eq:K2Kp3,eq:K2Kp1}.
For the $22$ component of $\widehat \cK$, defined in \Cref{eq:Kdfhatc}, we need \Cref{eq:K2Kp1}.

For the offdiagonal K matrices, again the two-pion subchannel in $K\pi\pi$ must be antisymmetric, and thus has $I=1$. 
It follows that
\begin{equation}
\cK_{23}^{K\pi,K\pi\pi,I=1/2;S} = 0\,.
\end{equation}
The decomposition for the $m=1/2$ state with $I=1$ two-pion subchannel is
\begin{multline}
    [(\pi \pi)_1 K]_{1/2} = \sqrt{\frac13}
    \left\{ \pi^+(p_1) \pi^0(p_2) K^0(p_3) - \pi^0(p_1) \pi^+(p_2) K^0(p_3) \right\}
    \\
    - \sqrt{\frac16}
    \left\{ \pi^+(p_1) \pi^-(p_2) K^+(p_3) - \pi^-(p_1) \pi^+(p_2) K^+(p_3) \right\}\,,
\end{multline}
while for the corresponding $K\pi$ state we have 
\begin{equation}
[\pi K]_{1/2} = \sqrt{\frac23} \pi^+(k_1) K^0(k_2) - \sqrt{\frac13} \pi^0(k_1) K^+(k_2) \,.
\end{equation}
The WZW vertex contributes to $\pi^+ K^0\leftarrow \pi^+\pi^0 K^0$ and
$\pi^0 K^+ \leftarrow \pi^+\pi^- K^+$, with an equal magnitude to that for
$K \overline K\leftarrow \pi^+\pi^0 \pi^-$, \Cref{eq:WZWamp}, 
and signs that can be determined by crossing.
Taking into account the flavor factors, we find
\begin{equation}
\cK_{23}^{K\pi,K\pi\pi,I=1/2;A} = \frac{\sqrt2 C_{WZW}}{3 \pi^2 F^5} [k_1 k_2 p_1 p_2]_\epsilon \left[1 + \cO(1/F^2) \right]\,. 
\end{equation}

Finally, we present results for the $I=1/2$ three-to-three K matrices, $\cK_{33}^{K\pi\pi,I=1/2;x,y}$.
The calculation is very similar to that for $I=3/2$, except now with nonzero contributions from all five classes of diagram shown in \Cref{fig:Mdfdiag}.
Contributions from individual types are collected in \Cref{tab:Kpp1Chptabc,tab:Kpp1Chptde},
with the totals given in \Cref{tab:Kpp1Chpttot}.

\begin{table}
    \centering
    \begin{tabular}{l||c|c|c} %|c|c}
         &  $(a)$ & $(b)$  & $(c)$  
         % & $(d)$  & $(e)$  
         \\[5pt]\hline\hline
     $c_{(0)}^{K\pi\pi,I=1/2;SS}$\vphantom{\rule{0pt}{18pt}}&
     $-\frac{4 M_K^2 M_\pi}{M_K+M_\pi}$ &
     $\frac{7 M_\pi^2 (M_\pi-2 M_K)}{3 (M_K+M_\pi)}$ &
     $-\frac{5 M_\pi^2}{3}$ % &
     \\[5pt]\hline
     $c_{(2)a}^{K\pi\pi,I=1/2;SS}$ \vphantom{\rule{0pt}{18pt}}  &  
     $-\frac{M_K^3+M_K^2 M_\pi+2 M_KM_\pi^2}{(M_K+M_\pi)^3}$ &
     $\frac{7 M_\pi^2 (M_K-M_\pi)}{4 (M_K+M_\pi)^3}$ &
     0 
     \\[5pt] \hline
     $c_{(2)b}^{K\pi\pi,I=1/2;SS}$ \vphantom{\rule{0pt}{18pt}}   &  
     $-\frac{47 M_K^2+76 M_K M_\pi+20 M_\pi^2}{144 (M_K+M_\pi)^2}$ &
     $\frac{5}{9}$ &
     0 
     \\[5pt]\hline
     $c_{(2)c}^{K\pi\pi,I=1/2;SS}$ \vphantom{\rule{0pt}{18pt}}   & 
     $\frac{85 M_K^3+93 M_K^2 M_\pi+156 M_K M_\pi^2+4 M_\pi^3}{144 (M_K+M_\pi)^3}$ &
     $-\frac{16 M_K^3+24 M_K^2 M_\pi+21 M_K M_\pi^2-29 M_\pi^3}{24 (M_K+M_\pi)^3}$ &
     $-\frac{5}{18}$  
     \\[5pt]\hline
     $c_{(2)}^{K\pi\pi,I=1/2;AA}$  \vphantom{\rule{0pt}{18pt}}   &  
     $\frac{7 M_K^2-4 M_K M_\pi-2 M_\pi^2}{72 (M_K+M_\pi)^2}$ &
     $-\frac{M_K-2 M_\pi}{3 (M_K+M_\pi)}$ &
     $-\frac{5}{36}$ 
     \\[5pt]\hline
     $c_{(2)a}^{K\pi\pi,I=1/2;SA}$  \vphantom{\rule{0pt}{18pt}}   &  
     $\frac{M_K^3-21 M_K^2 M_\pi+30 M_K M_\pi^2+4 M_\pi^3}{24 \sqrt{2} (M_K+M_\pi)^3}$ &
     $-\frac{4 M_K^3+18 M_K^2 M_\pi+45 M_K M_\pi^2-11 M_\pi^3}{12 \sqrt{2} (M_K+M_\pi)^3}$ &
     $\frac5{12\sqrt{2}}$   
     \\[5pt]\hline
     $c_{(2)b}^{K\pi\pi,I=1/2;SA}$  \vphantom{\rule{0pt}{18pt}}   &  
     $\frac{M_K^2+2 M_K M_\pi+4 M_\pi^2}{24 \sqrt{2} (M_K+M_\pi)^2}$ &
     $\frac{M_\pi-2 M_K}{6 \sqrt{2} (M_K+M_\pi)}$ &
     $\frac5{12\sqrt{2}}$ 
    \end{tabular}
    \caption{Contributions from diagrams in classes (a), (b), and (c) from \Cref{fig:Mdfdiag} to the threshold expansion coefficients of $\cK_{33}^{K\pi\pi, I=1/2}$.
    }
    \label{tab:Kpp1Chptabc}
\end{table}

\begin{table}
    \centering
    \begin{tabular}{l||c|c} 
         & $(d)$  & $(e)$  \\[5pt]\hline\hline
     $c_{(0)}^{K\pi\pi,I=1/2;SS}$\vphantom{\rule{0pt}{18pt}}&
     $\frac{M_\pi (M_K+2 M_\pi)^2}{12 (M_K-M_\pi)}$ &
     $-\frac{M_\pi (M_K-2 M_\pi)^2}{12 (M_K+M_\pi)}$ 
     \\[5pt]\hline
     $c_{(2)a}^{K\pi\pi,I=1/2;SS}$ \vphantom{\rule{0pt}{18pt}}  &  
     $\frac{(M_K-4 M_\pi) (M_K+2 M_\pi)}{48 (M_K-M_\pi)^2}$ &
     $-\frac{(M_K-2 M_\pi) (M_K+4 M_\pi)}{48 (M_K+M_\pi)^2}$ 
     \\[5pt] \hline
     $c_{(2)b}^{K\pi\pi,I=1/2;SS}$ \vphantom{\rule{0pt}{18pt}}   &  
     $\frac{(M_K-4 M_\pi) (M_K+2 M_\pi)}{48 (M_K-M_\pi)^2}$ &
     0 
     \\[5pt]\hline
     $c_{(2)c}^{K\pi\pi,I=1/2;SS}$ \vphantom{\rule{0pt}{18pt}}   & 
     $\frac{(2 M_K+M_\pi) (M_K+2 M_\pi)}{48 (M_K-M_\pi)^2}$ &
     $\frac{M_K-2 M_\pi}{16 (M_K+M_\pi)}$
     \\[5pt]\hline
     $c_{(2)}^{K\pi\pi,I=1/2;AA}$  \vphantom{\rule{0pt}{18pt}}   &  
     0 &
     0 
     \\[5pt]\hline
     $c_{(2)a}^{K\pi\pi,I=1/2;SA}$  \vphantom{\rule{0pt}{18pt}}   &  
     0 &
     $-\frac{M_K-2 M_\pi}{8 \sqrt{2} (M_K+M_\pi)}$
     \\[5pt]\hline
     $c_{(2)b}^{K\pi\pi,I=1/2;SA}$  \vphantom{\rule{0pt}{18pt}}   &  
     $\frac{M_K+2 M_\pi}{8 \sqrt{2} (M_\pi-M_K)}$ &
     0    
    \end{tabular}
    \caption{Contributions from diagrams in classes (d) and (e) from \Cref{fig:Mdfdiag} to the threshold expansion coefficients of $\cK_{33}^{K\pi\pi, I=1/2}$.
    }
    \label{tab:Kpp1Chptde}
\end{table}

\begin{table}
    \centering
    \begin{tabular}{l||c}
        Coefficient & Total \\[5pt]\hline\hline

        $c_{(0)}^{K\pi\pi,I=1/2;SS}$\vphantom{\rule{0pt}{18pt}}
        &
        $-\frac{
        M_K M_\pi
        \left(8 M_K^2+3 M_K M_\pi-14 M_\pi^2\right)}
        {2 (M_K-M_\pi)(M_K+M_\pi)}$
        \\[5pt]\hline

        $c_{(2)a}^{K\pi\pi,I=1/2;SS}$\vphantom{\rule{0pt}{18pt}}
        &
        $-\frac{
        4 M_K^5-4 M_K^4 M_\pi-3 M_K^3 M_\pi^2
        +12 M_K^2 M_\pi^3-10 M_K M_\pi^4+7 M_\pi^5}
        {4 (M_K-M_\pi)^2 (M_K+M_\pi)^3}$
        \\[5pt]\hline

        $c_{(2)b}^{K\pi\pi,I=1/2;SS}$\vphantom{\rule{0pt}{18pt}}
        &
        $\frac{
        2 M_K^4+M_K^3 M_\pi-6 M_K^2 M_\pi^2
        -5 M_K M_\pi^3+2 M_\pi^4}
        {8 (M_K-M_\pi)^2 (M_K+M_\pi)^2}$
        \\[5pt]\hline

        $c_{(2)c}^{K\pi\pi,I=1/2;SS}$\vphantom{\rule{0pt}{18pt}}
        &
        $-\frac{
        2 M_K^5+3 M_K^4 M_\pi-14 M_K^3 M_\pi^2
        -14 M_K^2 M_\pi^3+18 M_K M_\pi^4-7 M_\pi^5}
        {8 (M_K-M_\pi)^2 (M_K+M_\pi)^3}$
        \\[5pt]\hline

        $c_{(2)}^{K\pi\pi,I=1/2;AA}$\vphantom{\rule{0pt}{18pt}}
        &
        $\frac{4 M_\pi^2-3 M_K^2}
        {8 (M_K+M_\pi)^2}$
        \\[5pt]\hline

        $c_{(2)a}^{K\pi\pi,I=1/2;SA}$\vphantom{\rule{0pt}{18pt}}
        &
        $\frac{
        M_\pi\left(14 M_\pi^2-7 M_K M_\pi-9 M_K^2\right)}
        {8\sqrt{2}(M_K+M_\pi)^3}$
        \\[5pt]\hline

        $c_{(2)b}^{K\pi\pi,I=1/2;SA}$\vphantom{\rule{0pt}{18pt}}
        &
        $\frac{
        M_\pi\left(M_K^2-5 M_K M_\pi-8 M_\pi^2\right)}
        {8\sqrt{2}(M_K-M_\pi)(M_K+M_\pi)^2}$

    \end{tabular}
    \caption{Total LO contribution to the threshold expansion
    coefficients of
    $\cK_{33}^{K\pi\pi,I=1/2}$. }
    \label{tab:Kpp1Chpttot}
\end{table}

 \clearpage
\bibliographystyle{JHEP}
\bibliography{ref.bib}

@article{Adler:1969gk,
    author = "Adler, Stephen L.",
    title = "{Axial vector vertex in spinor electrodynamics}",
    doi = "10.1103/PhysRev.177.2426",
    journal = "Phys. Rev.",
    volume = "177",
    pages = "2426--2438",
    year = "1969"
}

@article{Bell:1969ts,
    author = "Bell, J. S. and Jackiw, R.",
    title = "{A PCAC puzzle: $\pi^0 \to \gamma \gamma$ in the $\sigma$ model}",
    doi = "10.1007/BF02823296",
    journal = "Nuovo Cim. A",
    volume = "60",
    pages = "47--61",
    year = "1969"
}

@article{Bardeen:1969md,
    author = "Bardeen, William A.",
    title = "{Anomalous Ward identities in spinor field theories}",
    doi = "10.1103/PhysRev.184.1848",
    journal = "Phys. Rev.",
    volume = "184",
    pages = "1848--1857",
    year = "1969"
}

@article{Bali:2021qem,
    author = {Bali, Gunnar S. and Braun, Vladimir and Collins, Sara and Sch{\"a}fer, Andreas and Simeth, Jakob},
    collaboration = "RQCD",
    title = "{Masses and decay constants of the {\ensuremath{\eta}} and {\ensuremath{\eta}}' mesons from lattice QCD}",
    eprint = "2106.05398",
    archivePrefix = "arXiv",
    primaryClass = "hep-lat",
    doi = "10.1007/JHEP08(2021)137",
    journal = "JHEP",
    volume = "08",
    pages = "137",
    year = "2021"
}

@article{ParticleDataGroup:2020ssz,
    author = "Zyla, P. A. and others",
    collaboration = "Particle Data Group",
    title = "{Review of Particle Physics}",
    doi = "10.1093/ptep/ptaa104",
    journal = "PTEP",
    volume = "2020",
    number = "8",
    pages = "083C01",
    year = "2020"
}

@article{HadronSpectrum:2009krc,
    author = "Peardon, Michael and Bulava, John and Foley, Justin and Morningstar, Colin and Dudek, Jozef and Edwards, Robert G. and Joo, Balint and Lin, Huey-Wen and Richards, David G. and Juge, Keisuke Jimmy",
    collaboration = "Hadron Spectrum",
    title = "{A Novel quark-field creation operator construction for hadronic physics in lattice QCD}",
    eprint = "0905.2160",
    archivePrefix = "arXiv",
    primaryClass = "hep-lat",
    reportNumber = "JLAB-THY-09-985",
    doi = "10.1103/PhysRevD.80.054506",
    journal = "Phys. Rev. D",
    volume = "80",
    pages = "054506",
    year = "2009"
}

@article{Alharazin:2026lno,
    author = "Alharazin, Herzallah and Raposo, Andr{\'e} Bai{\~a}o and Bulava, John and Dawid, Sebastian and Green, Jeremy R. and Morningstar, Colin and Romero-L{\'o}pez, Fernando and Salg, Miguel and Sharpe, Stephen R. and Stump, Andres",
    title = "{Three-body study of the Tcc(3875)+Tcc(3875)+ from lattice QCD}",
    eprint = "2602.17204",
    archivePrefix = "arXiv",
    primaryClass = "hep-lat",
    reportNumber = "DESY-26-021, HU-EP-26/10-RTG",
    doi = "10.22323/1.518.0085",
    journal = "PoS",
    volume = "LATTICE2025",
    pages = "085",
    year = "2026"
}

@article{Feng:2026ixm,
    author = {Feng, Yuchuan and Culver, Chris and D{\"o}ring, Michael and Mai, Maxim and Alexandru, Andrei and Lee, X., Frank},
    title = "{Coupled-channel approach to isotensor {\ensuremath{\pi}}{\ensuremath{\pi}}{\ensuremath{\pi}} scattering from lattice QCD}",
    eprint = "2601.16916",
    archivePrefix = "arXiv",
    primaryClass = "hep-lat",
    reportNumber = "JLAB-THY-26-4594",
    doi = "10.1103/dbm5-w8wx",
    journal = "Phys. Rev. D",
    volume = "114",
    number = "3",
    pages = "034506",
    year = "2026"
}

@article{Sakthivasan:2026bph,
    author = {Sakthivasan, Ajay S. and Feng, Yuchuan and D{\"o}ring, Michael and Mai, Maxim},
    title = "{The $a_1(1420)$ in a Unitary Coupled-Channel Three-Body Approach}",
    eprint = "2606.24709",
    archivePrefix = "arXiv",
    primaryClass = "hep-ph",
    reportNumber = "JLAB-THY-26-4812",
    month = "6",
    year = "2026"
}

@article{Bijnens:1993xi,
    author = "Bijnens, J.",
    title = "{Chiral perturbation theory and anomalous processes}",
    doi = "10.1142/S0217751X93001235",
    journal = "Int. J. Mod. Phys. A",
    volume = "8",
    pages = "3045--3105",
    year = "1993"
}

@inproceedings{Sharpe:2026mtt,
    author = "Sharpe, Stephen R.",
    title = "{Three-particle scattering amplitudes from lattice QCD}",
    booktitle = "{42th International Symposium on Lattice Field Theory}",
    eprint = "2601.04147",
    archivePrefix = "arXiv",
    primaryClass = "hep-lat",
    month = "1",
    year = "2026"
}

@article{Witten:1983tw,
    author = "Witten, Edward",
    title = "{Global Aspects of Current Algebra}",
    reportNumber = "PRINT-83-0262 (PRINCETON)",
    doi = "10.1016/0550-3213(83)90063-9",
    journal = "Nucl. Phys. B",
    volume = "223",
    pages = "422--432",
    year = "1983"
}

@article{ExtendedTwistedMass:2023hin,
    author = "Alexandrou, C. and others",
    collaboration = "Extended Twisted Mass",
    title = "{Pion transition form factor from twisted-mass lattice QCD and the hadronic light-by-light {\ensuremath{\pi}}0-pole contribution to the muon g-2}",
    eprint = "2308.12458",
    archivePrefix = "arXiv",
    primaryClass = "hep-lat",
    doi = "10.1103/PhysRevD.108.094514",
    journal = "Phys. Rev. D",
    volume = "108",
    number = "9",
    pages = "094514",
    year = "2023"
}

@article{Meyer:2013dxa,
    author = "Meyer, Harvey B.",
    title = "{Lattice QCD and the two-photon decay of the neutral pion}",
    eprint = "1303.0138",
    archivePrefix = "arXiv",
    primaryClass = "hep-lat",
    reportNumber = "MITP-13-017",
    doi = "10.1140/epja/i2013-13084-9",
    journal = "Eur. Phys. J. A",
    volume = "49",
    pages = "84",
    year = "2013"
}

@article{Feng:2012ck,
    author = "Feng, Xu and Aoki, Sinya and Fukaya, Hidenori and Hashimoto, Shoji and Kaneko, Takashi and Noaki, Jun-ichi and Shintani, Eigo",
    title = "{Two-photon decay of the neutral pion in lattice QCD}",
    eprint = "1206.1375",
    archivePrefix = "arXiv",
    primaryClass = "hep-lat",
    reportNumber = "KEK-CP-271, OU-HET-743-2012",
    doi = "10.1103/PhysRevLett.109.182001",
    journal = "Phys. Rev. Lett.",
    volume = "109",
    pages = "182001",
    year = "2012"
}

@article{Gerardin:2023naa,
    author = "G{\'e}rardin, Antoine and Verplanke, Willem E. A. and Wang, Gen and Fodor, Zoltan and Guenther, Jana N. and Lellouch, Laurent and Szabo, Kalman K. and Varnhorst, Lukas",
    title = "{Lattice calculation of the {\ensuremath{\pi}}0, {\ensuremath{\eta}} and {\ensuremath{\eta}}' transition form factors and the hadronic light-by-light contribution to the muon g-2}",
    eprint = "2305.04570",
    archivePrefix = "arXiv",
    primaryClass = "hep-lat",
    doi = "10.1103/PhysRevD.111.054511",
    journal = "Phys. Rev. D",
    volume = "111",
    number = "5",
    pages = "054511",
    year = "2025"
}

@article{Briceno:2025yuq,
    author = "Brice{\~n}o, Ra{\'u}l A. and Hansen, Maxwell T. and Jackura, Andrew W. and Edwards, Robert G. and Thomas, Christopher E.",
    title = "{Isotensor $\pi\pi\pi$ scattering with a $\rho$ resonant subsystem from QCD}",
    eprint = "2510.24894",
    archivePrefix = "arXiv",
    primaryClass = "hep-lat",
    reportNumber = "JLAB-THY-25-4592",
    month = "10",
    year = "2025"
}

@article{Yan:2025mdm,
    author = {Yan, Haobo and Mai, Maxim and Garofalo, Marco and Feng, Yuchuan and D{\"o}ring, Michael and Liu, Chuan and Liu, Liuming and Mei{\ss}ner, Ulf-G. and Urbach, Carsten},
    title = "{Emergence of the $\pi(1300)$ Resonance from Lattice QCD}",
    eprint = "2510.09476",
    archivePrefix = "arXiv",
    primaryClass = "hep-lat",
    month = "10",
    year = "2025"
}

@article{Yan:2024gwp,
    author = "Yan, Haobo and Mai, Maxim and Garofalo, Marco and Mei{\ss}ner, Ulf-G. and Liu, Chuan and Liu, Liuming and Urbach, Carsten",
    title = "{{\ensuremath{\omega}} Meson from Lattice QCD}",
    eprint = "2407.16659",
    archivePrefix = "arXiv",
    primaryClass = "hep-lat",
    doi = "10.1103/PhysRevLett.133.211906",
    journal = "Phys. Rev. Lett.",
    volume = "133",
    number = "21",
    pages = "211906",
    year = "2024"
}

@article{Hansen:2025oag,
    author = "Hansen, Maxwell T. and Romero-L{\'o}pez, Fernando and Sharpe, Stephen R.",
    title = "{Finite-volume formalism for N{\ensuremath{\pi}}{\ensuremath{\pi}} at maximal isospin}",
    eprint = "2509.24778",
    archivePrefix = "arXiv",
    primaryClass = "hep-lat",
    doi = "10.1007/JHEP02(2026)221",
    journal = "JHEP",
    volume = "02",
    pages = "221",
    year = "2026"
}

@article{Jackura:2025wbw,
    author = "Jackura, Andrew W. and Chambers, Nicholas C. and Brice{\~n}o, Ra{\'u}l A.",
    title = "{Symmetrizing relativistic three-body partial wave amplitudes}",
    eprint = "2507.14098",
    archivePrefix = "arXiv",
    primaryClass = "hep-ph",
    month = "7",
    year = "2025"
}

@article{Alotaibi:2025pxz,
    author = "Alotaibi, Athari and Hansen, Maxwell T. and Brice{\~n}o, Ra{\'u}l A.",
    title = "{Implementing the finite-volume three-pion scattering formalism across all non-maximal isospins}",
    eprint = "2508.11627",
    archivePrefix = "arXiv",
    primaryClass = "hep-lat",
    month = "8",
    year = "2025"
}

@article{Wess:1971yu,
    author = "Wess, J. and Zumino, B.",
    title = "{Consequences of anomalous Ward identities}",
    doi = "10.1016/0370-2693(71)90582-X",
    journal = "Phys. Lett. B",
    volume = "37",
    pages = "95--97",
    year = "1971"
}

@article{Dawid:2025zxc,
    author = {Dawid, Sebastian M. and Draper, Zachary T. and Hanlon, Andrew D. and H{\"o}rz, Ben and Morningstar, Colin and Romero-L{\'o}pez, Fernando and Sharpe, Stephen R. and Skinner, Sarah},
    title = "{QCD Predictions for Physical Multimeson Scattering Amplitudes}",
    eprint = "2502.14348",
    archivePrefix = "arXiv",
    primaryClass = "hep-lat",
    reportNumber = "MIT-CTP/5845",
    doi = "10.1103/6nql-yrhw",
    journal = "Phys. Rev. Lett.",
    volume = "135",
    number = "2",
    pages = "021903",
    year = "2025"
}

@article{Dawid:2025doq,
    author = {Dawid, Sebastian M. and Draper, Zachary T. and Hanlon, Andrew D. and H{\"o}rz, Ben and Morningstar, Colin and Romero-L{\'o}pez, Fernando and Sharpe, Stephen R. and Skinner, Sarah},
    title = "{Two- and three-meson scattering amplitudes with physical quark masses from lattice QCD}",
    eprint = "2502.17976",
    archivePrefix = "arXiv",
    primaryClass = "hep-lat",
    reportNumber = "MIT-CTP/5846",
    doi = "10.1103/bx16-lp3r",
    journal = "Phys. Rev. D",
    volume = "112",
    number = "1",
    pages = "014505",
    year = "2025"
}

@article{Luscher:1986pf,
    author = "Luscher, M.",
    title = "{Volume Dependence of the Energy Spectrum in Massive Quantum Field Theories. 2. Scattering States}",
    reportNumber = "DESY-86-034",
    doi = "10.1007/BF01211097",
    journal = "Commun. Math. Phys.",
    volume = "105",
    pages = "153--188",
    year = "1986"
}

@article{Luscher:1986n2,
 author = {L\"{u}scher, M.},
 title = "{Volume Dependence of the Energy Spectrum in Massive
 Quantum Field Theories. 2. Scattering States}",
 journal = "Commun.Math.Phys.",
 volume = "105",
 pages = "153-188",
 doi = "10.1007/BF01211097",
 year = "1986",
 reportNumber = "DESY 86/034",
 SLACcitation = "%%CITATION = CMPHA,105,153;%%",
}

@article{Luscher:1991n1,
 author = {L\"{u}scher, Martin},
 title = "{Two particle states on a torus and their relation to the
 scattering matrix}",
 journal = "Nucl.Phys.",
 volume = "B354",
 pages = "531-578",
 doi = "10.1016/0550-3213(91)90366-6",
 year = "1991",
 reportNumber = "DESY-90-131",
 SLACcitation = "%%CITATION = NUPHA,B354,531;%%",
}

@article{Luscher:1991n2,
 author = {L\"{u}scher, Martin},
 title = "{Signatures of unstable particles in finite volume}",
 journal = "Nucl.Phys.",
 volume = "B364",
 pages = "237-254",
 doi = "10.1016/0550-3213(91)90584-K",
 year = "1991",
 reportNumber = "DESY-91-052",
 SLACcitation = "%%CITATION = NUPHA,B364,237;%%",
}

@article{Kim:2005gf,
 author = "Kim, C. h. and Sachrajda, C. T. and Sharpe, Stephen R.",
 title = "{Finite-volume effects for two-hadron states in moving
 frames}",
 journal = "Nucl. Phys.",
 volume = "B727",
 year = "2005",
 pages = "218-243",
 doi = "10.1016/j.nuclphysb.2005.08.029",
 eprint = "hep-lat/0507006",
 archivePrefix = "arXiv",
 primaryClass = "hep-lat",
 reportNumber = "UW-PT-05-16, SHEP-0518",
 SLACcitation = "%%CITATION = HEP-LAT/0507006;%%"
}

@article{Morningstar:2011ka,
 author = "Morningstar, Colin and Bulava, John and Foley, Justin and Juge, Keisuke J. and
 Lenkner, David and Peardon, Mike and Wong, Chik Him",
 title = "{Improved stochastic estimation of quark propagation with Laplacian Heaviside smearing in lattice QCD}",
 eprint = "1104.3870",
 archivePrefix = "arXiv",
 primaryClass = "hep-lat",
 reportNumber = "DESY-11-035",
 doi = "10.1103/PhysRevD.83.114505",
 journal = "Phys. Rev. D",
 volume = "83",
 pages = "114505",
 year = "2011"
}

@article{Hansen:2012tf,
 author = "Hansen, Maxwell T. and Sharpe, Stephen R.",
 title = "{Multiple-channel generalization of Lellouch-L\"{u}scher
 formula}",
 journal = "Phys.Rev.",
 volume = "D86",
 pages = "016007",
 doi = "10.1103/PhysRevD.86.016007",
 year = "2012",
 eprint = "1204.0826",
 archivePrefix = "arXiv",
 primaryClass = "hep-lat",
 SLACcitation = "%%CITATION = ARXIV:1204.0826;%%",
}

@article{Hansen:2014eka,
 author = "Hansen, Maxwell T. and Sharpe, Stephen R.",
 title = "{Relativistic, model-independent, three-particle
 quantization condition}",
 journal = "Phys. Rev.",
 volume = "D90",
 year = "2014",
 number = "11",
 pages = "116003",
 doi = "10.1103/PhysRevD.90.116003",
 eprint = "1408.5933",
 archivePrefix = "arXiv",
 primaryClass = "hep-lat",
 SLACcitation = "%%CITATION = ARXIV:1408.5933;%%"
}

@article{Bruno:2014jqa,
 author = "Bruno, Mattia and others",
 title = "{Simulation of QCD with N$_{f} =$ 2 $+$ 1 flavors of non-perturbatively improved Wilson fermions}",
 eprint = "1411.3982",
 archivePrefix = "arXiv",
 primaryClass = "hep-lat",
 reportNumber = "DESY-14-216, FTUAM-14-48, HIM-2014-01, HU-EP-14-51, MITP-14-091, SFB-CPP-14-89, IFT-UAM-CSIC-14-117",
 doi = "10.1007/JHEP02(2015)043",
 journal = "JHEP",
 volume = "02",
 pages = "043",
 year = "2015"
}

@article{Hansen:2015zga,
 author = "Hansen, Maxwell T. and Sharpe, Stephen R.",
 title = "{Expressing the three-particle finite-volume spectrum in
 terms of the three-to-three scattering amplitude}",
 journal = "Phys. Rev.",
 volume = "D92",
 year = "2015",
 number = "11",
 pages = "114509",
 doi = "10.1103/PhysRevD.92.114509",
 eprint = "1504.04248",
 archivePrefix = "arXiv",
 primaryClass = "hep-lat",
 SLACcitation = "%%CITATION = ARXIV:1504.04248;%%"
}

@article{Briceno:2017tce,
 author = "Brice\~no, R. A. and Hansen, Maxwell T. and Sharpe,
 Stephen R.",
 title = "{Relating the finite-volume spectrum and the
 two-and-three-particle $S$ matrix for relativistic systems
 of identical scalar particles}",
 journal = "Phys. Rev.",
 volume = "D95",
 year = "2017",
 number = "7",
 pages = "074510",
 doi = "10.1103/PhysRevD.95.074510",
 eprint = "1701.07465",
 archivePrefix = "arXiv",
 primaryClass = "hep-lat",
 reportNumber = "JLAB-THY-17-2400",
 SLACcitation = "%%CITATION = ARXIV:1701.07465;%%"
}

@article{Hammer:2017uqm,
 author = "Hammer, Hans-Werner and Pang, Jin-Yi and Rusetsky, A.",
 title = "{Three-particle quantization condition in a finite
 volume: 1. The role of the three-particle force}",
 journal = "JHEP",
 volume = "09",
 year = "2017",
 pages = "109",
 doi = "10.1007/JHEP09(2017)109",
 eprint = "1706.07700",
 archivePrefix = "arXiv",
 primaryClass = "hep-lat",
 SLACcitation = "%%CITATION = ARXIV:1706.07700;%%"
}

@article{Hammer:2017kms,
 author = "Hammer, H. -W. and Pang, J. -Y. and Rusetsky, A.",
 title = "{Three particle quantization condition in a finite
 volume: 2. General formalism and the analysis of data}",
 journal = "JHEP",
 volume = "10",
 year = "2017",
 pages = "115",
 doi = "10.1007/JHEP10(2017)115",
 eprint = "1707.02176",
 archivePrefix = "arXiv",
 primaryClass = "hep-lat",
 SLACcitation = "%%CITATION = ARXIV:1707.02176;%%"
}

@article{Mai:2017bge,
 author = "Mai, M. and {D\"{o}ring}, M.",
 title = "{Three-body Unitarity in the Finite Volume}",
 journal = "Eur. Phys. J.",
 volume = "A53",
 year = "2017",
 number = "12",
 pages = "240",
 doi = "10.1140/epja/i2017-12440-1",
 eprint = "1709.08222",
 archivePrefix = "arXiv",
 primaryClass = "hep-lat",
 reportNumber = "JLAB-THY-17-2554",
 SLACcitation = "%%CITATION = ARXIV:1709.08222;%%"
}

@article{Blanton:2019igq,
 author = "Blanton, Tyler D. and Romero-L\'opez, Fernando and Sharpe,
 Stephen R.",
 title = "{Implementing the three-particle quantization condition
 including higher partial waves}",
 journal = "JHEP",
 volume = "03",
 year = "2019",
 pages = "106",
 doi = "10.1007/JHEP03(2019)106",
 eprint = "1901.07095",
 archivePrefix = "arXiv",
 primaryClass = "hep-lat",
 SLACcitation = "%%CITATION = ARXIV:1901.07095;%%"
}

@article{Hansen:2019nir,
 author = "Hansen, Maxwell T. and Sharpe, Stephen R.",
 title = "{Lattice QCD and Three-particle Decays of Resonances}",
 journal = "Ann. Rev. Nucl. Part. Sci.",
 volume = "69",
 year = "2019",
 pages = "65-107",
 doi = "10.1146/annurev-nucl-101918-023723",
 eprint = "1901.00483",
 archivePrefix = "arXiv",
 primaryClass = "hep-lat",
 SLACcitation = "%%CITATION = ARXIV:1901.00483;%%"
}

@article{Briceno:2019muc,
 author = {Brice\~no, Ra\'ul A. and Hansen, Maxwell T. and Sharpe,
 Stephen R. and Szczepaniak, Adam P.},
 title = "{Unitarity of the infinite-volume three-particle
 scattering amplitude arising from a finite-volume
 formalism}",
 journal = "Phys. Rev.",
 volume = "D100",
 year = "2019",
 number = "5",
 pages = "054508",
 doi = "10.1103/PhysRevD.100.054508",
 eprint = "1905.11188",
 archivePrefix = "arXiv",
 primaryClass = "hep-lat",
 reportNumber = "JLAB-THY-19-2945, CERN-TH-2019-078",
 SLACcitation = "%%CITATION = ARXIV:1905.11188;%%"
}

@article{Romero-Lopez:2019qrt,
 author = {Romero-L\'opez, Fernando and Sharpe, Stephen R. and
 Blanton, Tyler D. and Brice\~no, Ra\'ul A. and Hansen,
 Maxwell T.},
 title = "{Numerical exploration of three relativistic particles in
 a finite volume including two-particle resonances and
 bound states}",
 journal = "JHEP",
 volume = "10",
 year = "2019",
 pages = "007",
 doi = "10.1007/JHEP10(2019)007",
 eprint = "1908.02411",
 archivePrefix = "arXiv",
 primaryClass = "hep-lat",
 reportNumber = "JLAB-THY-19-3011, CERN-TH-2019-129",
 SLACcitation = "%%CITATION = ARXIV:1908.02411;%%"
}

@article{Blanton:2019vdk,
 author = "Blanton, Tyler D. and Romero-L\'opez, Fernando and Sharpe,
 Stephen R.",
 title = "{$I = 3$ three-pion scattering amplitude from lattice
 QCD}",
 journal = "Phys. Rev. Lett.",
 volume = "124",
 year = "2020",
 number = "3",
 pages = "032001",
 doi = "10.1103/PhysRevLett.124.032001",
 eprint = "1909.02973",
 archivePrefix = "arXiv",
 primaryClass = "hep-lat",
 SLACcitation = "%%CITATION = ARXIV:1909.02973;%%"
}

@article{Hansen:2020zhy,
 author = {Hansen, Maxwell T. and Romero-L\'opez, Fernando and Sharpe, Stephen R.},
 title = "{Generalizing the relativistic quantization condition to include all three-pion isospin channels}",
 eprint = "2003.10974",
 archivePrefix = "arXiv",
 primaryClass = "hep-lat",
 reportNumber = "CERN-TH-2020-045",
 doi = "10.1007/JHEP07(2020)047",
 journal = "JHEP",
 volume = "07",
 pages = "047",
 year = "2020"
}

@article{Blanton:2020gha,
 author = "Blanton, Tyler D. and Sharpe, Stephen R.",
 title = "{Alternative derivation of the relativistic three-particle quantization condition}",
 eprint = "2007.16188",
 archivePrefix = "arXiv",
 primaryClass = "hep-lat",
 doi = "10.1103/PhysRevD.102.054520",
 journal = "Phys. Rev. D",
 volume = "102",
 number = "5",
 pages = "054520",
 year = "2020"
}

@article{Hansen:2020otl,
 author = {Hansen, Maxwell T. and Brice\~no, Ra\'ul A. and Edwards, Robert G. and Thomas, Christopher E. and Wilson, David J.},
 collaboration = "Hadron Spectrum",
 title = "{Energy-Dependent $\pi^+ \pi^+ \pi^+$ Scattering Amplitude from QCD}",
 eprint = "2009.04931",
 archivePrefix = "arXiv",
 primaryClass = "hep-lat",
 reportNumber = "CERN-TH-2020-147, JLAB-THY-20-3242",
 doi = "10.1103/PhysRevLett.126.012001",
 journal = "Phys. Rev. Lett.",
 volume = "126",
 pages = "012001",
 year = "2021"
}

@article{Blanton:2020gmf,
 author = "Blanton, Tyler D. and Sharpe, Stephen R.",
 title = "{Relativistic three-particle quantization condition for nondegenerate scalars}",
 eprint = "2011.05520",
 archivePrefix = "arXiv",
 primaryClass = "hep-lat",
 doi = "10.1103/PhysRevD.103.054503",
 journal = "Phys. Rev. D",
 volume = "103",
 number = "5",
 pages = "054503",
 year = "2021"
}

@article{Mai:2021lwb,
 author = {Mai, Maxim and D\"oring, Michael and Rusetsky, Akaki},
 title = "{Multi-particle systems on the lattice and chiral extrapolations: a brief review}",
 eprint = "2103.00577",
 archivePrefix = "arXiv",
 primaryClass = "hep-lat",
 reportNumber = "JLAB-THY-21-3327",
 doi = "10.1140/epjs/s11734-021-00146-5",
 journal = "Eur. Phys. J. ST",
 volume = "230",
 number = "6",
 pages = "1623--1643",
 year = "2021"
}

@article{Blanton:2021mih,
 author = "Blanton, Tyler D. and Sharpe, Stephen R.",
 title = "{Three-particle finite-volume formalism for $\pi^+\pi^+K^+$ and related systems}",
 eprint = "2105.12094",
 archivePrefix = "arXiv",
 primaryClass = "hep-lat",
 doi = "10.1103/PhysRevD.104.034509",
 journal = "Phys. Rev. D",
 volume = "104",
 number = "3",
 pages = "034509",
 year = "2021"
}

@article{Mai:2021nul,
 author = {Mai, Maxim and Alexandru, Andrei and Brett, Ruair\'\i{} and Culver, Chris and D\"oring, Michael and Lee, Frank X. and Sadasivan, Daniel},
 collaboration = "GWQCD",
 title = "{Three-Body Dynamics of the a1(1260) Resonance from Lattice QCD}",
 eprint = "2107.03973",
 archivePrefix = "arXiv",
 primaryClass = "hep-lat",
 doi = "10.1103/PhysRevLett.127.222001",
 journal = "Phys. Rev. Lett.",
 volume = "127",
 number = "22",
 pages = "222001",
 year = "2021"
}

@article{Muller:2021uur,
    author = {M\"uller, Fabian and Pang, Jin-Yi and Rusetsky, Akaki and Wu, Jia-Jun},
    title = "{Relativistic-invariant formulation of the NREFT three-particle quantization condition}",
    eprint = "2110.09351",
    archivePrefix = "arXiv",
    primaryClass = "hep-lat",
    doi = "10.1007/JHEP02(2022)158",
    journal = "JHEP",
    volume = "02",
    pages = "158",
    year = "2022"
}

@article{Blanton:2021eyf,
 author = "Blanton, Tyler D. and Romero-L\'opez, Fernando and Sharpe, Stephen R.",
 title = "{Implementing the three-particle quantization condition for \ensuremath{\pi}$^{+}$\ensuremath{\pi}$^{+}$K$^{+}$ and related systems}",
 eprint = "2111.12734",
 archivePrefix = "arXiv",
 primaryClass = "hep-lat",
 reportNumber = "MIT-CTP/5360",
 doi = "10.1007/JHEP02(2022)098",
 journal = "JHEP",
 volume = "02",
 pages = "098",
 year = "2022"
}

@inproceedings{Romero-Lopez:2021zdo,
 author = "Romero-L\'opez, Fernando",
 title = "{Three-particle scattering amplitudes from lattice QCD}",
 booktitle = "{19th International Conference on Hadron Spectroscopy and Structure}",
 eprint = "2112.05170",
 archivePrefix = "arXiv",
 primaryClass = "hep-lat",
 reportNumber = "MIT-CTP/5375",
 month = "12",
 year = "2021"
}

@article{Jackura:2022gib,
    author = "Jackura, Andrew W.",
    title = "{Three-body scattering and quantization conditions from S-matrix unitarity}",
    eprint = "2208.10587",
    archivePrefix = "arXiv",
    primaryClass = "hep-lat",
    reportNumber = "JLAB-THY-22-3664",
    doi = "10.1103/PhysRevD.108.034505",
    journal = "Phys. Rev. D",
    volume = "108",
    number = "3",
    pages = "034505",
    year = "2023"
}

@article{Dawid:2023jrj,
    author = "Dawid, Sebastian M. and Islam, Md Habib E. and Brice\~no, Ra\'ul A.",
    title = "{Analytic continuation of the relativistic three-particle scattering amplitudes}",
    eprint = "2303.04394",
    archivePrefix = "arXiv",
    primaryClass = "nucl-th",
    doi = "10.1103/PhysRevD.108.034016",
    journal = "Phys. Rev. D",
    volume = "108",
    number = "3",
    pages = "034016",
    year = "2023"
}

@article{Baeza-Ballesteros:2023ljl,
    author = {Baeza-Ballesteros, Jorge and Bijnens, Johan and Husek, Tom\'a\v{s} and Romero-L\'opez, Fernando and Sharpe, Stephen R. and Sj\"o, Mattias},
    title = "{The isospin-3 three-particle K-matrix at NLO in ChPT}",
    eprint = "2303.13206",
    archivePrefix = "arXiv",
    primaryClass = "hep-ph",
    reportNumber = "LU TP 23-03, MIT-CTP/5541",
    doi = "10.1007/JHEP05(2023)187",
    journal = "JHEP",
    volume = "05",
    pages = "187",
    year = "2023"
}

@article{Draper:2023xvu,
    author = "Draper, Zachary T. and Hansen, Maxwell T. and Romero-L\'opez, Fernando and Sharpe, Stephen R.",
    title = "{Three relativistic neutrons in a finite volume}",
    eprint = "2303.10219",
    archivePrefix = "arXiv",
    primaryClass = "hep-lat",
    reportNumber = "MIT-CTP/5539",
    doi = "10.1007/JHEP07(2023)226",
    journal = "JHEP",
    volume = "07",
    pages = "226",
    year = "2023"
}

@article{Dawid:2023kxu,
    author = "Dawid, Sebastian M. and Islam, Md Habib E. and Briceno, Raul A. and Jackura, Andrew W.",
    title = "{Evolution of Efimov states}",
    eprint = "2309.01732",
    archivePrefix = "arXiv",
    primaryClass = "nucl-th",
    doi = "10.1103/PhysRevA.109.043325",
    journal = "Phys. Rev. A",
    volume = "109",
    number = "4",
    pages = "043325",
    year = "2024"
}

@article{Jackura:2023qtp,
    author = "Jackura, Andrew W. and Brice\~no, Ra\'ul A.",
    title = "{Partial-wave projection of the one-particle exchange in three-body scattering amplitudes}",
    eprint = "2312.00625",
    archivePrefix = "arXiv",
    primaryClass = "hep-ph",
    doi = "10.1103/PhysRevD.109.096030",
    journal = "Phys. Rev. D",
    volume = "109",
    number = "9",
    pages = "096030",
    year = "2024"
}

@article{Baeza-Ballesteros:2024mii,
    author = {Baeza-Ballesteros, Jorge and Bijnens, Johan and Husek, Tom\'a\v{s} and Romero-L\'opez, Fernando and Sharpe, Stephen R. and Sj\"o, Mattias},
    title = "{The three-pion K-matrix at NLO in ChPT}",
    eprint = "2401.14293",
    archivePrefix = "arXiv",
    primaryClass = "hep-ph",
    reportNumber = "MIT-CTP/5670",
    doi = "10.1007/JHEP03(2024)048",
    journal = "JHEP",
    volume = "03",
    pages = "048",
    year = "2024"
}

@article{Hansen:2024ffk,
    author = "Hansen, Maxwell T. and Romero-L\'opez, Fernando and Sharpe, Stephen R.",
    title = "{Incorporating $DD\pi$ effects and left-hand cuts in lattice QCD studies of the $T_{cc}^+(3875)$}",
    eprint = "2401.06609",
    archivePrefix = "arXiv",
    primaryClass = "hep-lat",
    reportNumber = "MIT-CTP/5667",
    doi = "10.1007/JHEP06(2024)051",
    journal = "JHEP",
    volume = "06",
    pages = "051",
    year = "2024"
}

@article{Briceno:2024txg,
    author = "Brice\~no, Raul A. and Jackura, Andrew W. and Pefkou, Dimitra A. and Romero-L\'opez, Fernando",
    title = "{Electroweak three-body decays in the presence of two- and three-body bound states}",
    eprint = "2402.12167",
    archivePrefix = "arXiv",
    primaryClass = "hep-lat",
    reportNumber = "MIT-CTP/5683",
    doi = "10.1007/JHEP05(2024)279",
    journal = "JHEP",
    volume = "05",
    pages = "279",
    year = "2024"
}

@article{Draper:2024qeh,
    author = "Draper, Zachary T. and Sharpe, Stephen R.",
    title = "{Three-particle formalism for multiple channels: the \ensuremath{\eta}\ensuremath{\pi}\ensuremath{\pi} + $ K\overline{K}\pi $ system in isosymmetric QCD}",
    eprint = "2403.20064",
    archivePrefix = "arXiv",
    primaryClass = "hep-ph",
    doi = "10.1007/JHEP07(2024)083",
    journal = "JHEP",
    volume = "07",
    pages = "083",
    year = "2024"
}

@article{Feng:2024wyg,
    author = {Feng, Yuchuan and Gil, Fernando and D{\"o}ring, Michael and Molina, Raquel and Mai, Maxim and Shastry, Vanamali and Szczepaniak, Adam},
    title = "{A unitary coupled-channel three-body amplitude with pions and kaons}",
    eprint = "2407.08721",
    archivePrefix = "arXiv",
    primaryClass = "nucl-th",
    reportNumber = "JLAB-THY-24-4107",
    doi = "10.1103/PhysRevD.110.094002",
    journal = "Phys. Rev. D",
    volume = "110",
    pages = "094002",
    year = "2024"
}

@article{Dawid:2024dgy,
    author = "Dawid, Sebastian M. and Romero-L{\'o}pez, Fernando and Sharpe, Stephen R.",
    title = "{Finite- and infinite-volume study of $DD\pi$ scattering}",
    eprint = "2409.17059",
    archivePrefix = "arXiv",
    primaryClass = "hep-lat",
    reportNumber = "MIT-CTP/5774",
    doi = "10.1007/JHEP01(2025)060",
    journal = "JHEP",
    volume = "01",
    pages = "060",
    year = "2025"
}

@article{Briceno:2024ehy,
    author = "Brice{\~n}o, Ra{\'u}l A. and Costa, Caroline S. R. and Jackura, Andrew W.",
    title = "{Partial-wave projection of relativistic three-body amplitudes}",
    eprint = "2409.15577",
    archivePrefix = "arXiv",
    primaryClass = "hep-ph",
    reportNumber = "JLAB-THY-24-4199",
    doi = "10.1103/PhysRevD.111.036029",
    journal = "Phys. Rev. D",
    volume = "111",
    number = "3",
    pages = "036029",
    year = "2025"
}

@article{Severt:2022jtg,
    author = "Severt, Daniel and Mai, Maxim and Mei{\ss}ner, Ulf-G.",
    title = "{Particle-dimer approach for the Roper resonance in a finite volume}",
    eprint = "2212.02171",
    archivePrefix = "arXiv",
    primaryClass = "hep-lat",
    doi = "10.1007/JHEP04(2023)100",
    journal = "JHEP",
    volume = "04",
    pages = "100",
    year = "2023"
}

\end{document}